\documentclass[12pt]{article}
\usepackage{setspace}
\usepackage{geometry}
\usepackage{color}
\usepackage{booktabs}
\usepackage{color}
\usepackage{amsfonts}
\usepackage{amssymb}
\usepackage{amsmath}
\usepackage[mathscr]{eucal}
\usepackage{amsfonts}
\usepackage{amsmath,amsxtra,latexsym,amsthm, amssymb, amscd}
\usepackage[colorlinks,citecolor=blue,filecolor=black,linkcolor=blue,urlcolor=black]{hyperref}
\usepackage{pgfplots}
\usepackage{url}
\usepackage[round]{natbib}
\usepackage{fancyhdr}
\usepackage{tcolorbox}
\usepackage{graphicx}
\usepackage{hyperref}
\usepackage{algpseudocode}
\usepackage{caption}
\usepackage{subcaption}
\usepackage{epstopdf}
\usepackage{soul}
\usepackage{titlesec}
\usepackage[T1]{fontenc}
\usepackage[mathscr]{eucal}
\usepackage{amsfonts}
\usepackage{amsmath,amsxtra,latexsym,amsthm, amssymb, amscd}
\usepackage[round]{natbib}
\usepackage{fancyhdr}
\usepackage{graphicx}
\usepackage{lipsum}
\usepackage{filecontents}
\usepackage[latin1]{inputenc}
\usepackage{lmodern}
\usepackage[none]{hyphenat}
\usepackage{fancyhdr}

\usepackage[inline,shortlabels]{enumitem}
\setlist[enumerate,1]{label=(\roman*)}

\newtheorem{ex}{Example}
\newcommand{\bex}{\begin{ex}}
\newcommand{\eex}{\end{ex}}
\newtheorem{remark}{Remark}
\newtheorem{theorem}{Theorem}
\newtheorem{definition}{Definition}
\newtheorem{lemma}{Lemma}

\newtheorem{proposition}{Proposition}
\newtheorem{assum}{Assumption}
\newtheorem{corollary}{Corollary}

\newcommand{\rr}{\mathbb{R}}

\newtheorem{example}{Example}

\begin{document}

\title{To Bubble or Not to Bubble: Asset Price Dynamics and Optimality in OLG Economies\footnote{We thank Gaetano Bloise, Tomohiro Hirano, Alexis Akira Toda, Olivier Bochet (Editor in Chief) and anonymous referees for their helpful comments and discussions.}}
\author{Stefano BOSI\thanks{
Universit\'e d'Evry, Universit\'e Paris-Saclay. Email: 
stefano.bosi@univ-evry.fr} \and Cuong LE VAN\thanks{CNRS, PSE, IPAG Business School, FTU Hanoi. Email: Cuong.Le-Van@univ-paris1.fr} \and Ngoc-Sang PHAM\thanks{%
EM Normandie Business School, M\'etis 
Lab. Email: npham@em-normandie.fr}}

\date{\today}
\maketitle

\begin{abstract}
We study an overlapping generations (OLG) exchange economy with an asset that yields dividends. First, we derive general conditions, based on exogenous parameters, that give rise to three distinct scenarios: (1) only bubbleless equilibria exist, (2) a bubbleless equilibrium coexists with a continuum of bubbly equilibria, and (3) all equilibria are bubbly. Under stationary endowments and standard assumptions, we provide a complete characterization of the equilibrium set and the associated asset price dynamics. In this setting, a bubbly equilibrium exists if and only if the interest rate in the economy without the asset is strictly lower than the population growth rate and the sum of per capita dividends is finite. Second, we establish necessary and sufficient conditions for Pareto optimality. Finally, we explore the relationship between asset price behaviors and the optimality of equilibria. 
\end{abstract}

\section{Introduction}

Asset valuation and its welfare consequences are long-standing questions in economics. In a seminal contribution, \citet{tirole85} studies rational bubbles on a dividend-paying asset in an overlapping-generations (OLG) economy and relates asset-price dynamics to the Pareto optimality of equilibrium allocations. Following the traditional literature \citep{tirole82,tirole85,sw97}, a rational bubble on a dividend-paying asset is said to exist when its market price exceeds its fundamental value, defined as the present discounted value of future dividends.\footnote{See \citet{bo13} and \citet{miao14} for surveys of bubbles in general, \citet{mv18} and \citet{ht24a} for surveys of rational bubbles, and \citet{ht24b} for a survey of rational bubbles attached to assets with positive dividends.} An equilibrium is called \textit{bubbly} (\textit{bubbleless}) if a bubble exists (does not exist). It is called \textit{asymptotically bubbly} if it is bubbly and  its bubble component does not converge to zero over time.

Proposition~1 in \citet{tirole85} identifies three classical equilibrium regimes. To state its main economic insights, let $G_n$ and $R^*$ denote, respectively, the gross population growth factor and the steady-state interest rate in the economy without the asset. With the constant dividend assumed by Tirole, the three regimes can be summarized as follows:
\begin{enumerate}
[(a)] \item No bubbly equilibrium exists if $1<G_n<R^*$.\footnote{Our notation $G_n$ corresponds to $1+n$ in \citet{tirole85} and to $n$ in \citet[Chapter~9]{ls18}.} 
\item A continuum of equilibria, including both bubbly and bubbleless equilibria, exists if $1<R^*<G_n$. 
\item There exists a unique equilibrium (and this is asymptotically bubbly) if $R^*<1<G_n$.\footnote{In this case, \citet[p.~1506]{tirole85} writes that ``bubbles are necessary for the existence  of an equilibrium in an economy in which there exists an (arbitrarily small) rent.'' See also \citet[Section~4.3]{araujo_etal-wariness2011} for a discussion     of the necessity of bubbles for equilibrium implementation.} \end{enumerate}

Tirole also connects these equilibrium regimes to welfare. Proposition~2 in \citet{tirole85} claims that, in the low interest rate case, asymptotically bubbleless equilibria are Pareto inefficient whereas the asymptotically bubbly equilibrium is Pareto optimal. A formal proof, however, is not provided.\footnote{Indeed, \citet[p.~1526]{tirole85} writes: ``By (a straightforward extension of) Theorem 5.6 in Balasko-Shell [3], the asymptotically bubbleless equilibria are inefficient and the asymptotically bubbly one is efficient.''} Tirole's equilibrium classification and welfare conclusions therefore provide the natural benchmarks for the two questions studied in this paper.

Much of the subsequent literature has focused on pure bubbles, namely assets that pay no dividends but nevertheless have positive prices. A smaller literature studies bubbles attached to dividend-paying assets. Recent work has clarified several aspects of Tirole's original analysis. In production economies, introducing a dividend-paying asset may generate equilibrium paths along which capital converges to zero \citep{bhlpp18,phamtoda2025,phamtodaEcta}. Such capital-collapse equilibria are not covered by Tirole's original classification, and \citet{phamtoda2025,phamtodaEcta} provide a global analysis that incorporates them. However, these analyses still require some restrictions (for instance, the saving function is increasing in the capital return).

Recent contributions have also provided new formal results for particular versions of Tirole's bubble-necessity regime. In an OLG exchange economy, \citet[Section~IV]{hiranotoda25} establish a Bubble Necessity Theorem under convergence assumptions on endowment growth and forward-rate functions. Related results are obtained by \citet[Section~5]{ht24b} under constant growth rates and homogeneous preferences. These results provide useful formal extensions of the classical bubble-necessity insight, but leave open how Tirole's three regimes can be characterized when preferences, endowments, and dividends vary over time and need not converge.

A second open issue concerns welfare. There is no general consensus that the presence of a bubble is equivalent to Pareto inefficiency. Indeed, \citet[p.~47]{sw97} observe that ``the connection between inefficiency and the existence of pricing bubbles is thus a rather loose one.''\footnote{See also \citet{sw93}, \citet{tirole12}, \citet{Geerolf18}, and the references therein. We thank Manuel Santos for sending us a copy of \cite{sw93} where they suggest, on pages 40-41, that it may not be correct that Pareto optimality implies the absence of bubbles.} This raises two related questions. First, how can Tirole's three classical equilibrium regimes be characterized under time-varying primitives? Second, how are bubble formation and Pareto optimality related?

We address these questions in a two-period OLG pure-exchange economy with a long-lived asset in fixed positive supply.\footnote{See \cite{sw97}, \cite{kocherlakota92}, \cite{hw00}, \cite{araujo_etal-wariness2011}, \cite{werner14}, \cite{bc19}, \cite{blp22} among others for rational bubbles of dividend-paying assets in models with infinitely-lived agents.  See \cite{pham2024} and  the references therein for a formal connection between OLG models and models with infinitely lived agents.}  The asset pays a possibly nonstationary stream of dividends, and preferences and young- and old-age endowments may also vary over time. The pure-exchange framework is deliberate. By abstracting from capital accumulation, the model isolates the intergenerational asset-trading mechanism and allows us to derive global results under time-dependent primitives.

Our first contribution is to revisit the three classical equilibrium regimes under these general nonstationary primitives. A central device is the benchmark economy without the long-lived asset. Its interest rate represents households' willingness to transfer resources across periods before the asset is introduced. The interaction among this benchmark interest rate, the growth of aggregate resources, and dividends plays a crucial role for characterizing the different regimes.

We first provide conditions under which every equilibrium is bubbleless, thereby extending the logic of Tirole's regime~(a) to a nonstationary environment. We also establish the existence of a bubbleless equilibrium when dividends are sufficiently small in present-value terms.

We then study co-existence, corresponding to Tirole's regime~(b). We identify conditions under which the same fundamentals support a bubbleless equilibrium and a continuum of bubbly equilibria. The result is constructive and does not rely on convergence to a steady state. Economically, co-existence arises when interest rates are low enough relative to population growth to sustain positive demand for the asset, while remaining sufficiently high relative to dividend growth for the fundamental component of the asset price to remain small. The result accommodates time-dependent, nonseparable preferences together with nonstationary endowments and dividends.

We next turn to bubble necessity, corresponding to Tirole's regime~(c). The central issue is whether demand for the long-lived asset can become asymptotically negligible. Our Condition~(B) is a local no-collapse condition on asset demand. It restricts equilibrium interest rates when households devote only a small fraction of their resources to the asset. Roughly speaking, if asset demand becomes small, equilibrium interest rates remain sufficiently low relative to the growth of the resources available to future buyers, while dividends are not too small relative to these interest rates. If asset demand were to collapse, the resulting low interest rates would make the fundamental value of dividends too large to be compatible with a finite equilibrium asset price. Asset demand must therefore remain economically significant.

Once this no-collapse property is established, the bubble classification becomes particularly simple. Whether every equilibrium is bubbly or bubbleless is determined by the cumulative size of dividends relative to the resources of the generations purchasing the asset. Sufficiently small dividends imply that every equilibrium is bubbly, whereas sufficiently large dividends imply that every equilibrium is bubbleless.

Our bubble-necessity result contains the recent result of \citet[Theorem 2]{hiranotoda25} as an important special case. We provide a more general characterization of when this classical mechanism operates. In particular, our approach replaces convergence requirements in \citet[Theorem 2]{hiranotoda25} with local restrictions on equilibrium behavior and yields a summability-based classification. It also covers boundary cases in which long-run growth-rate comparisons alone are inconclusive.

Under stationary endowments and time-independent separable preferences, we go further and provide a global characterization of the equilibrium set. We obtain a necessary and sufficient condition for the existence of a bubbly equilibrium, and characterize the long-run behavior of every equilibrium. Depending on fundamentals, there is either a unique bubbleless equilibrium, co-existence of a bubbleless equilibrium with a continuum of bubbly equilibria, or a unique bubbly equilibrium. We also characterize the boundary case in which the benchmark interest rate equals population growth. Thus, the stationary analysis provides an exchange-economy counterpart to Tirole's three-regime classification while allowing a general time-varying dividend sequence.

Our second contribution concerns Pareto optimality. Building on \citet{cass72}, \citet{bs80}, and \citet{OkunoZilcha1980}, we derive a sufficient condition for Pareto optimality and establish a converse under complementary regularity conditions. The analysis accommodates population growth and potentially unbounded endowments and allows us to study welfare and asset-price bubbles within the same framework.

Combining the pricing and welfare results yields a key economic message of the paper: bubble formation and Pareto optimality are distinct endogenous properties. A bubble concerns whether the equilibrium asset price can be fully accounted for by the present value of future dividends.  Pareto optimality instead concerns whether resources can be reallocated across generations so as to make some generations better off without making others worse off. Although both properties depend on equilibrium interest rates and saving behavior, neither generally determines the other.

The role of asset demand is particularly informative. We show that an equilibrium can be Pareto optimal, whether bubbly or bubbleless, when the asset continues to absorb a significant amount of resources. Conversely, when several equilibria coexist, their welfare can be ranked by the initial asset value: equilibria with larger initial asset values yield higher utility to every generation. Consequently, a continuum of bubbly equilibria may be Pareto inefficient. The mere presence of a bubble is therefore neither a sufficient nor a necessary indicator of Pareto inefficiency.

In the stationary low-interest-rate case, however, our results formally establish the classical welfare conclusion suggested by \citet{tirole85}. When a continuum of equilibria exists, the maximal equilibrium is asymptotically bubbly and Pareto optimal, whereas every other equilibrium---including the bubbleless equilibrium and all remaining bubbly equilibria---is Pareto dominated. Thus, Tirole's welfare intuition is fully validated in this important benchmark environment. Beyond this benchmark, however, the relationship between bubbles and welfare is more nuanced: a Pareto-optimal equilibrium may be either bubbly or bubbleless, while an asymptotically bubbly equilibrium may be either Pareto optimal or Pareto dominated.

Our results are also related to \citet{Sorger2026}, \citet{toda25} and its substantially revised version \citet{hiranotoda25efficiency}.\footnote{The first version of our paper \citep{blp25}, posted on arXiv on August 5, 2025, already contained the main results of the present paper, including those in Sections~\ref{section-pareto} and~\ref{section-bubble-pareto}. \citet{toda25} and \citet{Sorger2026} became publicly available later, and we were not aware of their results at that time.}
\citet{Sorger2026} studies the inevitability of stock-price bubbles in production economies, emphasizing differences in the growth of wage and dividend income.  Our analysis is complementary to \citet{Sorger2026}. His mechanism is technological: decreasing returns generate positive profits, and bubbles arise when wage income grows faster---or contracts more slowly---than dividend income, a property related to violations of the Inada conditions and the failure of balanced growth. Our mechanism is instead based on the interaction among benchmark interest rates, dividend growth, and the growth of endowments, rather than through production technology. We also address a broader equilibrium question: rather than focusing on unique bubbly equilibria, we characterize bubblelessness, co-existence, and bubble necessity and, under stationary endowments, the entire equilibrium set. Finally, we study Pareto optimality and welfare ranking across equilibria, showing that bubble formation and Pareto optimality are logically distinct.

\citet{hiranotoda25efficiency} focus on necessary and efficient land bubbles and infinite debt rollover. Our analysis is complementary: in a pure-exchange economy with general nonstationary primitives, we characterize both bubbly and bubbleless equilibria and show that bubble formation and Pareto optimality are distinct properties. Their Proposition B.2 provides a sufficient condition for Pareto optimality under constant population and bounded endowments, whereas our Theorem~\ref{pareto-theorem} allows for population growth and unbounded endowments and provides a converse under additional regularity conditions. We also rank coexisting equilibria by welfare, showing that larger initial asset values strictly Pareto dominate smaller ones. These results complement the global equilibrium analysis of \citet{phamtoda2025,phamtodaEcta} in production economies.

The remainder of the paper is organized as follows. Section~\ref{sec.1} presents the OLG exchange economy and the definition of an asset-price bubble. Section~\ref{section-bubble} studies bubbleless equilibria, co-existence, bubble necessity, and the stationary characterization. Section~\ref{section-pareto} studies Pareto optimality. Section~\ref{section-bubble-pareto} examines the relationship between asset-price bubbles and Pareto optimality. Section~\ref{conclusion} concludes. Technical proofs are presented in Appendices.

\section{\bf An OLG exchange economy}

\label{sec.1}

We consider a two-period overlapping-generations (OLG) exchange economy with a 
dividend-paying long-lived asset. Time is discrete, 
$t\in\{0,1,2,\ldots\}$, and there is a single consumption good.

At each date $t\geq 0$, a new generation of $N_t$ households enters the economy. 
Population grows at a constant gross rate: ${N_{t+1}}/{N_t}=G_n>0$ $\forall t\geq 0.$\footnote{We assume a constant population growth factor for expositional simplicity. Most of our results can be extended to the case in which $N_{t+1}/N_t$ varies over time but remains appropriately bounded.} 
Each household born at date $t$ lives for two periods, young and old. It receives an exogenous endowment $e_t^y\geq 0$ when young and $e_{t+1}^o\geq 0$ when old. 
Preferences of a household born at date $t$ are represented by the utility function
$U^t(c_t^y,c_{t+1}^o)$, where $c_t^y$ and $c_{t+1}^o$ denote its consumption when young and old, respectively.

There is a long-lived asset, which can be interpreted as a Lucas tree 
\citep{lucas78}. The asset is in positive fixed supply and pays a dividend in units of the consumption good. If a household purchases one unit of the asset at date $t$ at  price $q_t$, then at date $t+1$ it receives the dividend $D_{t+1}$ and can resell the asset at price $q_{t+1}$.

Let $z_t$ denote the asset holdings of a household born at date $t$. Its budget 
constraints are $c_t^y+q_t z_t\leq e_t^y$, 
$c_{t+1}^o\leq e_{t+1}^o+(q_{t+1}+D_{t+1})z_t$, together with $c_t^y\geq 0,  c_{t+1}^o\geq 0.$\footnote{One could also introduce a pure bubble asset $x_t$, as in 
\citet{tirole85}, by writing $c_t^y+q_tz_t+p_tx_t=e_t^y,  c_{t+1}^o=
e_{t+1}^o+(q_{t+1}+D_{t+1})z_t+p_{t+1}x_t$. Whenever $p_t>0$, the no-arbitrage condition ${(q_{t+1}+D_{t+1})}/{q_t} = {p_{t+1}}/{p_t}$ implies that the two assets earn the same return. Introducing an additional pure 
bubble asset therefore does not affect the main mechanisms studied below.}

At date $0$, the generation born at date $-1$ is already old and therefore consumes  only at date $0$. Its consumption is $c_0^o=e_0^o+(q_0+D_0)z_{-1}$,  where the initial asset holding $z_{-1}>0$ is given.
\subsection{Intertemporal equilibrium}


\begin{definition}
\label{def}An intertemporal equilibrium is a list $(z_{t},(c_{t}^{y},c_{t}^{o}),q_{t})_{t\geq
0}$ satisfying three conditions: (1) given $(q_{t},q_{t+1})$, the allocation 
$(z_{t},c_{t}^{y},c_{t+1}^{o})$ is a solution to the household's problem, 
(2) markets clear: $N_{t}z_{t}=N_{t+1}z_{t+1} \text{  }\forall t\geq -1$,  $N_{t}c_{t}^{y}+N_{t-1}c_{t}^{o} =N_{t}e_{t}^{y}+N_{t-1}e_{t}^{o}+D
_{t}z_{t-1}N_{t-1} \text{ }\forall t\geq 0$, and (3) $q_{t}\geq 0 $ $\forall t\geq 0$. 
\end{definition}

We impose standard assumptions.
\begin{assum}
\label{assum0}$z_{-1}>0$, $z_{-1}N_{-1}=1$, $N_{t}=G_n^{t}$ $\forall t\geq 0$, where $G_n>0$.
\end{assum}
\begin{assum}
\label{assum1new}The function $U^t: \rr^2_+\to \rr$ is increasing in each component,  quasi-concave, continuously differentiable on $\rr^2_{++}$, $\lim_{x_1\to 0^+}U^t_{1}(x_1,x_2)=\infty$ $\forall x_2>0$, 
 where $U^t_{i}$ denotes the partial derivative of $U^t$ with respect to the $i^{th}$ component.  
The sequence of endowments satisfies $e_{t}^{y}>0,e_{t+1}^{o}\geq 0$
$\forall t\geq 0$.
\end{assum}
 In equilibrium, we have $z_{t}G_n^{t}=z_{t}N_{t}=z_{-1}N_{-1}=1.$ 
 Denote $a_t\equiv q_tz_t$ the asset value per capital and $d_{t}\equiv {D _{t}}/{G_n^{t}}$ the dividend per capita. 
Observe that if a list $(z_{t},(c_{t}^{y},c_{t}^{o}),q_{t})_{t\geq
0}$ is an equilibrium, then $z_t=1/G_n^t$ and
\begin{subequations}\label{iff_equilibrium}
\begin{align}
c_{t}^{y}+q_{t}z_{t}& =e_{t}^{y}, \quad c_{t+1}^{o}=e_{t+1}^{o}+(q_{t+1}+D _{t+1})z_{t}, \\
\label{euler-bis}
\frac{q_{t+1}+D_{t+1}}{q_t}&=\frac{U^{t}_1\big(e_{t}^{y}-q_{t}z_{t},e_{t+1}^{o}+(q_{t+1}+D _{t+1})z_{t}\big)}{U^{t}_2\big(e_{t}^{y}-q_{t}z_{t},e_{t+1}^{o}+(q_{t+1}+D _{t+1})z_{t}\big)}.
\end{align}
\end{subequations}


\begin{definition}
\label{definition-interest}

(1) Consider an equilibrium. Define the interest rate $R_{t+1}$ between dates $t$ and $t+1$ by $
R_{t+1}\equiv (q_{t+1}+D _{t+1})/q_t$.

(2) Define the benchmark interest rate (i.e., the interest rate of the economy without asset) $R_{t+1}^{\ast}$ between dates $t$ and $t+1$   by $
R_{t+1}^{\ast }\equiv  {U^t_1(e_{t}^{y},e_{t+1}^{o})}/{U^t_2(e_{t}^{y},e_{t+1}^{o})}.$
\end{definition}
To obtain a relationship between $R_t$ and $R_t ^*$, we introduce an additional assumption.
\begin{assum}\label{derivative-ij}${U^t_{1}(x_1,x_2)}/{U^t_2(x_1,x_2)}$ is decreasing in $x_1$ and increasing in $x_2$.
\end{assum}
By Assumption \ref{derivative-ij} and the Euler condition (\ref{euler-bis}), we obtain the following result which will be used in next sections.
\begin{lemma}\label{RtRt*}Under Assumptions \ref{assum0}, \ref{assum1new}, \ref{derivative-ij}, in equilibrium, we have $R_{t}\geq R_{t}^{\ast }$ $\forall t$.
\end{lemma}





\subsection{Asset price bubbles}
Given an equilibrium, by the definition of $(R_t)$, we have the  asset pricing equation \begin{align}\label{assetpricing}
q_t=\frac{q_{t+1}+D_{t+1}}{R_{t+1}}.
\end{align}
 This leads to the traditional definition of fundamental value and bubble:
\begin{align}
\label{ft}
F_t&\equiv \sum_{s=1}^{\infty}\frac{D_{t+s}}{R_{t+1}\cdots R_{t+s}}, \quad &B_t=q_t-F_t=\lim_{T\to\infty}\frac{q_T}{R_{t+1}\cdots R_T} \text{ }\forall t\geq 0. 
\end{align}
$F_t$ and $B_t$ are called the fundamental value $F_t$ and the bubble component $B_t$ of the asset  at date  $t\geq 0$. 
Denote 
\begin{align}
a_{t}&\equiv \frac{q_{t}}{G_n^{t}}, &&d_{t}\equiv \frac{D _{t}}{G_n^{t}},&& f_t=\frac{F_t}{G_n^t}, & &b_t=\frac{B_t}{G_n^t}, &&Q_t=\frac{1}{R_1\cdots R_t}.
\end{align} 
\begin{definition}Consider an equilibrium. It is said to be bubbly if the equilibrium  asset price $q_0$ strictly exceeds its fundamental value $F_0$.\footnote{Here, we follow, among others, \citet[page 1172]{tirole82}, \citet[footnote 8]{tirole85},  \citet[pages 249-250]{kocherlakota92},  \citet[pages 27-29]{sw97},  \citet[page 259]{hw00},  \citet[Section 6]{lvp16}, \citet[Section 4]{bhlpp18},  \citet[Section II]{hiranotoda25}.} A bubbly 
equilibrium is said to be asymptotically bubbly if $\liminf_{t\to\infty}{a_t}/{e_t^y}>0,$ and asymptotically bubbleless if $\lim_{t\to\infty}{a_t}/{e_t^y}=0.$
\end{definition}
We have the following characterization, which is very general because it is only based on the asset pricing equation (\ref{assetpricing}).
\begin{lemma}
\label{prop1}In the case of strictly positive dividends ($D _{t}>0$ for
all $t$), the following statements are equivalent. 
A bubble exists (i.e., $q_0>F_0$) if and only if one of the following conditions holds:
\begin{enumerate}
\item \label{prop1-2}  $\lim_{t\rightarrow \infty }Q_{t}q_{t}>0$, i.e.,  $\lim_{t\rightarrow
\infty }\dfrac{G_n^{t}a_{t}}{R_{1}\cdots R_{t}}>0$.
\item \label{prop1-5}  $\sum_{t=1}^{\infty }D _{t}/q_{t} =\sum_{t=1}^{\infty }d_{t}/a_{t}\leq {({F_0}/{q_0})}/{(1-{F_0}/{q_0})}<\infty$.
\item \label{prop1-3}  The sequence $({f_t}/{a_t})=({F_t}/{q_t})$  is strictly decreasing and converges to $0$. 

\item \label{prop1-4}  The sequence $({b_t}/{a_t})=({B_t}/{q_t})$  is strictly increasing and converges to $1$.

\end{enumerate}
\end{lemma}
\begin{proof}
See    Appendix \ref{assetprice-basis}.
\end{proof}
The first statement directly follows the definition of bubble. Condition $\sum_{t=1}^{\infty}\frac{D_t}{q_t}<\infty$ was first established in Proposition 7 in \citet{montrucchio04}.\footnote{This simple characterization is useful in some models (see, for instance, \cite{lvp16}, \cite{blp18,bhlpp18,blp22}, \cite{hiranotoda25, ht24b}, \cite{Sorger2026}).} We give a new proof and strengthen the result by proving that $\sum_{t=1}^{\infty}{D_t}/{q_t}\leq \frac{{F_0}/{q_0}}{1-{F_0}/{q_0}}$.

Points \ref{prop1-3} and \ref{prop1-4} are novel. The existence of an asset price bubble means that the ratio of the fundamental value to the asset price ${F_t}/{q_t}$ decreases monotonically to zero, whereas ${B_t}/{q_t}$ increases monotonically to one when $t$ tends to infinity.  Thus the fundamental component is asymptotically negligible relative to the bubble component, i.e., $\lim_{t\to\infty}{F_t}/{B_t}=0$.

\section{\bf Results on asset price bubbles}\label{section-bubble}
\subsection{Equilibrium without asset price bubbles}
\label{nobubble_section}
Before exploring the equilibrium with bubbles, it is essential to understand necessary conditions of asset price bubbles.
\begin{proposition}[bubbleless equilibrium]
\label{necessitycondition} {\noindent}
\begin{enumerate}
\item \label{necessitycondition-point1}
Let Assumptions \ref{assum0}, \ref{assum1new} be satisfied. Then, every  equilibrium is bubbleless if
\begin{align}  \label{keycond}
\sum_{t=1}^{\infty}\frac{D_t}{G_n^te^y_t}=\infty \quad  \text{(non-negligible dividend condition)}.
\end{align}

\item Let Assumptions \ref{assum0}, \ref{assum1new}, \ref{derivative-ij} be satisfied. 
\begin{enumerate}
    \item Every  equilibrium is bubbleless if 
\begin{align}
 \label{bubble1}
\lim_{t\rightarrow \infty }\frac{G_n^te^y_t}{R_{1}^{\ast }\cdots
R_{t}^{\ast }}=0 \quad \text{(high interest rate condition)}.
\end{align}
\item\label{existence} There exists a bubbleless equilibrium if
\begin{align}  \label{existence-bubbleles}
\sum_{t\geq 1}\frac{D _{t}}{R_{1}^{\ast }\cdots R_{t}^{\ast }}<\infty \quad  \text{(not-too-low interest rate condition).}
\end{align}
\end{enumerate}
\end{enumerate}
\end{proposition}
\begin{proof}
See Appendix \ref{nobubble_section_proof}.
\end{proof}
To easily see the basic intuitions of conditions  (\ref{keycond}) and (\ref{bubble1}), focus on a simple case where $D_t=D_0G_d^t$, $e^y_t=e^y_0G ^t$ and $R^*_t=R^*$ $\forall t$. In this case, \eqref{keycond} becomes $G_d\geq G_nG$ while \eqref{bubble1} becomes $R^*>G_nG$. So, (\ref{keycond}) and (\ref{bubble1}) suggest that the existence of a bubble requires a low dividend growth rate ($G_d<G_nG$) and a low interest rate ($R^*\leq G_nG$).

Condition (\ref{keycond})  is related to  Corollary 1 in \cite{bhlpp18}, which shows that in an OLG model with production that every equilibrium is bubbleless if the sum of the ratio of dividend to aggregate output is infinite.\footnote{\cite{phamtoda2025} prove a similar result in their Lemma 3.1 and Corollary 3 in  \cite{blp22} shows a similar result in a model with infinitely-lived agents.}
The insight of condition (\ref{bubble1}) is in line with the main result in \cite{sw97}: there is no bubble if the
sum of discounted values of aggregate outputs is finite. However, the
condition in \cite{sw97} is based on endogenous variables. By contrast, our condition (\ref{bubble1}) is based on exogenous variables.

Proposition \ref{necessitycondition}\ref{existence} states that there exists a bubbleless equilibrium if  the present discounted value of dividends computed with the interest rates of the economy without asset is finite. A key in Proposition \ref{necessitycondition}\ref{existence} is $R_t\geq R^*_t$. Proposition \ref{necessitycondition}\ref{existence} is similar to Proposition 3.1  in \cite{phamtoda2025}. However, as we work with the Euler equation (\ref{euler-bis}), which is a non-autonomous one-dimensional system, we can obtain a similar result under weaker assumptions (our Assumption \ref{derivative-ij} is slightly weaker than Assumption 3 in \cite{phamtoda2025}). 



In the next sections, we study conditions under which there exists a bubbly equilibrium.
\subsection{A continuum of equilibria (with and without bubbles)}
\label{continuum-section}
Since $z_{t}N_{t}=1$ $\forall t\geq 0$, the equilibrium $(z_{t},(c_{t}^{y},c_{t}^{o}),q_{t})_{t\geq 0}$ is one-to-one represented by
the sequence of prices $(q_{t})_{t\geq 0}$ (or the sequence of asset value $(a_t)=(q_tz_t)$) which we also call an
equilibrium.  
The Euler equation (\ref{euler-bis}) motivates us to introduce the following function.
\begin{definition}\label{definition-functionK}
Let $t\geq 0$, $e^y_t>0,e^o_{t+1}> 0$.  Define the function $K_t: (0,e^y_t)\times [0,\infty)\to \rr$  by  $K_t(a,R)\equiv U^t_1(e_{t}^{y}-a,e_{t+1}^{o}+Ra)-RU^t_2(e_{t}^{y}-a,e_{t+1}^{o}+Ra).$
\end{definition}
We now present conditions under which there exists a continuum of equilibria.
  
\begin{theorem}[Continuum of equilibria: co-existence of bubbly and bubbleless equilibria]
\label{new5continuum}
Let Assumptions \ref{assum0}, \ref{assum1new}, \ref{derivative-ij} be satisfied and $e^o_t,d_t>0$ $\forall t$. Assume also that $U^t$ is strictly concave and the following conditions hold:
\begin{enumerate}
\item\label{continuum-condition1} There exists a sequence $(\epsilon_t)_{t\geq 0}$ such that, for any $t\geq 0$, 
\begin{enumerate}
\item\label{1a} $\epsilon_t\in (0,e^y_t)$, $\epsilon_t\leq \epsilon_{t+1}+d_{t+1}$.
\item\label{1b} $K_t(\epsilon_t,G_n)< 0$.
\end{enumerate}
\item \label{continuum-condition2} There exist $\lambda >0$ and $\gamma >1+\frac{1}{%
\lambda }$ satisfying  $\lambda d_t<\epsilon_t$ and
\begin{align}\label{Rdd-notlow}\text{(Not-too-low interest rate condition):  } \quad  R_{t+1}^{\ast }\geq  \Big(G_n \frac{d_{t+1}}{d_{t}}\Big)\gamma
  \text{ } \forall t\geq 0.
\end{align}

\end{enumerate}
Then, there exists at least one  bubbleless equilibrium and  there exists a continuum of bubbly equilibria (where the asset values and interest rates satisfy $a_t\in [\lambda d_t,\epsilon_t]$ and $R^*_t\leq R_{t}\leq G_n$ $\forall t$).

\end{theorem}
\begin{proof}See Appendix \ref{prooftheorem1new}.\end{proof}


Conditions \ref{continuum-condition1} and \ref{continuum-condition2} in Theorem \ref{new5continuum} imply the following ordering:\footnote{Since $U^t$ is concave, the function $K_t$ is increasing in the first component. So, condition \ref{1b}   implies that $K_t(0,G_n)<0$, i.e., $R^*_{t+1}< G_n$.} 
\begin{align}
\dfrac{D_{t+1}}{D_t}<    \gamma \dfrac{D_{t+1}}{D_t}\leq R^*_{t+1}< G_n
\end{align}

Let us now explain our constructive proof and its economic intuition.  Starting from any initial asset value $a_0$ in the interval specified in the theorem, the Euler equation determines an equilibrium gross interest rate $R_1$ (which need not be unique). Given $R_1$, the asset-pricing equation determines next period's asset value according to $a_1+d_1=a_0{R_1}/{G_n}$. Repeating this argument recursively constructs an equilibrium path $(a_t,R_{t+1})_{t\geq 0}$. Condition \ref{1b} provides an upper bound on equilibrium interest rates. In particular, along the constructed path, $R_{t+1}<G_n$. Economically, the fact that the equilibrium return remains below population growth allows the asset to be transferred to successively larger cohorts while keeping asset expenditure per young household within the upper bound $\epsilon_t$. Thus, the low interest rate condition keeps the constructed equilibrium path inside the desired corridor.
 
 At the same time, condition \eqref{Rdd-notlow} prevents equilibrium returns from being too low relative to dividend growth. Since $R_{t+1}\geq R_{t+1}^{*}$ and \eqref{Rdd-notlow}, the asset pricing equation yields \begin{align} \frac{a_{t+1}}{d_{t+1}} &= \frac{R_{t+1}}{G_n} \frac{d_t}{d_{t+1}} \frac{a_t}{d_t}-1\geq \gamma\frac{a_t}{d_t}-1. \label{asset-dividend-growth} \end{align} 
 Because $\gamma>1$ and the initial asset value is chosen so that $a_0/d_0>1/(\gamma-1)$, the asset-value-to-dividend ratio grows at least geometrically. Consequently, $\sum_{t\geq 1}{d_t}/{a_t}<\infty$, and Lemma \ref{prop1} implies that the constructed equilibrium is bubbly. 
 
 Finally, condition \eqref{Rdd-notlow} plays a second role. It implies that the present value of dividends discounted at the benchmark interest rates is finite. Hence, Proposition \ref{necessitycondition}\ref{existence} guarantees the existence of a bubbleless equilibrium. Theorem \ref{new5continuum} therefore identifies a co-existence regime: the same fundamentals support a bubbleless equilibrium and a continuum of bubbly equilibria.

To the best of our knowledge, Theorem \ref{new5continuum} is new to the literature. It provides general conditions under which bubbleless and bubbly equilibria coexist, while allowing for non-separable preferences, non-stationary endowments, and non-stationary dividends. Notably, neither the statement nor the proof of Theorem \ref{new5continuum} relies on convergence assumptions on the underlying variables, in contrast to some existing results in the literature \citep{tirole85, tirole12, phamtoda2025}. This makes the result applicable to environments in which endowments, dividends, and equilibrium variables may exhibit non-convergent dynamics.

A further contribution of Theorem \ref{new5continuum} is its constructive nature: its proof provides an explicit procedure for constructing bubbly equilibria alongside a bubbleless equilibrium. Thus, the theorem establishes not only the co-existence of bubbleless and bubbly equilibria, but also a method for generating the latter.\footnote{In a different environment, Proposition 7 of \citet{blp22} provides conditions for the existence of a continuum of bubbly equilibria in an exchange economy with heterogeneous infinitely lived agents and logarithmic utility. Their working-paper version \citep{blp21} presents additional examples.}


We illustrate Theorem \ref{new5continuum} by focusing on separable utility functions, where the saving function is increasing in the interest rate (which is similar to that in \cite{tirole85}, \cite{bhlpp18}, \cite{phamtoda2025,phamtodaEcta}). 
\begin{assum}\label{assum1} Assume that $U^t(x_1,x_2)=u(x_1)+ v(x_2)$ $\forall t,x_1,x_2$. The functions $u,v: \rr_+\to\rr$ are twice continuously differentiable, strictly increasing, strictly concave, $u^{\prime }(0)=v^{\prime }(0)=\infty $. The endowments satisfy  $e_{t}^{y}>0,e_{t+1}^{o}> 0$ $\forall t\geq 0$. The function $c \mapsto cv'(c)$ is increasing on $(0,\infty)$.\end{assum}

Under this assumption, $(q_t)$ is an equilibrium if and only if the sequence 
$(a_t,R_{t+1})_{t\geq 0}$ satisfies the following conditions:
\begin{align}\label{system1}
u^{\prime }(e_{t}^{y}-a_{t}) = R_{t+1}v^{\prime }(
e_{t+1}^{o}+R_{t+1}a_{t}), \text{ } 
a_{t+1}+d_{t+1} =a_{t}\frac{R_{t+1}}{G_n}, \text{ } 0<a_{t} <e_{t}^{y} \text{  }\forall t\geq 0.
\end{align}
Since $(R_{t+1})_{t\geq 0}$ is uniquely determined via $(a_t)$, 
we also call $(a_t)_{t\geq 0}$ an equilibrium.

\begin{definition}\label{defA0}We call the set of equilibria $\mathcal{A}_{0}$ the set of all values $a\geq 0$ such that there exists a sequence $(a_{t})_{t\geq 0}$ satisfying (\ref{system1}) and the initial asset value $a_0$ equals $a$.
\end{definition}

\begin{corollary}\label{theorem13-choose} Let Assumptions \ref{assum0} and \ref{assum1} 
 be satisfied.  
Assume that $e^o_t=e^o>0$ and there exist $\underline{e}>0$ and $\bar{e}>0$ such that $e^y_t\in [\underline{e},\bar{e}]$ $\forall t$. Denote $\underline{R}\equiv \dfrac{u^{\prime
}(\bar{e})}{ v^{\prime }(e^{o})}, \bar{R}\equiv \dfrac{u^{\prime
}(\underline{e})}{ v^{\prime }(e^{o})}$. Assume that there exists $G_d$ such that
\begin{subequations}
\begin{align}
\frac{D_{t+1}}{D_t}\equiv G_n\frac{d_{t+1}}{d_t}&\leq  G_d<\underline{R} \quad \text{ (low dividend growth)},\\
\underline{R}&\leq R_{t+1}^{\ast }\equiv \frac{u^{\prime
}(e_{t}^{y})}{v^{\prime }(e^{o})}\leq \bar{R}<G_n \quad \text{ (low interest rate).}
\end{align}
\end{subequations}
Under these conditions, there exists 
$\bar d>0$ such that, if $d_0<\bar d$, the assumptions in Theorem \ref{new5continuum} hold and the set of equilibria $\mathcal{A}_{0}$ is an interval $[\underline{a},\bar{a}]$. 
 For $a_0=\underline{a}$, the equilibrium is bubbleless. For $a_0>\underline{a}$, the equilibrium is bubbly.

\end{corollary}
\begin{proof}
See Appendix \ref{prooftheorem1new}.
\end{proof}

\begin{remark}
Corollary \ref{new5continuum-remark} in Online Appendix 1 shows a continuum of equilibria (with and without bubbles) for the case where $cv'(c)$ is not necessarily increasing.
\end{remark}


\subsection{Conditions under which every equilibrium is bubbly}
\label{onlybubbly-section}
We now explore conditions under which every equilibrium is bubbly (in other words, there is no bubbleless equilibrium). 
To present our conditions (based on exogenous variables), we make use of the Euler equation (\ref{euler-bis}). 
This equation motivates us to define two functions $V^t_1$ and $V^t_2$: $\rr_+^2\to \rr_+$ by $V^t_1(x_1,x_2)\equiv U^t_1(e^y_tx_1,e^y_tx_2)$ and $V^t_2(x_1,x_2)\equiv U^t_2(e^y_tx_1,e^y_tx_2)$.

Denote {$s_t\equiv {a_{t}}/{e^y_t}$} and {$g_{e,t+1}\equiv {e_{t+1}^{o}}/{e_{t}^{y}}$} the saving rate and the  endowment growth rate of household born at date $t$.  Equation \eqref{euler-bis} becomes $V^t_1\left(1-s_t,g_{e,t+1}+R_{t+1}s_t\right)-
R_{t+1}V^t_2\left(1-s_t,g_{e,t+1}+R_{t+1}s_t\right)=0.$
Given $\epsilon>0$ and $t\geq 0$, let us define 
\begin{align}\label{euler3}
\mathcal{R}_t(\epsilon)\equiv \Big\{X>0: X=\frac{V^t_1\left(1-\epsilon,g_{e,t+1}+X\epsilon\right)}{V^t_2\left(1-\epsilon,g_{e,t+1}+X\epsilon\right)},  0\leq e^y_tX\epsilon/G_n-d_{t+1}\leq e^y_{t+1}\Big\}.
\end{align}
This is the set of interest rates that are consistent with household optimality for a saving rate $\epsilon$ at date $t$ and with feasibility at date $t+1$. We are now ready to state our result.
\begin{theorem}\label{onlybubbly}
Let Assumption \ref{assum0} be satisfied and the function $U^t$ be quasi-concave, continuously differentiable, and strictly increasing in each component.

Assume also Condition (B):  there exist $\bar{\epsilon}\in (0,1)$, a positive sequence $(X_t)$, and  a date $T$  satisfying  the following conditions:
\begin{enumerate}
\item
\label{onlybubbly2}({\it Not-too-low dividend condition})   $\sum_{t=1}^{\infty}\dfrac{D_t}{X_1\cdots X_t}=\infty$.
\item
\label{onlybubbly3} {\it (Low interest rate conditions)}
\begin{enumerate}
\item  For any $t\geq T$, if $\epsilon\in (0,\bar{\epsilon})$ and  $X\in \mathcal{R}_t(\epsilon)$,  
then $X\leq X_{t+1}$.
\item $X_{t+1}\leq G_n\dfrac{e^y_{t+1}}{e^y_t} $ $\forall t\geq T$. 

\end{enumerate}
\end{enumerate}
Then, the following statements hold.
\begin{enumerate}
\item\label{onlybubblypart1} For any equilibrium, the ratio of asset value to endowment is uniformly bounded away from zero (i.e., $\liminf_{t\to\infty}\frac{a_t}{e^y_t}>0$).

\item \label{onlybubblypart2}  If $\sum_{t=1}^{\infty}{D_t}/{(G_n^te^y_t)}<\infty$, then every equilibrium is bubbly.  
\item \label{onlybubblypart22} If $\sum_{t=1}^{\infty}{D_t}/{(G_n^te^y_t)}=\infty$,  then every equilibrium is bubbleless.

\item \label{onlybubblypart3}Consequently, any equilibrium is bubbly if and only if \begin{align}
\label{onlybubbly1}
\sum_{t=1}^{\infty}\frac{D_t}{G_n^te^y_t}<\infty \quad \text{ (low dividend condition).}
\end{align} 
\end{enumerate}

\end{theorem}
\begin{proof}See Appendix \ref{onlybubblyproof}.\end{proof}

To facilitate the discussion of the economic intuition behind Theorem \ref{onlybubbly}, we first present a test for assessing the validity of Condition (B).

\begin{lemma}[Testing Condition (B)]\label{justifyB}
Let Assumption \ref{assum0} be satisfied and the function $U^t$ be quasi-concave, continuously differentiable, and strictly increasing in each component. 

Denote $\Phi_t(\epsilon_1,\epsilon_2)\equiv {V^t_1\left(1-\epsilon_1,g_{e,t+1}+\epsilon_2\right)}/{V^t_2\left(1-\epsilon_1,g_{e,t+1}+\epsilon_2\right)}$.
\begin{enumerate}
\item\label{holding-condition1}  Condition (B) holds if there exist $M,\bar{\epsilon}_0, R, \bar{\epsilon}_1, \bar{\epsilon}_2>0$ and $T_0$ such that 
\begin{subequations}
\begin{align}
\label{boundPhi} &\Phi_t(\epsilon,z)\leq M \text{ } \forall t\geq T_0, \forall \epsilon\in (0,\bar{\epsilon_0}), \forall z\in [0,(d_{t+1}+e^y_{t+1})G_n/e^y_t], \\
\label{B-holding}&\sum_{s\geq 1}\frac{D_s}{R^s}=\infty,  \text{ }\Phi_t(\epsilon_1,\epsilon_2)\leq R\leq \frac{G_n e^y_{t+1}}{e^y_t} \text{ }\forall t\geq T_0, \forall \epsilon_1\in (0,\bar{\epsilon}_1), \forall \epsilon_2\in (0,\bar{\epsilon}_2).
\end{align}
\end{subequations}
This statement still holds if we replace  condition $\sum_{s\geq 1}\frac{D_s}{R^s}=\infty$ in (\ref{B-holding}) by 
\begin{align}\label{B-holding1} R&<\limsup_{t\to\infty}D_t^{1/t} \text{ or }R\leq \frac{D_{t+1}}{D_t}.
\end{align}

\item \label{holding-condition2} Let $\limsup_{t\to\infty}\frac{D_{t+1}}{D_t}<\infty$. Condition (B) in  Theorem \ref{onlybubbly} is satisfied if there exist $M,\bar{\epsilon}_0$,
$\bar{\epsilon}_1, \bar{\epsilon}_2>0$ and $T_0$ such that \eqref{boundPhi} holds and 
\begin{align}\label{B-holding2}
\Phi_t(\epsilon_1,\epsilon_2)&\leq \min\Big\{\frac{D_{t+1}}{D_t},  \frac{G_ne^y_{t+1}}{e^y_t}\Big\} \text{  $ \forall t\geq T_0,$ $\epsilon_1\in (0,\bar{\epsilon}_1)$, $\epsilon_2\in (0,\bar{\epsilon}_2)$.}
\end{align}

\end{enumerate}
\end{lemma}

\begin{proof}See Appendix \ref{onlybubblyproof}.\end{proof}

We now explain the economic intuition of Condition (B) and Theorem  \ref{onlybubbly}.  Condition (B) is a local restriction at the boundary of zero asset demand. It requires that, whenever the saving rate $s_t\equiv a_t/e^y_t$ is small, every interest rate  $X$ consistent with the household Euler equation in \eqref{euler3} is bounded above by a reference interest rate $X_{t+1}$, and that this reference interest rate does not exceed the growth factor $G_n^{t+1}e^y_{t+1}/(G_n^te^y_t)$ of aggregate young resources.  Using the asset-pricing equation, we obtain
$$s_{t+1}=\frac{R_{t+1}}{G_n e_{t+1}^y/e_t^y}s_t -\frac{D_{t+1}}{G_n^{t+1}e_{t+1}^y}.$$
Hence, whenever Condition (B) implies $R_{t+1}\leq X_{t+1} \leq G_n\frac{e_{t+1}^y}{e_t^y}$,  we have $s_{t+1}\leq s_t$. Thus, the low-demand region is forward invariant: once the normalized asset demand becomes sufficiently small, it remains small thereafter. If an equilibrium entered this low-demand region, actual interest rates would remain below the reference rates, and the fundamental value would dominate $\sum_{t=1}^{\infty}\frac{D_t}{X_{1}\cdots X_t}=\infty$, contradicting a finite equilibrium price.  Condition (B) therefore prevents asset demand from collapsing and forces the saving rate $a^t/e^{y}_t$ to remain bounded away from zero.

Once this lower bound has been established, the bubble classification depends only on whether dividends are summable relative to aggregate young resources; see condition \eqref{onlybubbly1}. If $\sum_{t=1}^{\infty}\frac{D_t}{G_n^te^y_t}<\infty$, dividends become cumulatively negligible relative to the resources of the cohorts purchasing the asset, and every equilibrium is bubbly. If the series diverges, dividends remain sufficiently important to account for the entire asset price, and every equilibrium is bubbleless.

Recall from Proposition \ref{necessitycondition}\ref{necessitycondition-point1} that, without condition (B), the existence of a bubbly equilibrium requires that $\sum_{t=1}^{\infty}\frac{D_t}{G_n^te^y_t}<\infty$. With condition (B), we go further by showing that every equilibrium is bubbly if and only if the low dividend condition \eqref{onlybubbly1} holds.


Lemma \ref{justifyB} provides further intuition for Condition (B). In particular, condition \eqref{B-holding2} says that, when households devote only a small fraction of their endowment to the asset, the gross interest rate consistent with their Euler equation remains below both dividend growth and the growth of aggregate young endowments. Such low interest rates prevent asset demand from becoming asymptotically negligible; consequently, the equilibrium saving rate is bounded away from zero.



We now present applications of Theorem \ref{onlybubbly} and Lemma \ref{justifyB}. Online Appendix 2 provides further applications and extends Theorem \ref{onlybubbly} to cases where the saving rate need not be bounded away from zero (Theorem \ref{onlybubbly-general}).

\begin{corollary}[Theorem 2 in \citet{hiranotoda25}] \label{stronger}Let Assumption \ref{assum0} hold, and suppose that $U^t$ is quasi-concave, continuously differentiable, and strictly increasing in each component. Assume further that $G_n=1$ and the following conditions are satisfied.\\
\citet[Assumption 2]{hiranotoda25}: $G\equiv \lim_{t\to\infty}\frac{e^{y}_{t+1}}{e^y_t}>0$, $w\equiv \lim_{t\to\infty}\frac{e^{o}_{t}}{e^y_t}\geq 0$.\\
\citet[Assumption 3]{hiranotoda25}: the {\it forward-rate function} $f_t$, defined by  \begin{equation*}
f_{t}(x_1,x_2)\equiv \frac{U^{t}_{1}(e_{t}^{y}x_1,e_{t}^{y}x_2)}{U^{t}_{2}(e_{t}^{y}x_1,e_{t}^{y}x_2)}=\frac{V^{t}_{1}(x_1,x_2)}{V^{t}_{2}(x_1,x_2)},
\end{equation*}
uniformly converges in the sense that there exists a continuous function $f:\mathbb{R}_{++}\times \mathbb{R}_{+}\rightarrow 
\mathbb{R}_{+}$ such that $
\lim_{t\rightarrow \infty }\sup_{(x_1,x_2)\in K}|f_{t}(x_1,x_2)-f(x_1,x_2)|=0$
 for any nonempty compact set $K\subset \mathbb{R}_{++}\times \mathbb{R}_{+}$.
 
If \begin{align}\label{condition20ht}
f(1,Gw)<G_d\equiv 
\limsup_{t\to\infty}D_t^{1/t}<\lim_{t\to\infty}\frac{e^y_{t+1}}{e^y_t},
\end{align}
then all equilibria are bubbly and $\liminf_{t\to\infty}{a_t}/{e^y_t}>0$.

 \end{corollary}
 \begin{proof}See Appendix \ref{onlybubblyproof}.\end{proof}
 


Theorem 2 in \citet{hiranotoda25} relies on two convergence assumptions concerning endowment growth and the forward-rate function (their Assumptions 2 and 3). By contrast, Theorem \ref{onlybubbly} does not require these convergence assumptions and accommodates more general non-stationary preferences and endowments. Instead, our conditions impose local bounds on exogenous primitives. Moreover, our result goes beyond \citet[Theorem 2]{hiranotoda25} by providing a sharper classification: statements \ref{onlybubblypart22} and \ref{onlybubblypart3} provide conditions under which every equilibrium is bubbly or, conversely, every equilibrium is bubbleless.

Interestingly, Theorem~\ref{onlybubbly} also yields sharp conclusions when long-run growth-rate comparisons are inconclusive. In particular, even when dividends and aggregate young resources grow at the same exponential rate, their lower-order differences may lead to opposite outcomes: every equilibrium may be bubbly or every equilibrium may be bubbleless. Theorem~\ref{onlybubbly} distinguishes these cases through the summability of relative dividends, as the following result illustrates.

\begin{corollary}[Bubble classification at the growth-rate boundary]
\label{remark-onlybubbly}
Assume that $U^t(x_1,x_2)=\ln(x_1)+\beta\ln(x_2),$ where $\beta>0$. The benchmark interest rate is then $R_{t+1}^{\ast}   =\frac{e_{t+1}^o}{\beta e_t^y}.$ 
Suppose that $\limsup_{t\to\infty}\frac{R_{t+1}^{\ast}}
         {G_n e_{t+1}^y/e_t^y}    <1,$
(or, equivalently, $\limsup_{t\to\infty}
    \frac{e_t^o}{\beta G_n e_t^y}<1$), and that, for some $\alpha>0$, $    \frac{D_t}{G_n^t e_t^y}=\frac{1}{t^\alpha}.$ 
Then Condition~(B) in Theorem~\ref{onlybubbly} holds and
$\liminf_{t\to\infty}\frac{a_t}{e_t^y}>0$ in every equilibrium.
Moreover, every equilibrium is bubbly if $\alpha>1$ and bubbleless if $0<\alpha\leq1$.
\end{corollary}

A detailed proof of Corollary \ref{remark-onlybubbly} is presented in Online Appendix 2. In Corollary \ref{remark-onlybubbly}, we have  
$$\sum_{t=1}^{\infty}\frac{D_t}{G_n^t e_t^y}=
\sum_{t=1}^{\infty}\frac{1}{t^\alpha}
\begin{cases}
<\infty, & \text{if }\alpha>1,\\[2mm]
=\infty, & \text{if }0<\alpha\leq1.
\end{cases}
$$
To relate this result to \citet{hiranotoda25}, set $G_n=1$, $e_t^y=1$, and $e_t^o=e^o$ for all $t$, with $e^o/\beta<1$. Then \[ R=\frac{e^o}{\beta}<1, \qquad G_d=\limsup_{t\to\infty}D_t^{1/t} =1 =\lim_{t\to\infty}\frac{e_{t+1}^y}{e_t^y}=G. \] Hence $R<G_d=G$, so the strict inequality $G_d<G$ in their condition~(20) fails, while Corollary~\ref{remark-onlybubbly} determines whether every equilibrium is bubbly or bubbleless.

When $\alpha>1$, Corollary~\ref{remark-onlybubbly} is not covered by the earlier no-bubble conditions. In particular, none of conditions~\eqref{keycond}, \eqref{bubble1}, and \eqref{existence-bubbleles} in Proposition~\ref{necessitycondition} holds.

\subsection{Full characterization under stationary endowments}
\label{full-stationaryendowment}
We now seek to provide a full characterization of the set of equilibria and asset price behaviors when utility functions and endowments are time-independent. 

\begin{theorem}
\label{allsets} Let Assumptions \ref{assum0} and \ref{assum1} 
be satisfied. Consider stationary endowments, i.e., $%
e_{t}^{y}=e^{y}>0,e_{t}^{o}=e^{o}> 0$ for any  $t$. Denote $R^*\equiv \frac{%
u^{\prime}(e^y)}{ v^{\prime}(e^o)}$ the interest rate in the economy
without assets.

\begin{enumerate}
\item\label{part1} If $R^*>G_n$ (high interest rate condition) or $\sum_{t=1}^{\infty}\dfrac{D_t}{G_n^t}=\infty$ (not-too-low dividend condition), then
there exists a unique equilibrium and this equilibrium is bubbleless.

\item \label{part2} If $R^{\ast } <G_n$ and $\sum_{t=1}^{\infty}d_t=\sum_{t=1}^{\infty}\dfrac{D_t}{G_n^t}<\infty$, then one of the following cases must hold.
\begin{enumerate}
\item \label{set2} There exists a continuum of equilibria. The set of
equilibria  $\mathcal{A}_{0}$  (see Definition \ref{defA0}) is a compact interval $[\underline{a},\bar{a}]$.

 For $a_0\in (\underline{a},\bar{a}]$, the equilibrium is bubbly.

\label{set2ii}  For $a_0\in [\underline{a},\bar{a})$, the equilibrium satisfies $%
(a_t,b_t,R_t)\to (0,0,R^*)$.

For $a_0=\bar{a}$, the equilibrium satisfies $(a_{t},b_{t},R_{t})%
\rightarrow (\hat{a},\hat{a},G_n)$, where $\hat{a}>0$ is uniquely determined by  $u'(e^y-\hat{a})= G_n v'(e^o+G_n\hat{a})$. 

\item \label{set1} There exists a unique equilibrium. This equilibrium is bubbly and  $(a_{t},b_{t},R_{t})$ converges
to $(\hat{a},\hat{a},G_n)$ where $\hat{a}>0$ is uniquely determined by  $u'(e^y-\hat{a})= G_n v'(e^o+G_n\hat{a})$. 

\end{enumerate}

Moreover, the following claims hold.

 {\bf Claim 1}: If $R^{\ast } <G_n$ and $\sum_{t\geq 1}\frac{D
_{t}}{(R^{\ast })^{t}}<\infty $,  then the statement  \ref{set2} is true. 

{\bf Claim 2}: If   $R^{\ast } <G_n$, $\sum_{t=1}^{\infty}\dfrac{D_t}{G_n^t}<\infty$ and $R^{\ast
}<\limsup_{t\rightarrow \infty }D _{t}^{1/t}$, then the statement \ref{set1} is true. 


\item  \label{part4}If $R^*=G_n$ and $\sum_{t=1}^{\infty}{D_t}/{G_n^t}<\infty$, then there exists a unique equilibrium. This equilibrium is bubbleless and $(a_t,b_t,R_t)\to (0,0,G_n)$.
\end{enumerate}
\end{theorem}
\begin{proof}
See Appendix \ref{prooftheorem2}.
\end{proof}

Theorem \ref{allsets} explores the equilibrium set and the asymptotic properties of asset prices in all possible cases (see Online Appendix 4 for an explicit model with logarithmic utility function). This is a contribution with respect to the literature and the previous results in the present paper. We observe that the equilibrium set depends on the interplay between the interest rate of the economy without asset $R^{\ast }$, the population growth factor $G_n$ and the dividend growth rates.

We now discuss how our Theorem \ref{allsets} is related to the existing literature.  First, Theorem \ref{allsets} corresponds to Proposition 1 in \cite{tirole85}, who studies the asset price in an OLG model with dividend-paying asset and production.  However, Proposition 1 in \cite{tirole85} contains some concerns (see  \cite{phamtoda2025,phamtodaEcta} for a more detailed discussion). Our Theorem \ref{allsets}  provides a full characterization of the equilibrium set in an exchange economy with stationary endowments but with non-stationary dividends. Note that \cite{tirole85} assumes that $D_t={1}$ and did not study the case $R^*=G_n$.
 
Our added value is to provide the uniqueness and asymptotic properties of equilibrium under general dividends but a stronger assumption (namely, stationary endowment and exchange economy). \citet[Section 5]{ht24b} study the case where the utility is homogeneous of degree $1$ and $e_{t}^{y}=aG^{t},e_{t}^{o}=bG^{t},%
D _{t}=DG_{d}^{t}$, where $a,G,D,G_{d}$ are positive constants.  However, their approach cannot be directly applied to our setting because we only impose very minimal conditions on the dividend sequence and our utility function is not necessarily homogeneous of degree $1$. Moreover, \citet[Section 5]{ht24b} studies  neither the case $R^*=G_n$ nor $\limsup_{t\to\infty}D_t^{{1}/{t}}=G_n$ while our Theorem \ref{allsets} covers these cases.

Since three cases in Theorem \ref{allsets} are mutually exclusive,  we obtain an important result, establishing a necessary and sufficient condition for the existence of a bubble.
\begin{corollary}\label{iff_result}
Let Assumptions \ref{assum0}, \ref{assum1} 
be satisfied. Consider the case of stationary endowments, i.e., $%
e_{t}^{y}=e^{y}>0,e_{t}^{o}=e^{o}> 0$ for any  $t$. 

There exists a bubbly equilibrium if and only if the two following conditions hold: (1) $R^*<G_n$, and (2) $\sum_{t=1}^{\infty}d_t=\sum_{t=1}^{\infty}{D_t}/{G_n^t}<\infty$.


\end{corollary}



\section{\bf Pareto optimality}
\label{section-pareto}

In this section, we study Pareto optimality. Let us start by
providing a formal definition (see \cite{bs80} for instance).

\begin{definition}
Let $c^y_{-1}>0$ and $(d_t)_{t=0}^{\infty}$ be an exogenous non-negative sequence and $N_t=G_n^t>0$ $\forall t$. 
A feasible allocation path is a positive sequence $(c_{t}^{y},c_{t}^{o})_{t\geq 0}$ satisfying $
N_tc^y_t+N_{t-1}c^o_t=N_te^y_t+N_{t-1}e^o_t+D_t \quad (i.e., \text{ } c_{t}^{y}+\frac{c_{t}^{o}}{G_n}=e_{t}^{y}+\frac{e_{t}^{o}}{G_n}+d_{t}) \text{  } \forall t$.

A feasible allocation path is said to be Pareto optimal if there is no other feasible
allocation path $(c_{t}^{y\prime },c_{t}^{o\prime })_{t\geq 0}$ such that $
U^t(c_{t}^{y\prime },c_{t+1}^{o\prime }) \geq U^t\left(
c_{t}^{y},c_{t+1}^{o}\right) \text{  } \forall t\geq -1$, 
 with strict inequality for some $t$.\footnote{Here, we denote $c^{y'}_{-1}=c^y_{-1}$ and we consider $U^{-1}(x_1,x_2)=x_2$.}
\end{definition}


\begin{assum}\label{assumption-pareto}
The function $U^t$ is strictly concave, continuously differentiable, strictly increasing in each component. 
\end{assum}
As in Proposition 5.3 in \cite{bs80}, we have the following result (whose  detailed proof is presented in Online Appendix 5).

\begin{lemma}[Sufficient conditions for Pareto optimality]\label{lemmaPareto}Let Assumptions \ref{assum0}, \ref{assumption-pareto} be satisfied. Consider a feasible allocation $%
(c_{t}^{y},c_{t}^{o})_{t\geq 0} $. Define the sequence $(R_{t})_{t\geq 0}$ by 
\begin{equation}
R_{t+1}=\frac{U^t_1(c_{t}^{y},c_{t+1}^{o})}{U^t_2(c_{t}^{y},c_{t+1}^{o})} \text{  } \forall t\geq 0. \label{00131}
\end{equation}%
The path $(c_{t}^{y},c_{t}^{o})_{t\geq 0 }$ is Pareto optimal if $%
\liminf\limits_{t\rightarrow \infty }\frac{G_n^{t}}{R_{1}\cdots R_{t}}%
c_{t}^{y}=0$.
\end{lemma}

A natural question arises: can an equilibrium still be Pareto-optimal if the conditions stated in Lemma \ref{lemmaPareto} are not satisfied? It is well known that (see, for instance, \cite{OkunoZilcha1980}, page 802)  this question is, in general, difficult. To address this issue, we extend  \cite{OkunoZilcha1980} and \cite{bs80}.


Consider an equilibrium allocation $(c^y_t,c^o_t)_t$. Denote 
$Q_t\equiv {1}/{(R_1\cdots R_t)}$ and $P_t\equiv {G_n^t}/{(R_1\cdots R_t)}.$ In equilibrium, we observe that 
\begin{align}
Q_{t}c^y_t+Q_{t+1}c^o_{t+1}&=Q_{t}e^y_t+Q_{t+1}e^o_{t+1}, &
\frac{U^t_1(c^y_t,c^o_{t+1})}{U^t_2(c^y_t,c^o_{t+1})}&=\frac{Q_t}{Q_{t+1}}, & \frac{P_{t+1}}{P_t}&=\frac{G_n}{R_{t+1}}.
\end{align}

Denote {$e_t\equiv e^y_t+{e^o_t}/{G_n}+d_t$} the aggregate good supply per capita at date $t$.

We now introduce the notions of strictness and smoothness used by \cite{Benveniste1976}, \cite{OkunoZilcha1980}, which are closely related to the notion of Gaussian curvature used by \cite{bs80}.
\begin{definition}\label{boz-condition}
Given an equilibrium allocation $(c^y_t,c^o_t)_t$, the upper contour of the t-th generation is given by $
B_t(c) \equiv \big\{(c^{y'}_t,c^{o'}_{t+1})\in \rr_+^2:  U^t(c^{y'}_t,c^{o'}_{t+1})\geq  U^t(c^{y}_t,c^{o}_{t+1})\big\}.$\\
(i) We say that this allocation satisfies the "uniform strictness condition" if there exist $h\in (0,1]$ and $\mu>0$ such that, for any $t$,
\begin{align}\label{strict-uniform}
P_{t+1}(c^{o'}_{t+1}-c^{o}_{t+1})+G_nP_t(c^{y'}_t-c^{y}_t)\geq\frac{\mu}{G_nP_tc^y_t}\big(G_nP_t(c^{y'}_t-c^{y}_t)\big)^2 \\
\text{ }\forall (c^{y'}_t,c^{o'}_{t+1})\in B_t(c) \text{ satisfying  } (1-h)c^y_t<c^{y'}_t<c^{y}_t, c^{o'}_{t+1}>c^o_{t+1}.\notag
\end{align}
(ii) We say that this allocation satisfies the "uniform smoothness condition" if  for each $x\in (0,1)$, there exists ${\theta}_1(x),{\theta}_2(x)>0$ such that, for any $t$, if the couple
$(c^{y'}_t,c^{o'}_{t+1})$ satisfies 
\begin{align}
&P_{t+1}(c^{o'}_{t+1}-c^{o}_{t+1})+G_nP_t(c^{y'}_t-c^{y}_t)\geq\frac{{\theta}_2(x)}{P_{t+1}c^o_{t+1}}\big(P_{t+1}(c^{o'}_{t+1}-c^{o}_{t+1})\big)^2+\frac{{\theta}_1(x)}{G_nP_tc^y_t}\big(G_n P_t(c^{y'}_t-c^{y}_t)\big)^2,\notag\\
\label{smooth-uniform-1}
& xc^{y}_t<c^{y'}_t<c^{y}_t, c^{o}_{t+1}<c^{o'}_{t+1}<G_ne_{t+1},
\end{align}
then $(c^{y'}_t,c^{o'}_{t+1})\in B_t(c)$, i.e., $U^t(c^{y'}_t,c^{o'}_{t+1})\geq  U^t(c^{y}_t,c^{o}_{t+1})$.

\end{definition}
The notion of strictness in Definition \ref{boz-condition} is similar  to (but weaker than) that in \cite{OkunoZilcha1980}'s Definition 10.  Our notion of smoothness is quite different from the smoothness in Definition 11 in \cite{OkunoZilcha1980} (indeed, Definition 11 in \cite{OkunoZilcha1980} corresponds to our case with $\theta_2(x)=0$). 

Note that \cite{OkunoZilcha1980} did not  provide explicit conditions to ensure the uniform strictness and smoothness. Since the uniform strictness and smoothness conditions are quite implicit, a natural issue is to justify them. In Online Appendix 5, we show that conditions in Lemmas \ref{check-strict} and \ref{check-smooth} can be satisfied in many cases. For instance, condition \ref{check-strict-part1} in Lemma  \ref{check-strict} is satisfied if the function $u_t$ does not depend on $t$ and $0<\underline{c}^y\leq c^y_{t}\leq \bar{c}^y<\infty$ $\forall t$.

 We now state the main result of this section, which is similar to Theorem 3A and Theorem 3B in \cite{OkunoZilcha1980} and Proposition 5.6 in \cite{bs80}. 
\begin{theorem}\label{pareto-theorem}
Let Assumptions \ref{assum0} and \ref{assumption-pareto} be satisfied. 
Consider an equilibrium with the allocation $(c^y_t,c^o_t)_t$ and the interest rates  $(R_t)_{t}$. 
\begin{enumerate}
\item\label{pareto-theorem-part1} Assume that the equilibrium allocation $(c^y_t,c^o_t)_t$ satisfies the uniform strictness condition in Definition \ref{boz-condition}. Then, this allocation is Pareto optimal if 
\begin{align}\label{series1/P}\sum_{t\geq 1}\dfrac{1}{P_{t}e_t}=\infty \quad (i.e., \sum_{t\geq 1}\dfrac{R_1\cdots R_t}{G_n^te_t}=\infty).
\end{align}
\item \label{pareto-theorem-part2}  Assume that the allocation $(c^y_t,c^o_t)_t$ is Pareto optimal and satisfies the uniform smoothness condition in Definition \ref{boz-condition}. Assume also that $\liminf_{t\to\infty}\frac{c^y_t}{e_t}>0$, $\limsup_{t\to\infty}\frac{c^o_t}{G_ne_t}<1$, $\liminf_{t\to\infty}\frac{P_{t+1}c^o_{t+1}}{P_te_t}>0$. Then, we have
\begin{align}\label{series1/P}\sum_{t\geq 1}\dfrac{1}{P_{t}e_t}=\infty \quad (i.e., \sum_{t\geq 1}\dfrac{R_1\cdots R_t}{G_n^te_t}=\infty).
\end{align}
\end{enumerate}
\end{theorem}
\begin{proof}See Online Appendix 5.
\end{proof}

\cite{OkunoZilcha1980} provide an example of a Pareto-inefficient equilibrium satisfying condition (\ref{series1/P}), thereby illustrating the role of the uniform strictness condition in Theorem \ref{pareto-theorem}. \cite{bs80} introduce properties (C) and (C'), which require the Gaussian curvature of households' indifference surfaces through their equilibrium consumption to be bounded above and bounded away from zero, respectively.\footnote{See \cite{BonnisseauRakotonindrainy2016} for a geometric proof of the Balasko--Shell characterization of Pareto-optimal allocations in OLG exchange economies with a varying number
of commodities and consumers across periods and possibly incomplete and non-transitive preferences.} These properties are closely related to the uniform smoothness and strictness conditions of \cite{OkunoZilcha1980} and Definition \ref{boz-condition}, respectively.\footnote{See Footnote 8 in \cite{OkunoZilcha1980} and Footnote 9 in \cite{bs80}.}

Our assumptions nevertheless differ from those of \cite{OkunoZilcha1980} and \cite{bs80}. Their households consume multiple goods at each date, whereas we consider a single consumption good, and they impose upper and lower bounds on endowments, which we do not. Moreover, our uniform smoothness and strictness conditions admit explicit sufficient conditions that can be verified using elementary calculus; see Lemmas \ref{check-strict} and \ref{check-smooth} in Online Appendix 5. Most importantly, our economy includes a dividend-paying asset, allowing us to connect Pareto optimality to asset-price bubbles and to derive the new results presented in the next section.

\section{\bf Connection between asset price bubble and Pareto optimality}
\label{section-bubble-pareto}
In this section, we investigate the relationship between asset price bubbles and Pareto optimality. We begin by emphasizing that these are conceptually distinct notions and are characterized by different asymptotic conditions. Indeed, the existence of an asset price bubble is equivalent to $\lim_{t\to\infty}\frac{G_n^t a_t}{R_1\cdots R_t}>0,$ whereas Pareto optimality is, in many cases, equivalent to condition \eqref{series1/P}, namely, $\sum_{t\geq 1}\frac{R_1\cdots R_t}{G_n^t e_t}=\infty.$ 
Thus, although both properties depend on the long-run behavior of equilibrium prices and interest rates, there is no immediate reason for one to imply the other. The results below clarify the relationship between these two notions.

We first consider bubbleless Pareto-optimal equilibria.
\begin{proposition}
\label{efficient-bubbleless}

Let Assumptions \ref{assum0}, \ref{assum1new} and \ref{assumption-pareto} hold.
\begin{enumerate}
\item\label{efficient-bubbleless1}  An equilibrium is Pareto optimal and bubbleless if $%
\liminf\limits_{t\rightarrow \infty }\frac{G_n^{t}}{R_{1}\cdots R_{t}}%
e_{t}^{y}=0$.

\item\label{efficient-bubbleless2} Every equilibrium is Pareto optimal and bubbleless if the dividends are significant in the sense that $\limsup_{t\to\infty}{d_t}/{e^y_t}>0$.
\end{enumerate}
\end{proposition}
\begin{proof}See Appendix \ref{section-bubble-pareto-proof}.\end{proof}

The insight of Proposition \ref{efficient-bubbleless}\ref{efficient-bubbleless2} is that a significant level of dividends makes the market economy Pareto optimal. This is in line with  Propositions 5 and 8 in \cite{lvp16} in a model with infinitely lived agents. The difference is that we work in a nonstationary OLG exchange economy and study the Pareto optimality while they consider general equilibrium models with infinitely-lived agents and study the dynamic efficiency in the sense of \cite{malinvaud53}.  Proposition \ref{efficient-bubbleless}\ref{efficient-bubbleless2} is also in line with  Proposition 1 in \cite{rhee91} in an OLG model with land, where he proves that an economy is dynamically efficient if the income share of land does not vanish.

By combining Proposition \ref{necessitycondition} and Theorem \ref{pareto-theorem}, we obtain the following result which deepens  our understanding regarding the role of dividends and the benchmark interest rates. 

\begin{proposition}\label{bubbleless-pareto}
\begin{enumerate}
\item Let Assumptions \ref{assum0}, \ref{assum1new} and \ref{assumption-pareto} hold. Any equilibrium satisfying the uniform strictness condition is bubbleless and Pareto optimal if $\sum_{t=1}^{\infty}\frac{D_t}{e_tG_n^t}=\infty$.

\item Let Assumptions \ref{assum0}, \ref{assum1new}, \ref{derivative-ij} and  \ref{assumption-pareto} hold. Under the high-interest-rate condition \eqref{bubble1}, every equilibrium is bubbleless and Pareto optimal. If Assumption \ref{assum1} is also imposed, the equilibrium is unique.
\end{enumerate}
\end{proposition}
\begin{proof}See Appendix \ref{section-bubble-pareto-proof}.\end{proof}

The following result shows the importance of the asset value on the Pareto optimality.
\begin{proposition}[asset value and Pareto optimality]
\label{efficient-bubbleless-saving}
Let Assumptions \ref{assum0}, \ref{assum1new} and \ref{assumption-pareto} hold.
\begin{enumerate}
\item \label{efficient-bubbleless-saving2} 

An equilibrium is Pareto optimal if it satisfies the uniform strictness condition and the asset value is significant (in the sense that $\limsup_{t\to\infty}({a_t}/{e_t})>0$). 


\item \label{efficient-bubbleless-saving1} An equilibrium is Pareto optimal if it is bubbleless and the saving rate is bounded away from zero (i.e., $\liminf_{t\to\infty}({a_t}/{e^y_t})>0$).
\end{enumerate}
\end{proposition}

\begin{proof}See Appendix \ref{section-bubble-pareto-proof}.\end{proof}

Proposition \ref{efficient-bubbleless-saving}\ref{efficient-bubbleless-saving2} establishes the optimality of equilibrium, regardless of whether it is bubbly or bubbleless.  Moreover, Condition (B) in Theorem \ref{onlybubbly} implies that $\liminf_{t\to\infty}({a_t}/{e^y_t})>0$. If we assume Condition (B) and  $\limsup_{t\to\infty}({e^y_t}/{e_t})>0$ (i.e., young people have a significant endowment), then  $\limsup_{t\to\infty}({a_t}/{e_t})>0$. Consequently, under Condition (B) and $\limsup_{t\to\infty}({e^y_t}/{e_t})>0$, Proposition \ref{efficient-bubbleless-saving} implies that an equilibrium is Pareto optimal if it satisfies the uniform strictness condition or is bubbleless. 

Recall that the uniform strictness condition is easily satisfied in a large class of models (see  Lemma \ref{check-strict} in Online Appendix 5). Therefore, our results suggest that Condition (B) and the property that the asset value is significant  play a crucial role in determining Pareto optimality. Note that in Theorem \ref{pareto-assetbubble}\ref{pareto-assetbubble-2a} below, we show some cases where the equilibrium is not optimal and $\lim_{t\to\infty}({a_t}/{e^y_t})=\lim_{t\to\infty}({a_t}/{e_t})=0$ (by Claim 1 and point \ref{set2ii} of Theorem \ref{allsets}).

So far, we have presented some sufficient conditions for the Pareto optimality. We now show how an equilibrium need not be Pareto optimal. We show that this may happen when there exists a continuum of equilibria with bubbles. 
\begin{proposition}[Equilibria are bubbly and not Pareto optimal]
\label{paretorank}
Let Assumptions \ref{assum0} and \ref{assum1} be satisfied, and suppose that there exists a continuum of equilibria (as may occur, for instance, under the conditions of Theorem \ref{new5continuum} or Claim 1 of Theorem \ref{allsets}).

Then, the lifetime utility of households born at any date is strictly increasing in the initial asset value $(a_0)$. Hence, any equilibrium with $a_0<\bar a_0\equiv\max{a\in\mathbf A_0}$ 
is not Pareto optimal. In particular, there exists a continuum of bubbly equilibria that are not Pareto optimal.
\end{proposition}
\begin{proof}See Appendix \ref{section-bubble-pareto-proof}.\end{proof}

Note that the proof of Proposition \ref{paretorank} does not rely on Theorem \ref{pareto-theorem}. Moreover, Proposition \ref{paretorank} provides finer information by establishing a welfare ranking of households across different equilibria.

In the case of stationary endowments, combining Theorems \ref{allsets} and \ref{pareto-theorem} yields a fairly complete characterization of the equilibrium set and its welfare properties.
\begin{theorem}\label{pareto-assetbubble}
Suppose that the assumptions of Theorem \ref{allsets} hold and $u^{\prime\prime}(c)<0$ $\forall c>0$.
\begin{enumerate}
\item \label{pareto-assetbubble-1} If $R^*>G_n $, there exists a unique equilibrium. This equilibrium is bubbleless and Pareto optimal.

\item \label{pareto-assetbubble-2}  If $R^{\ast } <G_n $ and $\sum_{t\geq 1}\frac{D
_{t}}{(R^{\ast })^{t}}<\infty $, then there exists a continuum of equilibria. The set of equilibria $\mathcal{A}_0$ is a compact interval $[\underline{a},\bar{a}]$.

\begin{enumerate}
\item\label{pareto-assetbubble-2a} Any equilibrium with  $a_0<\bar{a}$ is not Pareto optimal. 

Any equilibrium with $a_0\in (\underline{a},\bar{a})$ is not Pareto optimal, bubbly but asymptotically bubbleless.

\item\label{pareto-assetbubble-2b} The maximal equilibrium $a_0=\bar{a}$  is asymptotically bubbly and Pareto optimal.

\end{enumerate}

\item\label{pareto-assetbubble-3} If   $R^{\ast } <G_n $, $\sum_{t=1}^{\infty}{D_t}/{G_n^t}<\infty$, and $R^{\ast
}<\limsup_{t\rightarrow \infty }D _{t}^{{1}/{t}}$, there exists a unique equilibrium. This equilibrium is asymptotically bubbly and Pareto optimal. 

\end{enumerate}
\end{theorem}
\begin{proof}See Appendix \ref{section-bubble-pareto-proof}.\end{proof}

According to Theorem \ref{pareto-assetbubble}, when the benchmark interest rate is low, i.e., $R^*<G_n$ and the dividends are low (i.e., $\sum_td_t<\infty$), only the asymptotically bubbly equilibrium can be Pareto optimal. This point is consistent with the traditional insight (see Proposition 2 in \cite{tirole85}, which claims that in the case of low interest rate $(R^*<G_n)$ and $d_t={d_0}/{G_n^t}$, only the asymptotically bubbly equilibrium is Pareto optimal).  \cite{tirole85} considers a specific form of dividend (i.e., $d_t={d_0}/{G_n^t}$) and consequently does not analyze the role of dividend growth. More importantly, \cite{tirole85}  does not provide a formal proof for his Proposition 2.

However, the following explicit model shows that when the interest rate in the economy without asset is low, an equilibrium which is Pareto optimal can be bubbly or bubbleless.
\begin{proposition}
\label{explicit-1}
$U^t(x_1,x_2)=\ln(x_1)+\beta \ln(x_2)$ where $\beta>0$, and  $e^o_t=0\text{  } \forall  t$.  There exists a unique equilibrium, which is determined by $\frac{q_t}{G_n^t}=a_t=\frac{\beta}{1+\beta} e^y_t$.\footnote{This equilibrium is similar to that in Section 5.1.1 in \cite{blp18}, Section 9.3.2 in \cite{blp17b}, or Proposition 1 in \cite{hiranotoda25}. See \cite{blp21}'s Section 4 for other explicit models with bubbles.} This equilibrium is Pareto optimal.
\begin{enumerate}
\item If $\sum_{t\geq 1}\frac{d_t}{e^y_t}<\infty$ (i.e., $\sum_{t=1}^{\infty}\frac{D_t}{G_n^te^y_t}<\infty$), it is  bubbly, with $\lim_{t\to\infty}\frac{b_t}{e^y_t}=\frac{\beta}{1+\beta}$.

\item If $\sum_{t\geq 1}\frac{d_t}{e^y_t}=\infty$ (i.e., $\sum_{t=1}^{\infty}\frac{D_t}{G_n^te^y_t}=\infty$), then it is bubbleless.
\end{enumerate}
\end{proposition}
\begin{proof}See Online Appendix 5.\end{proof}


In Proposition \ref{explicit-1}, the interest rate in the economy without the asset is zero at every date, i.e., $R_t^*=0$ for all $t$. Nevertheless, depending on the relative growth rates of dividends and endowments, the equilibrium characterized in Proposition \ref{explicit-1} may be either bubbly or bubbleless.\footnote{As shown in the proof of Corollary \ref{remark-onlybubbly}, the economy considered in Proposition \ref{explicit-1} satisfies Condition (B) of Theorem \ref{onlybubbly}.} In either case, however, the equilibrium allocation is Pareto optimal. This observation is consistent with Proposition \ref{efficient-bubbleless-saving}\ref{efficient-bubbleless-saving2} and complements the insight of \citet[Proposition 2]{tirole85}.

\citet[Proposition 2]{tirole85} shows, in his stationary setting, that a bubble that remains economically significant in the long run restores dynamic efficiency. The following example demonstrates that this conclusion need not extend to nonstationary economies: an equilibrium can exhibit an asymptotically significant bubble and yet fail to be Pareto optimal.
%
\begin{example}[An asymptotically bubbly but non-Pareto-optimal equilibrium]
\label{example_asymptotically_bubbly_nonoptimal}
Assume that the utility
function is time independent and given by $U(c_1,c_2)=u(c_1)+\frac12u(c_2)$, where $u(c)=c+\ln c.$
Let $G_n=1$, $D_t=1$, $e_t^y=6\cdot 2^t+1, e_{t+1}^o=3\cdot2^t-2$ $\forall t\geq 0$, and $e_0^o=1$. Then, there exists an equilibrium determined by the following:
\begin{subequations}
\begin{align}
\text{Asset price and supply: } q_t&=a_t=1+2^t, \quad R_{t+1}=2, \quad z_t=1\\
\text{Consumptions: } c_t^y&=c_{t+1}^o=5\cdot2^t.
\end{align}
\end{subequations}
This equilibrium is not Pareto optimal. It is bubbly and the ratio of the normalized bubble component to young endowment is $\lim_{t\to\infty}{b_t}/{e_t^y}
=1/6>0$.
\end{example}
\begin{proof}See Appendix \ref{section-bubble-pareto-proof}.
\end{proof}

The source of non-optimality in this example is not the presence of the bubble itself. Because $u^{\prime\prime}(c)=-1/c^2$, curvature vanishes as consumption grows, and the uniform strictness condition used in the stationary welfare result fails. The growing endowment sequence then permits an intergenerational reallocation that benefits every generation despite a bubble that remains significant relative to young endowments. Thus, asymptotic bubble significance alone does not guarantee Pareto optimality in a nonstationary economy.

\section{\bf Conclusion}\label{conclusion}

This paper separates two questions that are often conflated in overlapping-generations models: the existence of an asset price bubble and the Pareto efficiency of the associated equilibrium allocation. Under general nonstationary primitives, we provide conditions that distinguish three regimes: (a) all equilibria are bubbleless, (b) bubbly and bubbleless equilibria coexist, and (c) all equilibria are bubbly.

Under stationary endowments and time-independent separable preferences, we characterize the entire equilibrium set. A bubbly equilibrium exists if and only if the benchmark gross interest rate is below population growth and normalized dividends are summable. This characterization also resolves the boundary case in which the benchmark interest rate equals population growth.

The welfare analysis shows that bubble status alone does not determine Pareto efficiency. A bubbly equilibrium may be efficient or inefficient, and the same is true of a bubbleless equilibrium. In the stationary low-interest rate co-existence regime, however, the classical welfare conclusion suggested by \cite{tirole85} is valid: the asymptotically bubbly equilibrium is Pareto optimal, whereas every other equilibrium is Pareto dominated. In more general nonstationary economies, an asymptotically significant bubble need not restore Pareto optimality.



{\appendix
{\section{\bf Formal Proofs}\label{assetprice-basis}

\begin{proof}[{\bf Proof of Lemma  \ref{prop1}}]

 By iterating $q_t=(q_{t+1}+D_{t+1})/R_{t+1}$,\footnote{In this deterministic framework, the sequence of discount factors  $(R_t)$ is uniquely determined. The reader is referred to \cite{sw97}, \cite{aptm11}, \cite{pascoa11},  \cite{blp18} among others for the notion of bubbles in stochastic economies where discount factors (and state price processes) are not necessarily uniquely determined.} 
 we have
 \begin{align}\label{assetprice_decomposition}
     q_t=\sum_{s=t+1}^T\frac{Q_s}{Q_t}D_s+\frac{Q_T}{Q_t}q_T \text{ } \forall T> t\geq 0, \text{ where } Q_t\equiv \frac{1}{R_1\cdots  R_t},  Q_0\equiv 1.
 \end{align}


Consider the case $t=0$ in \eqref{assetprice_decomposition}. By letting $T$ tend to infinity and noting that $\sum_{s=1}^TQ_sD_s$ is increasing in $T$ and bounded by $q_0$, we obtain point \ref{prop1-2}.

According to \eqref{ft}, we can easily check that $
B_{t+1}=R_{t+1}B_t \text{ and }
F_{t+1}+D_{t+1}=R_{t+1}F_t.$ So,  there is a bubble at date $0$ if and only if  there is a bubble at date $t$.

By the definition of $F_t$ and $B_t$, we have  $q_t=F_t+B_t$. So, $\ref{prop1-3}\iff \ref{prop1-4}$.

{\bf Points \ref{prop1-2} and \ref{prop1-3} are equivalent}.  From (\ref{assetpricing}) and $F_{t+1}+D_{t+1}=R_{t+1}F_t$, we have
\begin{align}\label{dif-fq}
\frac{F_t}{q_t}-\frac{F_{t+1}}{q_{t+1}}=\Big(1-\frac{F_t}{q_t}\Big)\frac{D_{t+1}}{q_{t+1}}  \text{  } \forall t\geq 0.
\end{align}
So, we see that \ref{prop1-3} implies \ref{prop1-2} because $\frac{F_t}{q_t}>\frac{F_{t+1}}{q_{t+1}}$ implies that $q_t>F_t$. 

We now prove that point \ref{prop1-2} implies point \ref{prop1-3}. Assume that there is a bubble. By (\ref{dif-fq}), the sequence $(\frac{F_t}{q_t})$ is strictly decreasing. Moreover, the existence of a bubble means that  $\lim_{t\rightarrow \infty}Q_tq_t=\lim_{t\rightarrow \infty}\frac{q_t}{R_1\cdots R_t}>0$.  Then, there exists $x>0$ and $t_0>0$ such that $\frac{q_t}{R_1\cdots R_t}>x$ $\forall t\geq t_0$.

Take $t\geq t_0$. We have 
\begin{align*}
\frac{F_t}{q_t}&=\frac{1}{q_t}\sum_{s\geq 1}\frac{D_{t+s}}{R_{t+1}\cdots R_{t+s}}=\frac{R_1\cdots R_t}{q_t}\sum_{s\geq 1}\frac{D_{t+s}}{R_{1}\cdots R_{t+s}}< \frac{1}{x}\sum_{s\geq 1}\frac{D_{t+s}}{R_{1}\cdots R_{t+s}}.
\end{align*}
because $\frac{q_t}{R_1\cdots R_t}>x$ $\forall t\geq t_0$. Recall that $q_0\geq F_0=\sum_{t\geq 1}\frac{D_{t}}{R_{1}\cdots R_{t}}.$ It means that the series $\sum_{t\geq 1}\frac{D_{t}}{R_{1}\cdots R_{t}}$ converges. Thus, $\sum_{s\geq 1}\frac{D_{t+s}}{R_{1}\cdots R_{t+s}}$ converges to zero when $t$ tends to infinity. By consequence,
 $\lim_{t\to\infty}\frac{F_t}{q_t}=0$.

{\bf Point \ref{prop1-5}}.  By Proposition 7 in \cite{montrucchio04}, there is a bubble if and only if $\sum_{t=1}^{\infty}\frac{D_t}{q_t}<\infty$.  
However, as we want prove our finding $\sum_{t=1}^{\infty}\frac{D_t}{q_t}\leq \frac{\frac{F_0}{q_0}}{1-\frac{F_0}{q_0}}$, we present a new proof. Equation \eqref{dif-fq} and $q_t\geq F_t$ imply that $F_t/q_t$ is decreasing in $t$. In particular, we have $F_t/q_t\leq F_0/q_0$. From (\ref{dif-fq}) again, we get that 
\begin{align}\label{dif-fq2}
\frac{F_t}{q_t}-\frac{F_{t+1}}{q_{t+1}}=\Big(1-\frac{F_t}{q_t}\Big)\frac{D_{t+1}}{q_{t+1}}\geq \Big(1-\frac{F_0}{q_0}\Big)\frac{D_{t+1}}{q_{t+1}} \text{  } \forall t\geq 0.
\end{align}
Taking the sum over $t$, we have $
\frac{F_0}{q_0}>\frac{F_0}{q_0}-\frac{F_T}{q_T}\geq \Big(1-\frac{F_0}{q_0}\Big)\sum_{t=1}^T\frac{D_{t}}{q_{t}}.$  By letting $T$ tend to infinity, we obtain point \ref{prop1-5}


\end{proof}



\subsection{\bf Proofs of Section \ref{nobubble_section}}
\label{nobubble_section_proof}
\begin{proof}[{\bf  Proof of Proposition \ref{necessitycondition}}]

{\bf (i).}  If there exists a bubbly equilibrium, then, by Lemma \ref{prop1}, we have $\sum_{t=1}^{\infty }D _{t}/q_{t}<\infty$. Since $q_tz_t<e^y_t$ and $z_t=1/G_n^t$, we get that $\sum_{t=1}^{\infty}\frac{D_t}{G_n^te^y_t}<\sum_{t=1}^{\infty }D _{t}/q_{t}<\infty$, a contradiction.\\
{\bf (ii)a.} According to  Lemma \ref{prop1}\ref{prop1-2}, there is no bubble if and
only if $\lim_{t\rightarrow \infty }\dfrac{a_{t}G_n^{t}}{R_{1}\cdots R_{t}}=0$. Since $a_{t}\leq e_{t}^{y}$ and $R_{t}\geq R_{t}^{\ast }$  $\forall t$ (by  Lemma \ref{RtRt*}), we
have  
\begin{align*}
\dfrac{a_{t}G_n^{t}}{R_{1}\cdots R_{t}}<\dfrac{G_n^te^y_t}{R_{1}^{\ast
}\cdots R_{t}^{\ast }} \text{  } \forall  t.
\end{align*}
By our assumption (\ref{bubble1}), there is no bubble. In other words, every equilibrium is bubbleless. 
\\
{\bf (ii)b.} Consider the $T$-truncated economy which is defined as the economy $\mathcal{E}_{OLG}\equiv \mathcal{E}_{OLG}(U^t,(D
_{t})_{t},(e_{t}^{y},e_{t}^{o})_{t})$ except that there is no activity from date $T+1$ on, i.e., households born at date $T$ only consume $c^y_T=e^y_T$ and the budget constraints of household born at date $T-1$ are $c^y_{T-1}+q_{T-1}z_{T-1}\leq e^y_{T-1}, c^o_T\leq e^ o_T+D_Tz_{T-1}$, and $q_T=0$, $z_T=0$. By the standard argument, there is an equilibrium $%
(a_{t}^{T})_{t\leq T}$ for the $T$-truncated economy.

For every fixed $t\geq0$, the budget constraint implies $0\leq a_t^T\leq e_t^y
\text{ for every }T>t.$
Hence, by a standard diagonal argument, there exist an increasing sequence
$T_k\to\infty$ and a sequence $(a_t)_{t\geq0}$ such that  $\lim_{k\to\infty}
a_t^{T_k}= a_t\in[0,e_t^y] \text{ for every fixed }t\geq0.$ Moreover, $a_t<e_t^y$ for every $t$. Indeed, if $a_t=e_t^y$ for some
$t$, then $a_t^{T_k}\to e_t^y>0$ and $R_{t+1}^{T_k}=
G_n\frac{a_{t+1}^{T_k}+d_{t+1}}{a_t^{T_k}}$ is bounded, since $0\leq a_{t+1}^{T_k}\leq e_{t+1}^y$. This contradicts
the Euler equation and the Inada condition as
$e_t^y-a_t^{T_k}\to0$. Thus, $0\leq a_t<e_t^y$ $\forall t\geq0.$

We distinguish two cases. Suppose first that $a_t>0$ for every $t$.
Then
\[
R_{t+1}^{T_k}
=
G_n\frac{a_{t+1}^{T_k}+d_{t+1}}{a_t^{T_k}}
\longrightarrow
R_{t+1}\equiv G_n\frac{a_{t+1}+d_{t+1}}{a_t}
>0.
\]
Passing to the limit in the Euler equation shows that
$(a_t,R_{t+1})_{t\geq0}$ satisfies the equilibrium conditions and, in
particular, $a_{t+1}=a_t\frac{R_{t+1}}{G_n}-d_{t+1}.$

s
For every fixed $t$, iterating the asset-pricing equation in the
$T_k$-truncated economy and using $R_s^{T_k}\geq R_s^*$ gives
\[
0\leq
a_0^{T_k}
-\sum_{s=1}^{t}
\frac{G_n^s d_s}{R_1^{T_k}\cdots R_s^{T_k}}
\leq
\sum_{s=t+1}^{\infty}
\frac{G_n^s d_s}{R_1^*\cdots R_s^*}.
\]
Letting $k\to\infty$ and then $t\to\infty$, our assumption
$\sum_{s\geq1}\frac{G_n^s d_s}{R_1^*\cdots R_s^*}
=
\sum_{s\geq1}\frac{D_s}{R_1^*\cdots R_s^*}<\infty$
implies $a_0=\sum_{s=1}^{\infty}
\frac{G_n^s d_s}{R_1\cdots R_s}
=f_0.$
Hence the limiting equilibrium is bubbleless.

Suppose now that $a_\tau=0$ for some $\tau$, and let $\tau$ be the
first such date. We claim that $a_{\tau+1}+d_{\tau+1}=0.$
Otherwise, $
R_{\tau+1}^{T_k}= G_n\frac{a_{\tau+1}^{T_k}+d_{\tau+1}}{a_\tau^{T_k}}
\to +\infty.$
However, using the asset-pricing equation, the Euler equation can be
written as
\[
R_{\tau+1}^{T_k}
=
\frac{
U_1^\tau\!\left(
e_\tau^y-a_\tau^{T_k},
e_{\tau+1}^o+G_n(a_{\tau+1}^{T_k}+d_{\tau+1})
\right)}
{
U_2^\tau\!\left(
e_\tau^y-a_\tau^{T_k},
e_{\tau+1}^o+G_n(a_{\tau+1}^{T_k}+d_{\tau+1})
\right)}.
\]
If $a_{\tau+1}+d_{\tau+1}>0$, the right-hand side
converges to a point in $\mathbb{R}_{++}^2$, so continuity implies that
the right-hand side is bounded, a contradiction. Therefore,
$a_{\tau+1}=d_{\tau+1}=0$, and repeating the argument gives
$a_s=d_s=0$ $\forall s\geq\tau+1.$ Since $a_t>0$ for $t<\tau$, the limiting Euler and asset-pricing
equations hold up to date $\tau$. Iterating them yields
$$a_0= \sum_{s=1}^{\tau}
\frac{G_n^s d_s}{R_1\cdots R_s} + \frac{G_n^\tau a_\tau}{R_1\cdots R_\tau}
= \sum_{s=1}^{\infty}\frac{G_n^s d_s}{R_1\cdots R_s} =f_0,
$$
where the second equality follows from $a_\tau=0$ and
$d_s=0$ for all $s\geq\tau+1$. Hence the limiting equilibrium is
bubbleless also in this case.
 \end{proof}

\subsection{\bf Proofs for Section \ref{continuum-section}}
\label{prooftheorem1new}


\begin{proof}[\textbf{Proof of Theorem \ref{new5continuum}}]Since $U^t$ is strictly concave, the system \eqref{iff_equilibrium} and $z_t=1/G_n^t $ are necessary and sufficient to characterize an equilibrium.\\
{\bf Step 1}. Observe that condition \eqref{existence-bubbleles} is satisfied if  there exists $T$ such that $\frac{R_{t+1}^{\ast }}{G_n}\frac{d_{t}}{d_{t+1}}\geq \gamma >1$  $\forall t\geq T$. Indeed, the latter condition implies that \begin{align*}
\frac{R_{t+1}^{\ast }\cdots R_{t+s}^{\ast }}{G_n^{s}}\frac{d_{t}}{d_{t+s}}&
\geq \gamma ^{s}
\Rightarrow \sum_{s\geq 1}\frac{G_n^{s}}{R_{t+1}^{\ast }\cdots R_{t+s}^{\ast }}d_{t+s} \leq d_{t}\sum_{s\geq 1}\frac{1}{\gamma ^{s}}<\infty \text{  } \forall  t\geq
T.
\end{align*}
By consequence, Proposition \ref{necessitycondition}\ref{existence} implies that there exists a bubbleless equilibrium.
\\
{\bf Step 2}. We now construct a continuum of bubbly equilibria.  Given $\epsilon\geq 0$ and $t\geq 0$, we can define $R^{\epsilon}_{t+1}\equiv \max\{R\in [0,G_n]: K_t(\epsilon,R)\geq 0\}$.

Let $a_0$ be such that $\lambda d_0\leq a_0\leq \epsilon_0$.

Since $U^t$ is concave, the function $K_0$ is increasing in the first component and $a_0\leq \epsilon_0$, we have $K_0(a_0,G_n)\leq K_0(\epsilon_0,G_n)<0$ (because of our assumption $K_t(\epsilon_t,G_n)< 0$ $\forall t$). Note that $K_0(a_0,0)=U^0_1(e_{0}^{y}-a_{0},e_{1}^{o})>0$. So, by the intermediate value theorem, there exists $R_1\in (0,G_n)$ such that $K_0(a_0,R_1)=0$, i.e.,
$$K_0(a_0,R_1)\equiv U^0_1(e_{0}^{y}-a_{0},e_{1}^{o}+R_1a_{0})-
R_1U^0_2\left(e_{0}^{y}-a_{0},e_{1}^{o}+R_1a_{0}\right)=0.$$

Since $\epsilon_0\geq a_0$, we have $K_0(\epsilon_0,R_1)\geq K_0(a_0,R_1)= 0$. So, $K_0(\epsilon_0,R_1)\geq  0$. By the definition of $R_1^{\epsilon_0}$ and the fact that $R_1\in (0,G_n)$, we have $R_1\leq R_1^{\epsilon_0}\leq G_n$. 


 Then, we determine $a_1$ by  $
a_{1} =\frac{R_{1}}{G_n}a_{0}-d_{1}.$ We observe that 
\begin{align}
a_{1}& =\frac{R_{1}}{G_n}a_{0}-d_{1}\leq \frac{G_n}{G_n}
\epsilon_{0}-d_{1}=\epsilon_{0}-d_{1}\leq \epsilon_1 \text{ (by condition \ref{1a} in Theorem \ref{new5continuum}).}
\end{align}
We now give a lower bound of $a_1$. By using our assumption $R_1^*\geq \gamma \frac{D_{1}}{D_{0}}$ (i.e.,   $\frac{R_{1}^{\ast }}{G_n}\frac{d_{0}}{d_{1}}\geq \gamma$) and the definition of $a_1$, we have
\begin{equation*}
\frac{a_{1}}{d_{1}}=\frac{R_{1}}{G_n}\frac{d_{0}}{d_{1}}\frac{a_{0}}{d_{0}}%
-1\geq \frac{R_{1}^{\ast }}{G_n}\frac{d_{0}}{d_{1}}\frac{a_{0}}{d_{0}}-1\geq
\gamma \frac{a_{0}}{d_{0}}-1\geq \gamma \lambda -1>\lambda .
\end{equation*}
where we use  $R_t\geq R^*_t$ (see Lemma \ref{RtRt*}).
To sum up, we have $\lambda d_1\leq a_1\leq \epsilon_1,  
\frac{a_{1}}{d_{1}}\geq \gamma \frac{a_{0}}{d_{0}}-1$. Suppose that we can construct $(a_0,a_1,\ldots,a_t)$ with $$\lambda d_s\leq a_s \leq \epsilon_s,\quad \frac{a_{s}}{d_{s}}\geq \gamma \frac{a_{s-1}}{d_{s-1}}-1 \text{ } \forall s\leq t.$$
Let us look at date $t+1$. Since $a_t\leq \epsilon_t$, we have $K_t(a_t,G_n)\leq K_t(\epsilon_t,G_n)< 0$ (because of our assumption $K_t(\epsilon_t,G_n)< 0$ $\forall t$). Note that $K_t(a_t,0)=U^t_1(e_{t}^{y}-a_{t},e_{t+1}^{o})>0$. So, by the intermediate value theorem, there exists  $R_{t+1}$ such that $K_t(a_t,R_{t+1})=0$. By using the same argument, we have $R_{t+1}\leq R^{\epsilon_t}_{t+1}\leq G_n$ and 
\begin{align}
a_{t+1}& =\frac{R_{t+1}}{G_n}a_{t}-d_{t+1}\leq \frac{G_n}{G_n}%
\epsilon_{t}-d_{t+1}=\epsilon_t-d_{t+1}\leq \epsilon_{t+1},\\
\frac{a_{t+1}}{d_{t+1}}&=\frac{R_{t+1}}{G_n}\frac{d_{t}}{d_{t+1}}\frac{a_{t}}{d_{t}}%
-1\geq \Big(\frac{R_{t+1}^*}{G_n}\frac{d_{t}}{d_{t+1}}\Big)\frac{a_{t}}{d_{t}}
-1\geq \gamma \frac{a_{t}}{d_{t}}
-1\end{align}
We have constructed an equilibrium $(a_t)$ with  $a_{t}\in (0,\epsilon_t)\subset (0,e^y_t)$ with $\frac{a_{t+1}}{d_{t+1}}\geq \gamma \frac{a_t}{%
d_t}- 1$ and $a_t/d_t\geq \lambda>0$ $\forall t$.

We now prove that this equilibrium is bubbly. Define $x$ by $(\gamma -1)x=1$. We have 
$$ \frac{a_{t+1}}{d_{t+1}}-x\geq \gamma \left( \frac{a_{t}}{d_{t}}-x\right) 
\text{  }\forall t\geq 0, \text{ which implies that }  \frac{a_{t}}{d_{t}}-x\geq \gamma ^{t}(\frac{a_{0}}{d_{0}}-x) \text{ } \forall t.
$$

Note that $\frac{a_{0}}{d_{0}}\geq \lambda >\frac{1}{\gamma -1}=x$. Hence, $%
\frac{a_{0}}{d_{0}}-x>0$. By consequence, we have 
\begin{align*}
\frac{{a_{t}}}{{d_{t}}}& \geq x+\gamma ^{t}(\frac{a_{0}}{d_{0}}-x)>\gamma
^{t}(\frac{a_{0}}{d_{0}}-x) \Rightarrow \sum_{t\geq 1}\frac{d_{t}}{a_{t}} \leq \sum_{t\geq 0}\frac{1}{\gamma ^{t}(\frac{a_{0}}{d_{0}}-x)}=\frac{1}{\frac{a_{0}}{d_{0}}-x}\frac{1}{1-\frac{1}{\gamma }}.
\end{align*}%
Therefore, $\sum_{t\geq 1}{d_{t}}/{a_{t}}<\infty $. It means that this equilibrium is bubbly.


\end{proof}



\begin{proof}[{\bf Proof of Corollary \ref{theorem13-choose}}]
 Recall that $a_{t} \equiv \frac{q_{t}}{G_n^{t}}, d_{t} \equiv \frac{D _{t}}{G_n^t}$.  
We prepare our proof by two lemmas whose proofs can be found in Online Appendix 1.
\begin{lemma}
\label{5} Let Assumptions \ref{assum0}, \ref{assum1} 
be satisfied.  \\
(1) Define $\mathcal{D}_t\equiv \{a \in (0,e^y_t): au^{\prime}(e_{t}^{y}-a)< \lim_{c\to \infty }cv^{\prime }(c)\}$. For every $a\in \mathcal{D}_t$, there exists a unique value $R_{t+1}=g_t(a)$ such that $u^{\prime }(e_{t}^{y}-a)=R_{t+1}v^{\prime
}\left( e_{t+1}^{o}+R_{t+1}a\right)$. The function $g_t: \mathcal{D}_t\to \rr_+$ defined in that way is increasing and satisfies  
$\lim_{a\rightarrow 0}g_{t}(a)=R_{t+1}^{\ast }\equiv {u^{\prime}(e_{t}^{y})}/{ v^{\prime }(e_{t+1}^{o})}.$



(2)  The sequence $(a_t)$ is an equilibrium if and
only if it satisfies the system (\ref{system1}) and 
$R_{t+1} =g_{t}(a_{t})$.
\end{lemma}

Note that $ \lim_{c\to \infty }cv^{\prime }(c)$ may be infinity. Indeed, if $v(c)=\frac{c^{1-\sigma}}{1-\sigma}$ with $\sigma\in (0,1)$, then $\lim_{c\to \infty }cv^{\prime }(c)=\infty$. If $v(c)=\ln(c)+A\ln(B+c^{\sigma})$, where $A\geq 0, B\geq 0, \sigma\in (0,1)$, then $\lim_{c\to \infty }cv^{\prime }(c)<\infty$. For more properties of the function $g_t$, see our working paper version - \citet[pages 18-19]{blp25}.

\begin{lemma}
\label{interval1} Under Assumptions \ref{assum0} and \ref{assum1}, 
 the following statements hold. (i) The set $\mathcal{A}_{0}$ is a compact interval. (ii) The fundamental value function $f_{t}\left( a_{0}\right) $ is
decreasing in the initial value $a_{0}$ while the size of bubble $b_t(a_0)$ is strictly increasing. (iii) There exists at most one bubbleless solution. Moreover, if there are
two equilibria with initial asset values $a_{1,0}<a_{2,0}$, then any
equilibrium with initial asset value $a_{0}\in (a_{1,0},a_{2,0}] 
$ is bubbly. (iv) If condition (\ref{keycond}) or condition (\ref{bubble1}) holds, there exists a unique equilibrium and this is bubbleless.\footnote{Point (iv) is a direct consequence of point (iii) and Proposition \ref{necessitycondition}.}
\end{lemma}

We now come back to the proof of Corollary \ref{theorem13-choose}.\\
{\bf Step 1}. We show that we can choose $\bar{\epsilon}_1,\bar{\epsilon}_2,R>0$ such that 
\begin{align}\label{epsilon12}
\frac{u^{\prime}(e^y_t-\epsilon_1)}{ v^{\prime }(e^{o}+\epsilon_2)}<R<G_n
\text{ $\forall \epsilon_1\in (0,\bar{\epsilon}_1), \epsilon_2\in (0,\bar{\epsilon}_2), t\geq 0$.}
\end{align} 
Indeed, since $\frac{u^{\prime}(e_{t}^{y})}{  v^{\prime }(e^{o})}\leq \bar{R}<G_n$, we can choose $R,R'\in (\bar{R},G_n)$ with $R'<R$ and $\bar{\epsilon}_2>0$ such that $\frac{u^{\prime}(e_{t}^{y})}{  v^{\prime }(e^{o}+\epsilon_2))}<R'$ $\forall t, \forall \epsilon_2\in (0,\bar{\epsilon}_2)$. It implies that $e^y_t>(u^{\prime})^{-1}\big(R'  v^{\prime }(e^{o}+\epsilon_2)\big)>(u^{\prime})^{-1}\big(R  v^{\prime }(e^{o}+\epsilon_2)\big)$ since the inverse function $(u^{\prime})^{-1}$ of $u^{\prime}$ is decreasing.
So, we can take $\bar{\epsilon}_1>0$ such that $e^y_t-\epsilon_1>(u^{\prime})^{-1}\big(R  v^{\prime }(e^{o}+\epsilon_2)\big)$  $\forall \epsilon_1\in (0,\bar{\epsilon}_1)$. It means that we have (\ref{epsilon12}).\\
{\bf Step 2}.  We show that we can choose $\bar{\epsilon}>0$ small enough such that $g_t({\epsilon})<R$ and $\bar{\epsilon} u^{\prime }(\underline{e}-\bar{\epsilon})<  \lim_{c\rightarrow \infty
}cv^{\prime }(c) $ $\forall \epsilon\in (0,\bar{\epsilon}]$, where recall that $g_t$  is defined in Lemma \ref{5}. First, we can choose $\bar{\epsilon}>0$ such that $\bar{\epsilon}<\min (\epsilon_1,\underline{e})$ and 
\begin{align}
0<\frac{e^o\bar{\epsilon}u'(\underline{e}-\bar{\epsilon})}{  e^ov'(e^o)-\bar{\epsilon}u'(\underline{e}-\bar{\epsilon})}<\epsilon_2.
\end{align}

Now, let $\epsilon\in (0,\bar{\epsilon}]$ and $X=g_t(\epsilon)$. By the definition of $g_t(\epsilon)$, we have $X=\frac{u'(e^y_t-\epsilon)}{  v'(e^o+X\epsilon)}$ and hence 
\begin{align*}
u'(e^y_t-\epsilon)=X  v'(e^o+X\epsilon)=\frac{X}{e^o+X\epsilon} (e^o+X\epsilon)v'(e^o+X\epsilon)\geq \frac{X}{e^o+X\epsilon}  e^ov'(e^o)
\end{align*}
because the function $cv'(c)$ is increasing. Since $e^y_t\geq \underline{e}>0$, we have $u'(e^y_t-\epsilon)\leq u'(\underline{e}-\epsilon)$. Thus, $X  e^ov'(e^o)\leq u'(\underline{e}-\epsilon)(e^o+X\epsilon)$ and hence
\begin{align}\epsilon X\leq \frac{\epsilon u'(\underline{e}-\epsilon)e^o}{  e^ov'(e^o)-\epsilon u'(\underline{e}-\epsilon)}.
\end{align}
Since $\epsilon u'(\underline{e}-\epsilon)$ is increasing in $\epsilon$ and $\epsilon\leq \bar{\epsilon}$, we have
\begin{align}
X\epsilon \leq \frac{\bar{\epsilon} u'(\underline{e}-\bar{\epsilon})e^o}{  e^ov'(e^o)-\bar{\epsilon} u'(\underline{e}-\bar{\epsilon})}<\epsilon_2.
\end{align}
By (\ref{epsilon12}), we have $
X=\frac{u'(e^y_t-\epsilon)}{  v'(e^o+X\epsilon)}<R.$ It means that $g_t(\epsilon)<R<G_n$ $\forall t, \forall\epsilon\in (0,\bar{\epsilon}]$. 

Since $x u^{\prime }(\underline{e}-x)$ is continuous, increasing in $x$ and $\lim_{x\to 0} xu^{\prime }(\underline{e}-x)=0$, we can actually choose $\bar{\epsilon}$ low enough so that
$\bar{\epsilon} u^{\prime }(\underline{e}-\bar{\epsilon})<  \lim_{c\rightarrow \infty
}cv^{\prime }(c) $.\\
{\bf Step 3}. Take $\epsilon_t=\bar{\epsilon}>0$. We verify conditions \ref{continuum-condition1} and \ref{continuum-condition2} in Theorem \ref{new5continuum}.  First, we have $\epsilon_t\leq \epsilon_{t+1}+d_{t+1}$ because $\epsilon_t=\bar{\epsilon}>0$ and $d_t\geq 0$ $\forall t$.

Recall that Assumption \ref{assum1} 
allows us express $R_{t+1}$ as an increasing function $g_t(a_t)$ of $a_t$ (see Lemma \ref{5}). We verify that: $K_t(\epsilon_t,G_n)< 0$.  
Indeed, with our separable utility function, the function $K_t$ in Definition \ref{definition-functionK} becomes $K_t(a,R)=u'(e_{t}^{y}-a)-R v'(e_{t+1}^{o}+Ra).$ By the definition of $g_t$ (see Lemma \ref{5}), we have  $K_t(\epsilon_t,g_t(\epsilon_t))=0$. Since the function $K_t(\epsilon_t,R)$ is strictly decreasing in $R$, condition $g_t(\epsilon_t)< G_n$ implies that $K_t(\epsilon_t,G_n)<0$.

Note that by the definition of $g_t(\epsilon_t)$ and $R^{\epsilon}_{t+1}$, we have $g_t(\epsilon_t)=R^{\epsilon}_{t+1}$


 We now verify condition  \ref{continuum-condition2} in Theorem \ref{new5continuum}. Take  $\gamma \in (0,{\underline{R}}/{G_d})$ and $d_0,\lambda$, so that $\gamma>1+{1}/{\lambda}$, $\lambda d_0<\bar{\epsilon}$. Then $$\frac{R^*_{t+1}}{G_n\frac{d_{t+1}}{d_t}}\geq \frac{\underline{R}}{G_d}>\gamma \text{ and }\lambda d_t\leq  \lambda  d_0(\frac{G_d}{G_n})^{t}<\lambda d_0\leq \bar{\epsilon},$$ 
where we use ${d_{t+1}}/{d_t}\leq {G_d}/{G_n}$ $\forall t$ and $G_d<G_n$. 

\end{proof}
\subsection{\bf Proofs for Section \ref{onlybubbly-section}}
\label{onlybubblyproof}

\begin{proof}[{\bf Proof of Theorem \ref{onlybubbly}}]

{\bf Part \ref{onlybubblypart1}}. Denote $s_t\equiv {a_t}/{e^y_t}$ the saving rate of the household born at date $t$. We prove that $\liminf_{t\to\infty}s_t>0$ for any equilibrium. 

Let $\bar{\epsilon}\in (0,1)$, a positive sequence $(X_t)$, and  a date $T$ be those of Condition (B).

Take an equilibrium. Suppose that $\liminf_{t\to\infty}s_t=0$. Then there exists $t_0\geq T$ such that $s_{t_0}<\bar{\epsilon}<1$. So, $0<a_{t_0}<e^y_{t_0}$. By consequence, we have the Euler condition.
\begin{align}\frac{V^{t_0}_1\left(1-s_{t_0},g_{e,t_0+1}+R_{t_0+1}s_{t_0}\right)}{V^{t_0}_2\left(1-s_{t_0},g_{e,t_0+1}+R_{t_0+1}s_{t_0}\right)}=
R_{t_0+1}
\end{align}
By the asset pricing equation, we have $a_{t_0+1}=R_{t_0+1}e^y_{t_0}s_{t_0}/G_n-d_{t_0+1}$. Since $a_{t_0+1}\in [0,e^y_{t_0+1}]$, we have $0\leq e^y_{t_0} R_{t_0+1}s_{t_0}/G_n-d_{t_0+1}\leq e^y_{t_0+1}$. 
By  condition (B\ref{onlybubbly3}), we have $R_{t_0+1}\leq X_{t_0+1}$. Since $X_{t_0+1}\leq  G_n{e^y_{t_0+1}}/{e^y_{t_0}}$, we have $R_{t_0+1}\leq   G_n{e^y_{t_0+1}}/{e^y_{t_0}}$. Combining with  $a_{t_0+1}+d_{t_0+1} =a_{t_0}{R_{t_0+1}}/{G_n}$, we get
\begin{align}
a_{t_0+1}\leq a_{t_0}\frac{R_{t_0+1}}{G_n}&\leq a_{t_0}\frac{e^y_{t_0+1}}{e^y_{t_0}} 
\Rightarrow s_{t_0+1}\equiv \frac{a_{t_0+1}}{e^y_{t_0+1}}\leq  \frac{a_{t_0}}{e^y_{t_0}}=   s_{t_0}<\bar{\epsilon}.
\end{align}
Therefore, by induction, we have, 
$s_{t+1}\leq   s_{t} <\bar{\epsilon}$ and $R_{t+1}\leq X_{t+1}$ $\forall t\geq t_0$. This implies that $R_{t_0+1}\cdots R_t\leq X_{t_0+1}\cdots X_t$ $\forall t>t_0$. 
We now look at the fundamental value
\begin{align}
F_0&=\sum_{t\geq 1}\frac{D_t}{R_1\cdots R_t}=\sum_{t=1}^{t_0}\frac{D_t}{R_1\cdots R_t}+\sum_{t=t_0+1}^{\infty}\frac{D_{t}}{R_1\cdots R_{t_0}R_{t_0+1}\cdots R_{t}}.
\end{align}
Consider the second term $A_0\equiv \sum_{t=t_0+1}^{\infty}\frac{D_{t}}{R_1\cdots R_{t_0}R_{t_0+1}\cdots R_{t}}$.
We have 
\begin{align}
A_0&= \frac{1}{R_1\cdots R_{t_0}}\sum_{t=t_0+1}^{\infty}\frac{D_{t}}{R_{t_0+1}\cdots R_{t}}\geq \frac{1}{R_1\cdots R_{t_0}}\sum_{t=t_0+1}^{\infty}\frac{D_{t}}{X_{t_0+1}\cdots X_t}\\
&=\frac{X_{1}\cdots X_{t_0}}{R_1\cdots R_{t_0}}\sum_{t=t_0+1}^{\infty}\frac{D_{t}}{X_{1}\cdots X_t}=\infty
\end{align}
because of our assumption (B\ref{onlybubbly2}), i.e., $\sum_{t=1}^{\infty}\frac{D_t}{X_{1}\cdots X_t}=\infty$. This implies that $F_0=\infty$. Since $q_0\geq F_0$, we have  $q_0=\infty$, a contradiction. So, we have $\liminf_{t\to\infty}s_t>0$.\\
{\bf Parts \ref{onlybubblypart2} and \ref{onlybubblypart22}}.  Let the condition in the first statement be satisfied. According to   Theorem \ref{onlybubbly}\ref{onlybubblypart1}, any equilibrium satisfies $\liminf_{t\to\infty}s_t>0$.  By combining with Lemma \ref{prop1}\ref{prop1-5}, condition $\liminf_{t\to\infty}s_t>0$ and   our assumption  $\sum_{t=1}^{\infty}\frac{D_t}{G_n^te^y_t}<\infty$, we obtain that this equilibrium is bubbly.

If $\sum_{t=1}^{\infty}\frac{D_t}{G_n^te^y_t}=\infty$, then $\sum_{t=1}^{\infty}\frac{d_t}{a_t}\geq \sum_{t=1}^{\infty}\frac{D_t}{G_n^te^y_t}=\infty$. Lemma \ref{prop1}\ref{prop1-5} implies that this equilibrium is bubbleless.\\
{\bf Part \ref{onlybubblypart3}} is a direct consequence of Parts \ref{onlybubblypart2} and \ref{onlybubblypart22}.
\end{proof}

\begin{proof}[{\bf Proof of Lemma \ref{justifyB}}]
{\bf Point \ref{holding-condition1}}.
Take $\bar{\epsilon}\in (0,1)$ be such that $\bar{\epsilon}<\min\{\bar{\epsilon}_0,\bar{\epsilon}_1\}$ and $\bar{\epsilon}M<  \bar{\epsilon}_2$. 
Then, we define $T=T_0$ and $X_t=R$ $\forall t$. Obviously, we have  $X_{t+1}\leq G_n\dfrac{e^y_{t+1}}{e^y_t} $ $\forall t\geq T_0$. 

Now, we prove Condition (B). Take any $t\geq T$, $\epsilon\in (0,\bar{\epsilon})$ and  $X\in \mathcal{R}_t(\epsilon)$, i.e.,
\begin{align}\label{checkRepsilon}X=\frac{V^t_1\left(1-\epsilon,g_{e,t+1}+X\epsilon\right)}{V^t_2\left(1-\epsilon,g_{e,t+1}+X\epsilon\right)} \text{ and } 0\leq e^y_tX\epsilon/G_n-d_{t+1}\leq e^y_{t+1}.
\end{align}
We must prove that $X\leq X_{t+1}=R$.
%
Since $\epsilon\leq \bar{\epsilon}<\bar{\epsilon}_0$ and $X\epsilon\leq m_{t+1}\equiv (d_{t+1}+e^y_{t+1})G_n/e^y_t$, condition \eqref{boundPhi} implies that $\Phi_t(\epsilon,X\epsilon)\leq M$. It means that $X\leq M$. Thus, $X\epsilon\leq M\bar{\epsilon}<\bar{\epsilon}_2$. So, condition \eqref{B-holding} implies that $X=\Phi_t(\epsilon,X\epsilon)\leq R$.

It is easy to see that $\sum_{t\geq 1}\frac{D_t}{X_1\cdots X_t}=\sum_{t\geq 1}\frac{D_t}{R^t}=\infty.$ 

The last step is to prove that $\sum_{t\geq 1}\frac{D_t}{R^t}=\infty$ if $R<\limsup_{t\to\infty}D_t^{1/t}$ or $R\leq \frac{D_{t+1}}{D_t}$ $\forall t\geq T$.
 
Assume that $R\leq \frac{D_{t+1}}{D_t}$ $\forall t\geq T$, we have $$R^{s}\leq\frac{ D_{T+1}}{D_T}\cdots \frac{D_{T+s}}{D_{T+s-1}}=\frac{D_{T+s}}{D_T} \text{ } \forall s\geq 1.$$
 This implies that $\frac{D_{T+s}}{R^{T+s}}\geq \frac{D_T}{R^T}$. Taking the sum over $s$, we get that $\sum_{t\geq 1}\frac{D_t}{X_1\cdots X_t}=\sum_{t\geq 1}\frac{D_t}{R^t}=\infty$.
 
 Assume now that $R<\limsup_{t\to\infty}D_t^{1/t}$. Then, there exists an infinite and increasing sequence $(t_k)_{k\geq 1}$ such that $R<D_{t_k}^{\frac{1}{t_k}}$  $\forall k$, $\forall t_1\geq T$. This implies that $D_{t_k}>R^{t_k}$. So, $\sum_{t\geq 1}\frac{D_t}{R^t}\geq \sum_{k\geq 1}\frac{D_{t_k}}{R^{t_k}}=\infty$. 
 \\
{\bf Point \ref{holding-condition2}}. 
Define $T\equiv T_0$ and take a sequence $(X_t)$ such that $X_{t+1}\equiv \min\big\{\frac{D_{t+1}}{D_t},  \frac{G_n e^y_{t+1}}{e^y_t}\big\}$ $\forall t\geq T_0$. Then  $X_{t+1}\leq G_n\dfrac{e^y_{t+1}}{e^y_t} $ $\forall t\geq T$ and $\sum_{t=1}^{\infty}\frac{D_t}{X_1\cdots X_t}=\infty.$
 
Take $\bar{\epsilon}\in (0,1)$ be such that $\bar{\epsilon}<\min\{\bar{\epsilon}_0,\bar{\epsilon}_1\}$ and $\bar{\epsilon}M<\bar{\epsilon}_2$.

Now, we prove Condition (B). Take any $t\geq T$, $\epsilon\in (0,\bar{\epsilon})$ and  $X\in \mathcal{R}_t(\epsilon)$, i.e., \eqref{checkRepsilon} holds.
Since $\epsilon\leq \bar{\epsilon}<\bar{\epsilon}_0$ and $X\epsilon\leq m_{t+1}\equiv (d_{t+1}+e^y_{t+1})G_n/e^y_t$, condition \eqref{boundPhi} implies that $\Phi_t(\epsilon,X\epsilon)\leq M$. It means that $X\leq M$. Thus, $X\epsilon\leq M\bar{\epsilon}<\bar{\epsilon}_2$. So, condition \eqref{B-holding} implies that $X=\Phi_t(\epsilon,X\epsilon)\leq X_{t+1}$. 
\end{proof}
\begin{proof}[{\bf Proof of Corollary \ref{stronger}}]
 
Recall that \cite{hiranotoda25} assume $G_n=1$. So, it suffices to check   Condition (B) and $\sum_{t=1}^{\infty}\frac{D_t}{e^y_t}<\infty$, then, by applying  our Theorem \ref{onlybubbly}\ref{onlybubblypart2}, we get the result.\\
{\bf Step 1}. We prove that: $G_d\equiv \limsup_{t\to\infty}D_t^{1/t}<G\equiv \lim_{t\to\infty}\frac{e^y_{t+1}}{e^y_t}$ implies our assumption (\ref{onlybubbly1}) for the case $G_n=1$, i.e., $\sum_{t=1}^{\infty}\frac{D_t}{e^y_t}<\infty.$ Let $a,b$ be such that $G_d<a<b<G$. There exists a date $t_0$ such that $D_t^{1/t}<a<b<\frac{e^y_{t+1}}{e^y_t}$ $\forall t\geq t_0$. Therefore, 
\begin{align}
\frac{e^y_t}{e^y_{t_0}}&=
\frac{e^y_t}{e^y_{t-1}}\cdots 
\frac{e^y_{t_0+1}}{e^y_{t_0}}>b^{t-t_0} \text{ } \forall t> t_0.
\end{align}
So, we have $\frac{D_t}{e^y_t}<\frac{a^t}{e^y_{t_0}b^{t-t_0}}$. Since $a<b$, we have $\sum_{t> t_0}\frac{D_t}{e^y_t}<\infty$.
\\
{\bf Step 2}. 
Let us check Condition (B) by verifying conditions in Lemma \ref{justifyB}\ref{holding-condition1}.\\
{\bf Step 2.1}. We prove condition \eqref{boundPhi}. Since $\lim_{t\to\infty}d_t/e^y_t=0$ and $\lim_{t\to\infty}\frac{e^y_{t+1}}{e^y_t}=G\in (0,\infty)$, we have $\lim_{t\to\infty}(d_{t+1}+e^y_{t+1})G_n/e^y_t=G_nG\in (0,\infty)$. Note also that $\lim_{t\to\infty}g_{e,t+1}=\lim_{t\to\infty}{e_{t+1}^{o}}/{e_{t}^{y}}=Gw$. So, $\{(1-\epsilon,g_{e,t+1}+z): \epsilon\in (0,\epsilon_1), z\in [0,(d_{t+1}+e^y_{t+1})G_n/e^y_t]\}$ is  a subset of a compact set $K\subset \mathbb{R}_{++}\times \mathbb{R}_{+}$ when $\epsilon>0$ is small enough and $T$ is high enough.  Combining with the assumption that $f_t$ uniformly converges to $f$, there exists an integer $T>0$ and $\bar{\epsilon}_0, M>0$ such that 
$\Phi_t(\epsilon,z)\leq M \text{ } \forall t\geq T, \forall \epsilon\in (0,\bar{\epsilon_0}), \forall z\in [0,(d_{t+1}+e^y_{t+1})G_n/e^y_t]$.\\
{\bf Step 2.2}. We now prove \eqref{B-holding}. 
Since $f(1,G_e)<G_d$ and the function $f$ is continuous, there exist $\bar{\epsilon}\in (0,\bar{\epsilon}_0)$, $\gamma>0$, $G_1,G_2,G_3\in (0,G_d)$ such that 
\begin{align}\label{step21}
f(1-\epsilon,g)<G_1<G_2<G_3<G_d
\end{align} $\forall \epsilon \in [0,\bar{\epsilon}], 0\leq g\leq G_e+\gamma$.

Since $G_e\equiv \lim_{s\to\infty} \frac{e^0_{s+1}}{e^y_s}$, we can choose $t_0$ such that $\frac{e^0_{s+1}}{e^y_s}<G_e+\frac{\gamma}{2}$ $\forall t\geq t_0$.

Since $G_2<G_d<G\equiv \lim_{t\to\infty}\frac{e^y_{t+1}}{e^y_t}$, we can choose  $t_1$ such that $G_2<\frac{e^y_{t+1}}{e^y_t}$ $\forall t\geq t_1$

Take $\epsilon_f\in (0,G_2-G_1)$.

By Assumption 3 in \cite{hiranotoda25}, there exists $t_2$ such that 
\begin{align}\label{step22}
\sup_{(\epsilon,g)\in [0,\bar{\epsilon}]\times [0,G_e+\gamma]}|f_t(1-\epsilon,g)-f(1-\epsilon,g)|<\epsilon_f \text{ } \forall t\geq t_2.
\end{align} 
\noindent

Let us take $R=G_2$, $\bar{\epsilon}_1\equiv \bar{\epsilon}$, $\bar{\epsilon}_2< \gamma/2$, and $T_0\equiv \max\{T,t_0,t_1,t_2\}$.  By the definition of $R$, we have $R=G_2\leq e^y_{t+1}/e^y_t$ $\forall t\geq T$ and $R<\limsup_{t\to\infty}D_t^{1/t}$.

Now, take any $t\geq T,$ $\epsilon_1\in (0,\bar
{\epsilon}_1)$, $\epsilon_2\in (0,\bar{\epsilon}_2)$. We have $$g_{e,t+1}+\epsilon_2=\frac{e^0_{s+1}}{e^y_s}+\epsilon_2 \leq G_e+\frac{\gamma}{2}+\frac{\gamma}{2}=G_e+\gamma.$$
It implies that $g_{e,t+1}+\epsilon_2\in  [0,G_e+\gamma]$. Then, condition (\ref{step22}) implies that \begin{align}&&|\frac{V^t_1\left(1-\epsilon_1,g_{e,t+1}+\epsilon_2\right)}{V^t_2\left(1-\epsilon_1,g_{e,t+1}+\epsilon_2\right)}-f(1-\epsilon_1,g_{e,t+1}+\epsilon_2)|<\epsilon_f.
\end{align} 
Since $g_{e,t+1}+\epsilon_2\leq G_e+ \gamma$, condition (\ref{step21}) implies that $f(1-\epsilon,g_{e,t+1}+\epsilon_2)<G_1$. Combining with  $\epsilon_f<G_2-G_1$, we have 
\begin{align}&&\frac{V^t_1\left(1-\epsilon_1,g_{e,t+1}+\epsilon_2\right)}{V^t_2\left(1-\epsilon_1,g_{e,t+1}+\epsilon_2\right)}\leq \epsilon_f + G_1\leq G_2=R.
\end{align}
\end{proof}

\subsection{\bf Proof of Theorem \protect\ref{allsets}}
\label{prooftheorem2} 
Here under stationary endowments, the function $g_t$, defined by Lemma \ref{5}, does not depend on $t$. So, we write $g$ instead of $g_t$. We summarize our equilibrium system. 
\begin{subequations}
\label{system-stationary}
\begin{align}
u^{\prime }(e^{y}-a_{t})& = R_{t+1}v^{\prime }\left(
e^{o}+R_{t+1}a_{t}\right), & R_{t+1} &=g(a_{t}),\\
a_{t+1}+d_{t+1}& =a_{t}\frac{R_{t+1}}{G_n},&  
0&<a_{t} <e^{y} \text{  } \forall  t\geq 0.
\end{align}\end{subequations}

We are inspired by the strategy of \cite{tirole85}, \cite{bhlpp18}, \cite{phamtoda2025}. 
We need several steps. The following claim is immediate.
\begin{lemma}
\label{3cases}Consider a solution to the system (\ref{system-stationary}). 
Only three mutually exclusive cases hold: \textbf{Case A}. $R_{t}\geq R_{t-1} \text{  } \forall  t$; \textbf{Case B}: There exists $t$ such that $R_t<R_{t-1}$ and $R_t\leq G_n$; \textbf{Case C}: There exists $t$ such that $R_t<R_{t-1}$ and for any $t_0$ satisfying $R_{t_0}<R_{t_0-1}$, we have $R_{t_0}> G_n$.
\end{lemma}
\begin{lemma}
\label{tl2g2}Consider the system (\ref{system-stationary}). Consider an equilibrium $a_{0}$. Assume that $R_{t}<R_{t-1}$ and $R_{t}\leq
G_n$ for some $t$. Then $R_{t}$ converges to $R^{\ast }$, $a_{t}$ converges
to zero, and $
R^{\ast } <G_n.$
\end{lemma}

\begin{lemma}
\label{additional} Consider the system (\ref{system-stationary}). Consider an equilibrium $a_{0}$. If $a_t$ converges to $0$, then $R_t$ converges to $%
g(0)=R^*$.
\end{lemma}
Lemma \ref{additional} directly follows the definition of $g$, i.e., $R_{t}=g(a_{t-1}) \text{  } \forall  t$. Lemma \ref{tl2g2} is related to \citet[Lemma 2]{tirole85} and its proof is presented in Online Appendix 3.  The following result is a key, which can be viewed as a generalized version of   \citet[Lemma 3]{tirole85}.

\begin{lemma}
\label{tl3g2}

Consider the system (\ref{system-stationary}). Assume that 
\begin{align} 
\label{sumdt}\sum_{t\geq 1}d_t<\infty
\end{align}
Consider an equilibrium. One of the following cases must hold.

\begin{enumerate}
\item \label{c1} The equilibrium is bubbly, $(a_t,b_t,R_t)$ converges to $%
(0,0,R^*)$ and $R^*<G_n$.
\item \label{c2}  $(a_{t},b_{t},R_{t})$ converges
to $(\hat{a},\hat{b},G_n)$ with $\hat{a}\geq \hat{b}$ and $\hat{a}$ is uniquely determined by  $u'(e^y-\hat{a})= G_n v'(e^o+G_n\hat{a})$ (i.e., $G_n=g(\hat{a})$).
%
%
\item \label{c3} This equilibrium is bubbleless, (i.e., $b_{t}=0 \text{  } \forall  t$), and $(a_{t},b_t,R_{t})=(a_t,0,R_t)$ converges to $(0,0,R^{\ast })$.
\end{enumerate}
\end{lemma}
\begin{proof}[{Proof of Lemma \ref{tl3g2}}]

We consider three cases: A, B, C in Lemma \ref{3cases}.\\
\textbf{Case A}. If $R_{t}\geq R_{t-1} \text{  } \forall  t$. Then $R_{t}$ converges.\\
\textbf{Case A.1}. If $R_{t_0}>G_n$ for some $t_0$, then $R_{t}\geq R_{t_0}>G_n$ $\forall  t\geq t_0$. Then, there is no bubble. Indeed, if $b_0>0$, then  $b_t$ converges to infinity (because $b_{t+1}=b_{t}\frac{R_{t+1}}{{G_n}}$ $\forall t$), which is a contradiction.

So, there is no bubble. Since there is no bubble, we have 
\begin{equation*}
a_{t}=f_{t}=\sum_{s=1}^{\infty }\frac{G_n}{R_{t+1}}\cdots \frac{G_n}{R_{t+s}}%
d_{t+s}\leq \sum_{s=1}^{\infty }d_{t+s} \text{  } \forall  t\geq t_{0}.
\end{equation*}%
Using condition (\ref{sumdt}), we obtain that $a_{t}$ converges to $0$. Hence, $R_{t}=g(a_{t-1})$ converges to $g(0)=R^{\ast }$. To sum up, we have 
$(a_t,b_t,R_t)=(a_t,0,R_t)\to (0,0,R^*) \text{ with } R^*>G_n.$  We are in the case \ref{c3} of Lemma \ref{tl3g2}.\\
\textbf{Case A.2}:  $R_t\leq G_n \text{  } \forall  t$, then we have   $R_t\geq R_{t-1}$ and $R_t\leq G_n \text{  } \forall  t$. This implies that $R_t$ converges to some value $R$ with $0\leq R\leq G_n$. There are two subcases.\\
{\bf Case A.2.1}: $R<G_n$. Then, $b_{t}$ converges to zero (because $b_{t+1}=\frac{%
R_{t+1}}{G_n}b_{t}$) and $a_{t}$ converges to zero because $\lim_{t\to\infty }R_t<G_n$ and  $
a_{t+1}=\frac{R_{t+1}}{G_n}a_{t}-d_{t+1}<\frac{R_{t+1}}{G_n}a_{t} \text{  } \forall  t.$ So, we have $(a_t,b_t,R_t)\to (0,0,R^*) \text{ with } R^*<G_n$. We are in the case \ref{c1} or the case \ref{c3} in Lemma \ref{tl3g2}.\\
{\bf Case A.2.2}: $R=G_n$. Since $b_{t+1}=b_{t}\frac{R_{t+1}}{{G_n}}$ and $R_{t}\leq G_n \text{  } \forall  t$, the
sequence $b_{t}$ is decreasing and hence converges to some value $\hat{b}$. Look at the sequence $(a_t)$. There are two cases.

First, if $R_t<G_n$ $\forall t$. We have $
a_{t+1}=\frac{R_{t+1}}{G_n}a_{t}-d_{t+1}\leq \frac{R_{t+1}}{G_n}a_{t}<a_t \text{  } \forall  t.$ 
So, $a_t$ must converge to some value $\hat{a}$ with $\hat{a}\geq \hat{b}$ because $a_t\geq b_t$ $\forall t$. We are in the case \ref{c2} of Lemma \ref{tl3g2}.


Second, if there exists $T$ such that $R_T=G_n$ for some $T$, then $R_t=G_n$ $\forall t\geq T$ (because $(R_t)$ is increasing and $\lim_{t\to\infty}R_t=R=G_n$). Therefore, for any $t>T$, we have 
\begin{equation*}
f_{t}=\sum_{s=1}^{\infty }\frac{G_n^{s}d_{t+s}}{R_{t+1}\cdots R_{t+s}}=\sum_{s\geq 1}d_{t+s}.
\end{equation*}
So, by assumption (\ref{sumdt}), $f_t$ converges to zero. It means that $\lim_{t\to\infty}a_t=\lim_{t\to\infty}b_t$.

To sum up, we have $(a_t,b_t,R_t)\to (\hat{a},\hat{a},G_n).$  We are in the case \ref{c2} of Lemma \ref{tl3g2}.\\
\textbf{Case B}: There exists $t$ such that $R_{t}<R_{t-1}$ and $R_{t}\leq G_n$. Then, by using Lemma \ref{tl2g2}, we have $\lim_{t\to\infty}R_t<G_n$ and $\lim_{t\to\infty}a_t=0$. Therefore, $(a_t,b_t,R_t)\to (0,0,R^*) \text{ with } R^*<G_n$.
So, the equilibrium is either in the case \ref{c1} or in the case \ref{c3}
of our proposition. \\
%
%
\textbf{Case C}: There exists $t$ such that $R_t<R_{t-1}$, and for any $t_0$
satisfying $R_{t_0}<R_{t_0-1}$, we have $R_{t_0}> G_n$. In this case, we have $%
R_{t}>G_n$ since $R_t<R_{t-1}$. We claim $R_{t+1}>G_n$. Indeed, if $R_{t+1}\leq
G_n $, we have $R_{t+1}\leq G_n<R_t$. Since $R_{t+1}<R_t$, we have $R_{t+1}>G_n$, a contradiction.

By induction, we have $R_{t+\tau}> G_n,$ $\forall \tau\geq 0$. This implies that $%
\liminf_{t\to\infty}R_t\geq G_n$.\\
\textbf{Case C.1}: $b_{0}>0$. Since $b_t\leq a_t\leq e^y$, the sequence $(b_t)$ is uniformly bounded from above,  $R_{t+\tau }>G_n$ $\forall  \tau \geq 0$ and $b_{t+1}=b_t\frac{R_{t+1}}{G_n}$, we must have $\limsup_{t\rightarrow \infty }R_{t}\leq G_n$ (otherwise, $\limsup_{t\rightarrow \infty }b_{t}=\infty $ which is impossible). So, $\lim_{t\rightarrow \infty }R_{t}=G_n$. 

Again, since $R_{t+\tau }>G_n$ $\forall  \tau \geq 0$, we have
\begin{equation*}
f_{T}=\sum_{s=1}^{\infty }\frac{G_n^{s}d_{T+s}}{R_{T+1}\cdots R_{T+s}}<\sum_{s\geq 1}d_{T+s} \text{ }\forall T> t.
\end{equation*}
By assumption (\ref{sumdt}), we obtain that $f_{t}$ converges to zero.
Since $(b_{t+\tau })_{\tau }$ converges (because it is increasing thanks to $R_{t+\tau }>G_n$), we have $\lim_{t\to\infty}b_{t}>0$. Then $\lim_{t\to\infty}a_{t}=\lim_{t\to\infty}b_{t}>0$. Since $\lim_{t\to\infty}R_{t}=G_n$, we must have $\lim_{t\to\infty}a_{t}=\hat{a}$. Summing up, we have $(a_t,b_t,R_t)\to (\hat{a},\hat{a},G_n).$ 
So, the equilibrium is in the case \ref{c2} of Lemma \ref{tl3g2}.
\\
\textbf{Case C.2}: If $b_{0}=0$, then $b_{h}=0 \text{  } \forall  h\geq 0$. Since $%
R_{t+\tau }>G_n \text{  } \forall  \tau \geq 0$, we have
\begin{equation*}
f_{\tau}=\sum_{s=1}^{\infty }\frac{G_n^{s}d_{\tau+s}}{R_{\tau+1}\cdots R_{\tau+s}}<\sum_{s\geq 1}d_{\tau+s}.
\end{equation*} By combining with our condition (\ref{sumdt}), $f_{t}$
converges to zero. Then $a_{t}=f_{t}+b_{t}$ also converges to zero. To sum
up, we have $R_{t-1}>R_{t}$ and $R_{t+\tau }>G_n \text{  } \forall  \tau \geq 0$. Summing up, $(a_t,b_t,R_t)\to (0,0,R)$ with $R\geq G_n$. The
equilibrium is in the case \ref{c3} of Lemma \ref{tl3g2}. \end{proof}

Lemmas \ref{tl5g2}, \ref{tl7g} below can be viewed as a generalized version of \protect\cite{tirole85}'s Lemmas 5 and Lemma 7 respectively. Their proofs are presented in Online Appendix 3.

\begin{lemma}
\label{tl5g2}

Assume that $\sum_{t\geq 1}d_t<\infty$ and $R^*<G_n$. 
Consider the system (\ref{system-stationary}). There exists at most one bubbly equilibrium $a_{0}$ such that the interest rate $R_{t}$
converges to $G_n$. So, there exists at most one asymptotically bubbly
equilibrium.
\end{lemma}

\begin{lemma}
\label{tl7g}
Consider the system (\ref{system-stationary}). Assume that there exists an equilibrium $(a^b_t)$  satisfying $a^b_t\to a^b\in [0,e^y)$ and $R^b_t\to R^b<G_n$. Then, there exists $\bar{b}_{0}$ such that for any $b_{0}\in (0,\bar{b}_{0})$%
, the sequence $(a_{t})$ defined by 
$a_{0} =a_{0}^{b}+b_{0}, R_{t+1} =g(a_{t}), a_{t+1} =\frac{R_{t+1}}{G_n}a_{t}-d_{t+1}$ 
is a bubbly equilibrium, and we have $\lim_{t\rightarrow \infty }R_{t}<G_n$.
\end{lemma}

\textbf{We now prove Theorem \ref{allsets}}. \\
{\bf Part \ref{part1}} is a  consequence of Lemma \ref{RtRt*} and  Proposition \ref{necessitycondition}.\\
{\bf Part \ref{part2}}. We consider two cases. \\
\textbf{Case 1:} The equilibrium set $\mathcal{A}_0$ (see Definition \ref{defA0})  is singleton.  If $a_{t}$ does not converge to $\hat{a}$ (note that $\hat{a}>0$ because $R^*<G_n$), then by Lemma  \ref{tl3g2}, $%
a_{t}$ must converge to zero. Hence, $R_{t+1}=g(a_{t})$ converges to $%
g(0)=R^{\ast }$. By our assumption $R^{\ast }<G_n$ and Lemma \ref{tl7g}, we
can construct another equilibrium. This is a contradiction.

Therefore, we have  $a_{t}\rightarrow \hat{a}$. Since $\hat{a}>0$, only case \ref{c2} in Lemma \ref{tl3g2} holds. Since $a_{t}\rightarrow \hat{a}>0$, we can take $x>0$ and $t_0$ such that $a_t\geq x$ $\forall t\geq t_0$. Then, 
$\sum_{t\geq t_0}\frac{d_t}{a_t}\leq \sum_{t\geq t_0}\frac{d_t}{x}<\infty$. So, Lemma \ref{prop1} implies that this equilibrium is bubbly. By Lemma \ref{prop1}\ref{prop1-4}, we have $\lim_{t\to\infty}b_t/a_t=1$. Hence $\lim_{t\to\infty}b_t=\lim_{t\to\infty}a_t=\hat{a}$.\\
\textbf{Case 2:} The equilibrium set $\mathcal{A}_0$ is not singleton. By Lemma \ref%
{interval1}, the equilibrium set is a compact interval, denoted by $[%
\underline{a},\bar{a}]$. For any $a_0>\underline{a}$, the equilibrium is bubbly (by point 2 of Lemma \ref{interval1}).

For $a_0=\bar{a}$, the equilibrium is bubbly and $a_{t}\rightarrow \hat{a}$.
Indeed, if $a_{t}$ does not converge to $\hat{a}$, then by Lemma \ref%
{tl3g2}, $a_{t}$ must converge to zero. By our assumption $R^{\ast }<G_n$ and
Lemma \ref{tl7g}, we can construct another equilibrium with $a_0^{\prime }>a_0=\bar{a}
$. This is a contradiction. So, $\lim_{t\to\infty}a_t=\hat{a}>0$. 
Then, we have $b_t=\frac{R_1\cdots R_t}{G_n^t}b_0>0$. As in the case 1, we have 
 $\lim_{t\to\infty}b_t=\lim_{t\to\infty}a_t=\hat{a}$.

So, by Lemma \ref{tl5g2}, $a_0=\bar{a}$ is the unique bubbly equilibrium satisfying $\lim_{t\to\infty}R_{t}= G_n$. By consequence, for any bubbly equilibrium $a_0\in (\underline{a},\bar{a}) $, we have $\lim_{t\to\infty}R_t\not= G_n$. Therefore, Lemma \ref{tl3g2}  implies that $(a_t,b_t,R_t)$ converges to $%
(0,0,R^*)$ $\forall a_0\in (\underline{a},\bar{a})$.

Last, look at the minimal equilibrium $a_0=\underline{a}$. Take an equilibrium $a'_0$ with $a_0'\in (\underline{a},\bar{a})$. According to the proof of point 2 of  Lemma \ref{interval1}, we have $a_t'>a_t, R_t'>R_t$ $\forall t$. Since we have proved that $(a_t',b_t',R_t')$ converges to $(0,0,R^*)$, we get that $(a_t,b_t,R_t) \to 
(0,0,R^*)$.\\
%
%
%
%
{\bf Proof of Claim 1 in Theorem \ref{allsets}}.  Assume that $R^{\ast } <G_n$ and $\sum_{t\geq 1}\frac{D_{t}}{(R^{\ast })^{t}}<\infty $. Since $\sum_{t\geq 1}\frac{D _{t}}{(R^{\ast
})^t}<\infty $, Proposition \ref{necessitycondition}\ref{existence} implies that
there exists a bubbleless equilibrium. So, case \ref{set2} must hold.\\ %
{\bf Proof of Claim 2 in Theorem \ref{allsets}}. Assume that  $R^{\ast } <G_n$, $R^{\ast}<\limsup_{t\rightarrow \infty }D _{t}^{1/t}$ and $\sum_{t=1}^{\infty}\dfrac{D_t}{G_n^t}<\infty$. Condition $\sum_{t=1}^{\infty}\dfrac{D_t}{G_n^t}<\infty$ implies that $\limsup_{t\rightarrow \infty }D _{t}^{1/t}\leq G_n$ and $\lim_{t\to\infty}d_t=0$. 

Note that $\frac{V^t_1\left(1-\epsilon_1,g_{e,t+1}+\epsilon_2\right)}{V^t_2\left(1-\epsilon_1,g_{e,t+1}+\epsilon_2\right)}=\frac{u'(e^y(1-\epsilon_1))}{ v'(e^y(\frac{e^o}{e^y}+\epsilon_2))}$ and $R^*\equiv \frac{%
u^{\prime}(e^y)}{ v^{\prime}(e^o)}$. So, applying Lemma \ref{justifyB}\ref{holding-condition1},  Condition (B) in  Theorem \ref{onlybubbly} holds.  Theorem \ref{onlybubbly}\ref{onlybubblypart1} then implies that every equilibrium satisfies $\liminf_{t\to\infty}a_t/e^y>0$. So,  case \ref{set2} of Theorem \ref{allsets} cannot happen and, hence,  
case \ref{set1} must hold.\\
{\bf Part \ref{part4} of  Theorem \ref{allsets}}. Conditions $R^*=G_n$ and $\sum_{t=1}^{\infty}d_t=\sum_{t=1}^{\infty}\dfrac{D_t}{G_n^t}<\infty$ imply that 
$\sum_{t=1}^n\frac{D_t}{R_1^*\cdots R_t^*}=\sum_{t=1}^n\frac{D_t}{(R^*)^t}=\sum_{t=1}^n\frac{D_t}{G_n^t}<\infty.$
Applying Proposition \ref{necessitycondition}\ref{existence}, there exists a bubbleless equilibrium.

Let $(a_t)$ be an equilibrium. Since $\sum_{t=1}^{\infty}d_t<\infty$ and $R^*=G_n$,   Lemma \ref{tl3g2} indicates that only case 2 or case 3 in Lemma \ref{tl3g2} happens. In both cases, we have $\lim_{t\to\infty}R_t=G_n$. If case 3 in Lemma \ref{tl3g2} happens, we directly obtain $(a_{t},b_t,R_{t}) \to (0,0,R^{\ast })$. If case 2 in Lemma \ref{tl3g2} happens, we have $\lim_{t\to \infty}a_t=\hat{a}$. However, since $R^*=G_n$, by the definition of $\hat{a}$, we have $\hat{a}=0$, and, hence, $\lim_{t\to \infty}b_t=0$.

Consider an equilibrium, by Lemma \ref{RtRt*}, we always have $R_t\geq R^*=G_n$. By consequence, the bubble component is $
b_0=\lim_{t\to\infty}\frac{G_n^t}{R_1\cdots R_t}a_t\leq \lim_{t\to\infty}\frac{G_n^t}{G_n^t}a_t\leq \lim_{t\to\infty}a_t=0.$ 
So, there is no bubble. Combining with Lemma \ref{interval1}(iii), we get that there exists a unique equilibrium and this is bubbleless. 

\subsection{\bf Proofs for Section \ref{section-bubble-pareto}}
\label{section-bubble-pareto-proof}
\begin{proof}[{\bf Proof of Proposition \ref{efficient-bubbleless}}]
{\bf Part 1}. According to Lemma \ref{prop1}, an equilibrium is bubbly if and only
if $\lim_{t\rightarrow \infty }\frac{G_n^{t}a_{t}}{R_{1}\cdots R_{t}}>0$.
Since $a_{t}\leq e_{t}^{y}$, we have  
$\dfrac{G_n^{t}a_{t}}{R_{1}\cdots R_{t}} \leq \dfrac{G_n^{t}e_{t}^{y}}{R_{1}\cdots R_{t}} \text{ } \forall t$,  which implies that
$\lim_{t\rightarrow \infty }\dfrac{G_n^{t}a_{t}}{R_{1}\cdots R_{t}}\leq \liminf\limits_{t\rightarrow \infty }\dfrac{G_n^{t}e_{t}^{y}}{R_{1}\cdots R_{t}}.$ Therefore, an equilibrium is bubbleless because $\liminf\limits_{t%
\rightarrow \infty }\frac{G_n^{t}}{R_{1}\cdots R_{t}}e_{t}^{y}=0$.

Since $c^y_t\leq e^y_t$, we have $
\liminf\limits_{t\rightarrow \infty }\frac{G_n^{t}}{R_{1}\cdots R_{t}}
c_{t}^{y}\leq \liminf\limits_{t\rightarrow \infty }\frac{G_n^{t}}{R_{1}\cdots R_{t}}
e_{t}^{y}=0$. So, we have $\liminf\limits_{t\rightarrow \infty }\frac{G_n^{t}}{R_{1}\cdots R_{t}}
c_{t}^{y}=0$. By Lemma \ref{lemmaPareto}, this equilibrium is Pareto optimal.

{\bf Part 2}. Consider an equilibrium. Recall that $\sum_{t=1}^{\infty}\frac{G_n^td_t}{R_1\cdots R_t}=f_0\leq a_0<\infty$. This implies that $\lim_{t\rightarrow \infty }\frac{G_n^{t}d_t}{R_{1}\cdots R_{t}}=0$. 
Our assumption $\limsup_{t\to\infty}\frac{d_t}{e^y_t}>0$ implies that there exists a sequence $(t_k)_{k\geq 1}$ and $x>0$ such that $d_{t_k}\geq xe^y_{t_k}$ $\forall k\geq 1$.

We have 
\begin{align}
\frac{G_n^{t_k}e_{t_k}^{y}}{R_{1}\cdots R_{t_k}}%
=\frac{G_n^{t_k}d_{t_k}}{R_{1}\cdots R_{t_k}}%
\frac{e_{t_k}^{y}}{d_{t_k}}\leq \frac{G_n^{t_k}d_{t_k}}{R_{1}\cdots R_{t_k}}%
\frac{1}{x}
\end{align}
Since $\lim_{t\rightarrow \infty }\frac{G_n^{t}d_t}{R_{1}\cdots R_{t}}=0$, we have $\lim_{k\to\infty}\frac{G_n^{t_k}e_{t_k}^{y}}{R_{1}\cdots R_{t_k}}=0$. It means that $\liminf\limits_{t\rightarrow \infty }\frac{G_n^{t}}{R_{1}\cdots R_{t}}e_{t}^{y}=0$. According to Part 1, this equilibrium is Pareto optimal and bubbleless.

\end{proof}

\begin{proof}[{\bf Proof of Proposition \ref{bubbleless-pareto}}] 
\begin{enumerate}
    \item 
Note that $\Big(\sum_{t\geq 1}\frac{D_t}{R_1\cdots R_t}\Big)\Big(\sum_{t\geq 1}\frac{R_1\cdots R_t}{G_n^te_t}\Big)\geq \sum_{t=1}^{\infty}\frac{D_t}{G_n^te_t}=\infty.$ 
Since $\sum_{t\geq 1}\frac{D_t}{R_1\cdots R_t}\leq q_0<\infty$, we have  $\sum_{t\geq 1}\frac{R_1\cdots R_t}{G_n^te_t}=\infty$. By Theorem \ref{pareto-theorem}, this equilibrium is Pareto optimal. Moreover, condition $\sum_{t=1}^{\infty}\frac{D_t}{e_tG_n^t}=\infty$ implies that $\sum_{t=1}^{\infty}\frac{D_t}{G_n^te^y_t}=\infty$ because $e^y_t<e_t$. By Proposition \ref{necessitycondition}, this equilibrium is bubbleless.

\item Take any equilibrium. By Lemma \ref{RtRt*}, we have $R_t\geq R^*_t$ $\forall t$.  So, condition $\lim_{t\rightarrow \infty }\frac{G_n^te^y_t}{R_{1}^{\ast }\cdots
R_{t}^{\ast }}=0$ implies that $\lim_{t\rightarrow \infty }\frac{G_n^te^y_t}{R_{1}\cdots
R_{t}}=0.$ Proposition \ref{efficient-bubbleless}\ref{efficient-bubbleless1} implies that this equilibrium is Pareto optimal and bubbleless. Since there is no bubbly equilibrium, Lemma \ref{interval1} implies that there is a unique equilibrium.
\end{enumerate}
 \end{proof}

\begin{proof}[{\bf Proof of Proposition \ref{efficient-bubbleless-saving}}]
\begin{enumerate}
    \item 
$\limsup_{t\to\infty}\frac{a_t}{e_t}>0$ implies that there exist $x>0$ and an infinite and increasing sequence of time $(t_k)_{k\geq 1}$ such that ${a_{t_k}}/{e_{t_k}}>x$ $\forall t$.

Recall that $\infty>a_0> \frac{G_n^{t}}{R_{1}\cdots R_{t}}a_t$ $\forall t$. Hence, we have
\begin{align}\frac{R_1\cdots R_{t_k}}{G_n^{t_k}e_{t_k}}= \frac{a_{t_k}}{e_{t_k}}\frac{R_1\cdots R_{t_k}}{G_n^{t_k}a_{t_k}}>\frac{a_{t_k}}{e_{t_k}}\frac{1}{a_0}>x\frac{1}{a_0}>0\text{ } \forall t.
\end{align}
%
By consequence, $\sum_{k}\frac{R_1\cdots R_{t_k}}{G_n^{t_k}e_{t_k}}=\infty$. Therefore, $\sum_{t\geq 1}\frac{R_1\cdots R_t}{G_n^te_t}=\infty$. Applying  Theorem \ref{pareto-theorem}\ref{pareto-theorem-part1}, this equilibrium is Pareto optimal.

\item 
Consider an equilibrium.  The no-bubble condition means that $\lim\limits_{t\rightarrow \infty }\frac{G_n^{t}}{R_{1}\cdots R_{t}}a_t=0$. 

Since $\liminf_{t\to\infty}\frac{a_t}{e^y_t}>0$, there exists $x>0$ and $t_0$ such that $\frac{a_t}{e^y_t}>x$ $\forall t\geq t_0$. Then, for all $t\geq t_0,$ we have 
\begin{align*}
\frac{G_n^{t}e^y_t}{R_{1}\cdots R_{t}}
< \frac{G_n^{t}a_t}{R_{1}\cdots R_{t}}\frac{1}{x}, \text{ which implies that }
\liminf\limits_{t\rightarrow \infty }\frac{G_n^{t}e^y_t}{R_{1}\cdots R_{t}}\leq \lim\limits_{t\rightarrow \infty }\frac{G_n^{t}a_t}{R_{1}\cdots R_{t}}\frac{1}{x}.
\end{align*}
Combining with  $\lim\limits_{t\rightarrow \infty }\frac{G_n^{t}}{R_{1}\cdots R_{t}}a_t=0$, we get that $\lim\limits_{t\rightarrow \infty }\frac{G_n^{t}}{R_{1}\cdots R_{t}}e^y_t=0$. By Proposition  \ref{efficient-bubbleless}, this equilibrium is Pareto optimal.
\end{enumerate}
\end{proof}
\begin{proof}[{\bf Proof of Proposition \ref{paretorank}}]

First, we prove the following claim on the  welfare ranking when there exists a continuum of equilibria): Let Assumptions \ref{assum0}, \ref{assum1} 
 be satisfied. For two equilibria with initial asset values $a_0>a_0'$, we denote $U_t$ and $U_t'$ the utility of households born at date $t$ in the equilibrium $a_0$ and $a_0'$ respectively. Then we have $U_t>U_t'$ for any date $t\geq -1$. It means that the utility of each generation is increasing in the initial asset value.

Let us prove this claim. The utility of households born at date $t$ is $
U_t=u(e_{t}^{y}-a_{t})+ v(e^o_{t+1}+R_{t+1}a_t).$ Taking the derivative with respect to $a_t$ and using the Euler equation as well as $\frac{\partial R_{t+1}}{\partial a_t}>0$ (see Lemma \ref{5}) we have $\frac{\partial U_t}{\partial a_t}>0$.
Since $a_t$ is strictly increasing in $a_0$, the utility $U_t$ is also strictly increasing in $a_0$.  By combining with Lemma \ref{interval1}, we obtain our result.

\end{proof}

\begin{proof}[{\bf Proof of Theorem \ref{pareto-assetbubble}}]
{\bf Point \ref{pareto-assetbubble-1}}. According to Theorem \ref{allsets}, there exists a unique equilibrium and this equilibrium is bubbleless. By Lemma \ref{RtRt*}, we have, in equilibrium, $R_t\geq R^*$  $\forall t$. By combining with $R^*_t=R^*>G_n$, we have  $\liminf_{t\to\infty}R_t>G_n$. Therefore, Lemmas \ref{RtRt*} and \ref{lemmaPareto} imply that this equilibrium is Pareto optimal.\\
{\bf Point \ref{pareto-assetbubble-2}}. Point \ref{pareto-assetbubble-2a} is a consequence of Theorem \ref{allsets} and Proposition \ref{paretorank}. 
Let us prove point \ref{pareto-assetbubble-2b} by using Theorems \ref{allsets},  \ref{pareto-theorem} and  
Lemma \ref{check-strict}\ref{check-strict-part1}. First, by Theorem \ref{allsets}, this equilibrium satisfies $(a_{t},b_{t},R_{t})$ converges
to $(\hat{a},\hat{a},G_n)$ where $\hat{a}>0$ is uniquely determined by  $u'(e^y-\hat{a})= G_n v'(e^o+G_n\hat{a})$ (i.e., $G_n=g(\hat{a})$).

Since $b_0=\lim_{t\to\infty}\frac{G_n^t}{R_1\cdots R_t}a_t=\lim_{t\to\infty}\frac{G_n^t}{R_1\cdots R_t}\hat{a}>0$, we have $\lim_{t\to\infty}\frac{G_n^t}{R_1\cdots R_t}=\frac{b_0}{\hat{a}}\in (0,\infty)$. Then, combining with $\lim_{t\to\infty}d_t=0$,  we have
 \begin{align*}
\lim_{t\to\infty}\frac{R_1\cdots R_t}{G_n^t(e^y+\frac{e_o}{G_n}+d_t)}=\frac{R_1\cdots R_t}{G_n^t(e^y+\frac{e_o}{G_n})}=\frac{\hat{a}}{b_0(e^y+\frac{e^o}{G_n})}>0.
\end{align*}
By consequence, we have $\sum_{t\geq 1}\frac{R_1\cdots R_t}{G_n^t(e^y+\frac{e^o}{G_n}+d_t)}=\infty$.

To conclude that this equilibrium is Pareto optimal, it suffices to verify the uniform strictness condition. We do so by using Lemma \ref{check-strict}\ref{check-strict-part1}. Indeed, recall that $c^y_t>0$ $\forall t$ and $\lim_{t\to\infty}c^y_t=e^y-\hat{a}>0$. Since the function $u$ is in $C^2$ and $u^{\prime\prime}(e^y-\hat{a})\in (-\infty,0)$, there exists $h>0$ such that
$$\inf_{t\geq 0}\Big\{\frac{c^y_t}{u^{\prime}(c^y_t)}\inf_{x\in [(1-h)c^y_t,c^y_t]} \big(-\frac{1}{2}u^{\prime\prime}(x)\big)\Big\}>0.$$
So, by Lemma \ref{check-strict}\ref{check-strict-part1}, the equilibrium allocation satisfies the uniform strictness condition. 
\\
{\bf Point \ref{pareto-assetbubble-3}}. By Claim 2 in Theorem \ref{allsets}, there exists a unique equilibrium. This is asymptotically bubbly and  $(a_{t},b_{t},R_{t})$ converges
to $(\hat{a},\hat{a},G_n)$. By using the same argument as above, this equilibrium is Pareto optimal.

\end{proof}

\begin{proof}[{\bf Proof of Example \ref{example_asymptotically_bubbly_nonoptimal}}]
{\bf Step 1.} We first verify that the proposed allocation is an equilibrium. Since $\frac{q_{t+1}+D_{t+1}}{q_t}= \frac{1+2^{t+1}+1}{1+2^t}
=2$, 
the equilibrium gross interest rate is $R_{t+1}=2$. Moreover, we can easily verify that $c_t^y+q_tz_t=e^y_t$ and $c^o_{t+1}=5\cdot2^t=e_{t+1}^o+(q_{t+1}+D_{t+1})z_t$.
Since $u'(c)=1+1/c$, we have
$$\frac{U_1(c_t^y,c_{t+1}^o)}   {U_2(c_t^y,c_{t+1}^o)}=
\frac{u'(5\cdot2^t)}{\frac12u'(5\cdot2^t)}=2=R_{t+1}.
$$
Strict concavity implies household optimality, and $z_t=1$ clears the asset market.\\
{\bf Step 2.} We next compute the bubble component. Since $R_t=2$ and $D_t=1$, we have $F_t=\sum_{s=1}^{\infty}{1}/{2^s}=1,  B_t=q_t-F_t=2^t.$ 
Since $G_n=1$, we have $b_t=B_t=2^t$, and hence $\lim_{t\to\infty}{b_t}/{e_t^y}=1/6$.\\
%
{\bf Step 3.} We show that the equilibrium is not Pareto optimal. Define another allocation $\widetilde c_t^y, \widetilde c_t^o$ by 
$ \widetilde c_t^y=c_t^y-x_t,  \widetilde c_t^o=c_t^o+x_t, \text{ where }x_t=\frac35\,2^t-\frac12>0 \text{ }\forall t\geq 0.$
This is obviously a feasible allocation. The initial old generation is strictly better off because $x_0=1/10>0$.

For a generation born at date $t$, the utility gain is
\begin{subequations}
\begin{align}
U_t'-U_t&=u(5\cdot2^t-x_t)-u(5\cdot2^t) +\frac12\left[u(5\cdot2^t+x_{t+1})-u(5\cdot2^t)\right]\\
&=-x_t+\frac12x_{t+1} +\ln\left(1-\frac{x_t}{5\cdot2^t}\right)
+\frac12\ln\left(1+\frac{x_{t+1}}{5\cdot2^t}\right)\\
 &\geq -x_t+\frac12x_{t+1} +\ln\left(1-\frac{x_t}{5\cdot2^t}\right)=\frac{1}{4} +\ln\left(1-\frac{x_t}{5\cdot2^t}\right)
\end{align}
\end{subequations}
Since $-x_t+\frac12x_{t+1}=\frac14$
and
$\frac{x_t}{5\cdot2^t}=\frac3{25}-\frac{1}{10\cdot2^t}
<\frac3{25},$ we obtain $U_t'-U_t>\frac14+\ln\big(\frac{22}{25}\big)>0.$ Thus every generation is strictly better off under the alternative feasible allocation. The equilibrium is not Pareto optimal.
\end{proof}


\clearpage
\setcounter{page}{1}
\renewcommand{\thepage}{A\arabic{page}}

{\small
\begingroup
\setcounter{footnote}{1}
\renewcommand{\thefootnote}{\fnsymbol{footnote}}

\centerline{\Large Online Appendix to ``To Bubble or Not to Bubble:}
\centerline{\Large Asset Price Dynamics and Optimality in OLG Economies''}
\centerline{ }

\centerline{\large
Stefano BOSI\footnote{%
Universit\'e d'Evry, Universit\'e Paris-Saclay.
Email: stefano.bosi@univ-evry.fr}
\text{  }
Cuong LE VAN\footnote{%
CNRS, PSE, IPAG Business School, FTU Hanoi.
Email: Cuong.Le-Van@univ-paris1.fr}
\text{  }
Ngoc-Sang PHAM\footnote{%
EM Normandie Business School, M\'etis Lab.
Email: npham@em-normandie.fr}}

\centerline{\large\today}

\endgroup
\setcounter{equation}{0} 
\numberwithin{equation}{section}
\subsection*{Online Appendix 1: Additional results and proofs for Section \ref{continuum-section}}


We present another corollary of Theorem  \ref{new5continuum}. 
\begin{corollary}\label{new5continuum-remark}  
Assume that the utility is  $U^t(x_1,x_2)=\frac{x_1^{1-\sigma}}{1-\sigma}+\beta \frac{x_2^{1-\sigma}}{1-\sigma}$ where $\sigma>0,\beta>0$, while endowments satisfy $\frac{e_{t+1}^{o}}{e^y_t}=G_e>0$ and $e^y_t\leq e^y_{t+1}$ $\forall t$. In this case, the benchmark interest rate  $R^*_t=R^*={G_e^{\sigma}}/{\beta}$ $\forall t$.   Assume that dividends grow at a constant rate: $D_t=D_0G_d^t$, where $D_0,G_d>0$ $\forall t$.

Then, Assumptions in Theorem  \ref{new5continuum} are satisfied if the interest rate and dividends are low in the sense that  $$G_d<R^*<G_n$$  and $D_0$ is low enough.
\end{corollary}
Note that $\sigma$ may be lower or higher than $1$ (so, Assumption \ref{assum1} may not hold because $cv'(c)$ may not be increasing). 

\begin{proof}[{\bf Proof of Corollary \ref{new5continuum-remark}}]

Assume that $U^t(x_1,x_2)=\frac{x_1^{1-\sigma}}{1-\sigma}+\beta \frac{x_2^{1-\sigma}}{1-\sigma}$ where $\sigma>0,\beta>0$. The Euler condition becomes $(e^y_t-a_t)^{-\sigma}=\beta R_{t+1}(e_{t+1}^{o}+R_{t+1}a_{t})^{-\sigma}$
 and the function $K_t(a,R)=(e^y_t-a)^{-\sigma}-\beta R (e_{t+1}^{o}+Ra)^{-\sigma}$. Observe that 
\begin{align}\label{functionH_definition}
K_t(a,R)\lesseqgtr 0 \Leftrightarrow H_t(a,R)\equiv \frac{e_{t+1}^{o}}{e^y_t}R^{\frac{-1}{\sigma}}+\frac{a}{e^y_t}\Big(R^{1-\frac{1}{\sigma}}+\beta^{\frac{1}{\sigma}}\Big)-\beta^{\frac{1}{\sigma}}\lesseqgtr 0.
\end{align}
Neither $K_t$ nor $H_t$ depends on dividends. 

We now explain how to choose parameters so that conditions in Theorem \ref{new5continuum} hold.  

We have $D_t=D_0G_d^t$, i.e., $d_t=d_0d^t$ $\forall t$ with $D_0=d_0, G_d=G_nd$.



Condition $R^*<G_n$ implies that $G_eG_n^{\frac{-1}{\sigma}}-\beta^{\frac{1}{\sigma}}<0$. So, we can choose $\epsilon>0$ small enough such that $G_eG_n^{\frac{-1}{\sigma}}+\epsilon(G_n^{1-\frac{1}{\sigma}}+\beta^{\frac{1}{\sigma}})-\beta^{\frac{1}{\sigma}}<0$.

Define $\epsilon_t\equiv \epsilon e^y_t$. By the definition of $\epsilon$ and the function $H$  in (\ref{functionH_definition}), we have $H_t(\epsilon_t,G_n)<0$. 

 


We look at conditions \ref{1a}, \ref{1b} in Theorem  \ref{new5continuum}. Condition \ref{1a} becomes ${\epsilon}{e^y_t}<{\epsilon}{e^y_{t+1}}+d_{t+1}$. This is satisfied because we assume that $e^y_t\leq e^y_{t+1}$ $\forall t$. Condition \ref{1b} becomes $H_t(\epsilon_t,G_n)<0$ which is satisfied as we have just explained. 
Lastly, we verify the not-too-low interest rate condition (\ref{Rdd-notlow}). With the above settings, condition (\ref{Rdd-notlow}) becomes $R^*>G_nd\gamma.$ 
This is satisfied if we take $\gamma\in (0,\frac{R^*}{G_nd})$. We choose $\lambda$ and the dividend growth rate $d$ small enough so that $\gamma>1+\frac{1}{\lambda}$ and $\lambda d_t<\epsilon_t\equiv \epsilon e^y_t$.

So,  all assumptions in Theorem  \ref{new5continuum} are satisfied under conditions in Corollary \ref{new5continuum-remark}.

\end{proof}

\begin{proof}[{{\bf Proof of Lemma \ref{5}}}]
Let $e^y_t>0,e^o_{t+1}> 0$. Let $a_{t}\in (0,e_{t}^{y})$.  Consider the function $K: [0,\infty)\to \rr$ defined by $K(R)\equiv u^{\prime }(e_{t}^{y}-a_{t})-
Rv^{\prime }\left( e_{t+1}^{o}+Ra_{t}\right)$ $\forall R\in [0,\infty)$. We have 
\begin{align*}
K^{\prime }(R)& =- v^{\prime }\left(
e_{t+1}^{o}+Ra_{t}\right) - R a_{t}v^{\prime \prime }\left(
e_{t+1}^{o}+R a_{t}\right).
\end{align*}
Since $e^o_{t}>0$ $\forall t$, we have 
$$K^{\prime }(R) =- v^{\prime }\left(
e_{t+1}^{o}+Ra_{t}\right) - R a_{t}v^{\prime \prime }\left(
e_{t+1}^{o}+R a_{t}\right)  < - v^{\prime }\left( e_{t+1}^{o}+R a_{t}\right) -
(e_{t+1}^{o}+R a_{t})v^{\prime \prime }\left(
e_{t+1}^{o}+R a_{t}\right) \leq 0,
$$
where the last inequality holds because $cv^{\prime }(c)$ is increasing in $c$.  Therefore, the function $K$ is
strictly decreasing on $(0,\infty)$.  Since $e^o_{t+1}>0$, we have $K(0)=u^{\prime }(e_{t}^{y}-a_{t})>0$. 

We now look at $\lim_{x\rightarrow \infty }K(x)$. We have 
\begin{align*}
&\lim_{x\rightarrow \infty }xv^{\prime }\left( e_{t+1}^{o}+xa_{t}\right) 
=\lim_{x\rightarrow \infty }\frac{x\left(
e_{t+1}^{o}+xa_{t}\right) v^{\prime }\left( e_{t+1}^{o}+xa_{t}\right)}{e_{t+1}^{o}+xa_{t}}\\
&  =\frac{1}{a_{t}}\lim_{x\rightarrow \infty }\left(
e_{t+1}^{o}+xa_{t}\right) v^{\prime }\left( e_{t+1}^{o}+xa_{t}\right) =\frac{
1}{a_{t}}\lim_{c\rightarrow \infty }cv^{\prime }(c).
\end{align*}
This implies that $\lim_{x\rightarrow \infty }K(x)=u^{\prime }(e_{t}^{y}-a_{t})-\frac{1
}{a_{t}}\lim_{c\rightarrow \infty }cv^{\prime }(c).$
Therefore, there exists a unique $R_{t+1}$ satisfying  $u^{\prime
}(e_{t}^{y}-a_{t})= R_{t+1}v^{\prime }\left(
e_{t+1}^{o}+R_{t+1}a_{t}\right)$ if and only if $a_tu^{\prime
}(e_{t}^{y}-a_{t})<\lim_{c\to\infty}cv^{\prime}(c)$.

Taking the derivative of both sides of the equation $u^{\prime
}(e_{t}^{y}-a_{t})= R_{t+1}v^{\prime }\left(
e_{t+1}^{o}+R_{t+1}a_{t}\right) $ with respect to $a_{t}$, we get that 
\begin{align*}
 -u^{\prime \prime }(e_{t}^{y}-a_{t})=& R_{t+1}v^{\prime \prime }\left(
e_{t+1}^{o}+R_{t+1}a_{t}\right) \left(a_t\frac{\partial R_{t+1}}{\partial a_{t}%
}+R_{t+1}\right) + \frac{\partial R_{t+1}}{\partial a_{t}}v^{\prime
}\left( e_{t+1}^{o}+R_{t+1}a_{t}\right).
\end{align*}%
This implies that
$$ \left[ R_{t+1}a_tv^{\prime \prime }\left(
e_{t+1}^{o}+R_{t+1}a_{t}\right) +v^{\prime }\left(
e_{t+1}^{o}+R_{t+1}a_{t}\right) \right] \frac{\partial R_{t+1}}{\partial
a_{t}}=-u^{\prime \prime }(e_{t}^{y}-a_{t})- R_{t+1}^{2}v^{\prime \prime
}\left( e_{t+1}^{o}+R_{t+1}a_{t}\right)>0.
$$
Again, by Assumption \ref{assum1}, we have $R_{t+1}a_tv^{\prime \prime }\left( e_{t+1}^{o}+R_{t+1}a_{t}\right) +v^{\prime
}\left( e_{t+1}^{o}+R_{t+1}a_{t}\right)=-K'(R_{t+1})>0$, 
which implies that, ${\partial R_{t+1}}/{\partial
a_{t}}>0$. Therefore, $R_{t+1}$ is strictly increasing in $a_{t}$. 
\end{proof}
\begin{proof}[{\bf Proof of Lemma \ref{interval1}}]

We follow the strategy of  \cite{tirole85_appendix}, \cite{bhlpp18_appendix,blp22_appendix}.\\ 
{\bf Point (ii)}. Let $a_0'>a_0$ be two elements in $\mathcal{A}_{0}$, and $(a_t'), (a_t)$ be two associated equilibrium sequences. We have $R_1'=g_0(a_0')\geq g_0(a_0)=R_1$. Then, we have $a_1'=\frac{R_1'}{G_n}a_0'-d_1>\frac{R_1}{G_n}a_0-d_1=a_1$. By induction, we have $a_t'>a_t$ and $R_t'\geq R_t$ $\forall t$. Thus, we can compare the fundamental values
\begin{align*}
f_{0}'& =\sum_{s=1}^{\infty }\frac{G_n}{%
R'_{1}}\cdots \frac{G_n}{R'_{s}}d_{s} \leq \sum_{s=1}^{\infty }\frac{G_n}{%
R_{1}}\cdots \frac{G_n}{R_{s}}d_{s}=f_0\\
b_{0}'&= a'_{0}-f'_{0}>a_{0}-f_{0}=b_0.
\end{align*}
\noindent
{\bf Point (iii)} is a direct consequence of the above proof and the fact that $\mathcal{A}_{0}$ is an interval.\\
{\bf Point (i)}. Firstly, we prove that $\mathcal{A}_{0}$ is an interval. Let us consider two equilibria $(a_{1,t},R_{1,t+1})_{t\geq 0}$ and $(a_{2,t},R_{2,t+1})_{t\geq 0}$ with initial asset values $a_{1,0}<a_{2,0}$. Take $a_0=\lambda a_{1,0}+(1-\lambda)a_{2,0}\in (a_{1,0},a_{2,0})$ with $\lambda\in (0,1)$. We have to prove that there exists a sequence $(a_{t})_{t\geq 0}$ satisfying (\ref{system1}). Clearly, $a_0\in (0,e^y_0)$. From $a_0$, we can define $R_1=g_1(a_0)$, thanks to Lemma \ref{5} and the fact that  $a_{0}u^{\prime
}(e_{0}^{y}-a_{0})<a_{2,0}u^{\prime
}(e_{0}^{y}-a_{2,0})< \lim_{c\rightarrow \infty }cv^{\prime }(c)$.

 Since $a_0\in (a_{1,0},a_{2,0})$, we have $R_1\in [R_{1,1},R_{2,1}]$. Then, we define $a_1$ by $a_1+d_1= \frac{R_1}{G_n}a_0$. We see that
\begin{align*}
a_{1,1}=\frac{R_{1,1}}{G_n}a_{1,0}-d_1<a_1=\frac{R_1}{G_n}a_0-d_1<\frac{R_{2,1}}{G_n}a_{2,0}-d_1=a_{2,1}.
\end{align*}
By induction, we construct that the equilibrium sequence $(a_t)$. So, $a_0\in \mathcal{A}_{0}$. It means that $\mathcal{A}_{0}$ is an interval.

It remains to prove that $\mathcal{A}_{0}$ is closed. This is a direct consequence of
Lemmas \ref{tl10g} and \ref{new2} below.

\begin{lemma}
\label{tl10g} The  equilibrium set $\mathcal{A}_{0}$ in Definition  (\ref{defA0})  is
closed on the right: if $(a_{0}^{m})_{m\geq 0}$ is a strictly increasing
sequence with  $a_{0}^{m}\in \mathcal{A}_{0} \text{ } \forall m\geq 0$, then $a_{0}\equiv
\lim_{m\rightarrow \infty }a_{0}^{m}$  belongs to $\mathcal{A}_{0}$.
\end{lemma}
This result can be viewed as an adapted version of \citet[Lemma 10]{tirole85_appendix}.
\begin{proof}[{Proof of Lemma \ref{tl10g}}] By the definition, we have 
\begin{align}
u^{\prime }(e_{t}^{y}-a^m_{t})& = R^m_{t+1}v^{\prime }\left(
e_{t+1}^{o}+R^m_{t+1}a^m_{t}\right) \\
R_{t+1}^{m}& =g_{t}(a_{t}^{m}), \quad a_{t+1}^{m} =\frac{R_{t+1}^{m}}{G_n}a_{t}^{m}-d_{t+1}, \quad  0<a_{t}^{m}<e_{t}^{y} \text{  } \forall  t\geq 0.
\end{align}
Since the sequence $(a_0^m)_m$ is increasing in $m$, we have $R_1^m=g_1(a^m_0)$ is increasing in $m$. This implies that $a_1^m=\frac{R_{1}^{m}}{G_n}a_{0}^{m}-d_{1}$ is increasing in $m$. By induction, $a_t^m$ and $R_t^m$ are increasing in $m$. Define $a_t\equiv \lim_{m\to\infty}a_t^m, R_t\equiv \lim_{m\to\infty}R_t^m$. To prove that $a_0$ is in the set $\mathcal{A}_{0}$ in Definition  (\ref{defA0}), it remains to prove that $a_t\in (0,e^y_t)$ $\forall t$. 

It is easy to see that $a_t\geq a^m_t>0$.

We now prove that $a_t<e^y_t$. We have $u^{\prime }(e_{t}^{y}-a^m_{t}) = R^m_{t+1}v^{\prime }\left(
e_{t+1}^{o}+R^m_{t+1}a^m_{t}\right)\leq  R^m_{t+1}v^{\prime }\left(
e_{t+1}^{o}\right).$

If $\lim_{m\to \infty}a^m_t=e^y_t$, then $u^{\prime }(e_{t}^{y}-a^m_{t})$
converges to infinity. This implies that $R^m_{t+1}$ converges to infinity.

For $m\geq 1$, we have
\begin{align}
b^m_{t+1}=b^m_0\frac{R^m_1\cdots R^m_tR^m_{t+1}}{G_n^{t+1}}\geq b^1_0\frac{%
R_1^*\cdots R_t^*R^m_{t+1}}{G_n^{t+1}}.
\end{align}
where $R^*_t$ is the interest rate of the economy without assets.

Since $a^1_0>a^0_0$, point (ii) of Lemma \ref{interval1} implies that $b^1_0>b^0_0\geq 0$. So, $b_{0}^{1}>0$. Combining with $\lim_{m\rightarrow \infty
}R_{t+1}^{m}=\infty $, we obtain that $\lim_{m\rightarrow \infty
}b_{t+1}^{m}=\infty $. However, this is impossible because $b_{t}^{m}\leq
a_{t}^{m}<e_{t}^{y}$ $\forall t$. \end{proof}

\begin{lemma}
\label{new2} The  equilibrium set $\mathcal{A}_{0}$ in Definition  (\ref{defA0})  is
closed on the left: if $(a_{0}^{m})_{m\geq 1}$ is a strictly decreasing
sequence with  $a_{0}^{m}\in \mathcal{A}_{0} \text{  } \forall  m\geq 1$, then $a_{0}\equiv
\lim_{m\rightarrow \infty }a_{0}^{m}$  belongs to $\mathcal{A}_{0}$.

\end{lemma}

\begin{proof}[{Proof of Lemma \ref{new2}}]
By definition, we have%
\begin{align}
R_{t+1}^{m}& =g_{t}(a_{t}^{m}), \quad 
a_{t+1}^{m} =\frac{R_{t+1}^{m}}{G_n}a_{t}^{m}-d_{t+1}, \quad 
 0<a_{t}^{m}<e_{t}^{y} \text{  } \forall  t\geq 0.
\end{align}
As in the proof of Lemma \ref{tl10g}, we can define $a_t\equiv \lim_{m\to\infty}a_t^m, R_t\equiv \lim_{m\to\infty}R_t^m$. 
It is easy to see that $R_{t}^{m}\geq R_{t}\geq R_{t}^{\ast }$ $\forall m,\forall t$, where the sequence $\left( R_{t}\right) $ corresponds to the
initial condition $a_{0}$ and $R_{t}^{\ast }$ is the interest rate of the
economy without the asset.

It is obvious that $a_t\leq a^m_t<e^y_t$. So, it remains to prove that $%
a_{t}>0 \text{  } \forall  t\geq 0$.

Fix a date $t$. We have  
\begin{equation*}
a_{t}^{m}=\frac{G_n}{R_{t+1}^{m}}(a_{t+1}^{m}+d_{t+1})\geq \frac{G_n}{R_{t+1}^{m}%
}d_{t+1}.
\end{equation*}%
Let $m\rightarrow \infty $, we get that $a_{t}\geq \frac{G_n}{R_{t+1}}%
d_{t+1}>0 $. \end{proof}

\end{proof}


\subsection*{Online Appendix 2: Additional results and proofs for Section \ref{onlybubbly-section}}

\begin{proof}[{\bf Proof of Corollary \ref{remark-onlybubbly}}]


The Euler equation now is $(e^y_t-a_t)^{-1}-\beta X(e_{t+1}^{o}+X a_{t})^{-1}=0$ and equation (\ref{euler3}) becomes 
\begin{align}\label{euler4}
X(1-\frac{\epsilon}{\beta(1-\epsilon)})=\frac{g_{e,t+1}}{\beta(1-\epsilon)}.
\end{align}
Let us check Condition (B)  in Theorem \ref{onlybubbly}. 
Since $\limsup_{t\to \infty}\frac{R^*_{t+1}}{G_n\frac{e^y_{t+1}}{e^{y}_t}}<1$, then we can choose $\bar{\epsilon}\in (0,\beta/(1+\beta))$ small enough, $\delta\in (0,1)$ close enough to 1 and a date $T$ so that $\frac{R^*_{t+1}}{G_n\frac{e^y_{t+1}}{e^{y}_t}}<\frac{\beta-\bar{\epsilon}(1+\beta)}{\beta}\delta <1$ $\forall t\geq T$. 

We next define $X_{t+1}\equiv \delta G_n\frac{e^y_{t+1}}{e^y_t}$. 
Then, we have $X_{t+1}\leq  G_n\frac{e^y_{t+1}}{e^y_t}$.  Since $\frac{D_t}{G_n^te^y_t}=\frac{1}{t^{\alpha}}$ where $\alpha>0$,  and $X_{t+1}\equiv \delta G_n\frac{e^y_{t+1}}{e^y_t}$, where $\delta\in (0,1)$,  we have  
\begin{align}
\sum_{t=1}^{\infty}\frac{D_t}{X_1\cdots X_t}=\sum_{t=1}^{\infty}\frac{D_t}{\delta^tG_n^te^y_t}e^y_ 0=e^y_ 0\sum_{t=1}^{\infty}\frac{1}{\delta^t t^{\alpha}}=\infty. 
\end{align} 

For any $\epsilon\in (0,\bar{\epsilon})$, we have 
\begin{align}
\label{key2}\frac{e_{t+1}^{o}}{e_{t}^{y}}\frac{1}{\beta-{\epsilon}(1+\beta)}\leq \frac{e_{t+1}^{o}}{e_{t}^{y}}\frac{1}{\beta-\bar{\epsilon}(1+\beta)}&\leq  \delta G_n\frac{e^y_{t+1}}{e^y_t}=X_{t+1} \text{ }\forall t\geq T.
\end{align}

Let $\epsilon\in (0,\bar{\epsilon})$ and $X\in \mathcal{R}_t(\epsilon)$, then we have, thanks to (\ref{key2}), 
$$X=g_{e,t+1}\frac{1}{\beta-\epsilon(1+\beta)}=\frac{e_{t+1}^{o}}{e_{t}^{y}}\frac{1}{\beta-\epsilon(1+\beta)}\leq X_{t+1}.$$
So, Condition (B) is satisfied.

By construction $\frac{D_t}{G_n^te^y_t}=\frac{1}{t^{\alpha}}$ with $\alpha>0$, we have $\sum_{t=1}^{\infty}\frac{D_t}{G_n^te^y_t}=\sum_{t=1}^{\infty}\frac{1}{t^{\alpha}}<\infty$ if $\alpha>1$ and $=\infty$ if $\alpha\in (0,1]$. Applying Theorem \ref{onlybubbly}\ref{onlybubblypart2}, we obtain our desired result.

\end{proof}


\subsubsection*{Other applications of Theorem \ref{onlybubbly}}

\begin{corollary}[bounded economy]\label{holding-condition1b} Let Assumption \ref{assum0} be satisfied and the function $U^t$ be quasi-concave, continuously differentiable, and strictly increasing in each component. Assume also that one of the two following situations be satisfied.
\begin{enumerate}
\item The utility and endowments are time-independent (i.e.,  $U^t(x_1,x_2)=U(x_1,x_2)$ and $e^y_t=e^y>0, e^o_t=e^o$ $\forall t$) and $\limsup_{t\to\infty}d_t<\infty$. The benchmark interest rate $R^*$ is low in the sense that
\begin{align}\label{onlybubbly-stationary}R^*\equiv \frac{U_1(e^y,e^o)}{U_2(e^y,e^o)}<\min\big\{\limsup_{t\to\infty}D_t^{1/t},  G_n\big\}<\infty.
\end{align}

\item The utility $U^t(x_1,x_2)=u(x_1) + v(x_2)$ $\forall x_1,\forall x_2$. The endowments satisfy $e^o_t=e^o\geq 0$, $e^y_t\in [\underline{e},\bar{e}]\subset \rr_{++}$ $\forall t$, and $\limsup_{t\to\infty}d_t<\infty$. The benchmark interest rate $R^*$ is low in the sense that
\begin{align}\label{onlybubbly-stationary2}\bar{R}\equiv \frac{u^{\prime}(\underline{e})}{ v^{\prime}(e^o)}<\min\big\{\limsup_{t\to\infty}D_t^{1/t},  G_n\big\}<\infty.
\end{align}
\end{enumerate}
Then, Condition (B) holds. Therefore, every equilibrium is bubbly if and only if $\sum_{t=1}^{\infty}\frac{D_t}{G_n^t}<\infty$. Every equilibrium is bubbleless if and only if $\sum_{t=1}^{\infty}\frac{D_t}{G_n^t}=\infty$.
\end{corollary}

\begin{proof}[{\bf Proof of Corollary \ref{holding-condition1b}}]
{\bf Situation 1}. 
Since $R^* <\min\big\{ \limsup_{t\to\infty}D_t^{1/t},  G_n\big\}\equiv R_m$ 
and the partial derivative $U_1,U_2$ are continuous, there exists $R$, $\bar{\epsilon}_1$ and $\bar{\epsilon}_2$ such that $$\frac{U_1\big(e^y(1-\epsilon_1), e^y(\frac{e^o}{e^y}+\epsilon_2)\big)}{U_2\big(e^y(1-\epsilon_1), e^y(\frac{e^o}{e^y}+\epsilon_2)\big)}<R<R_m$$ $\forall \epsilon_1\in (0,\bar{\epsilon}_1)$, $\epsilon_2\in (0,\bar{\epsilon}_2)$. So, condition \eqref{B-holding} holds. 

Since endowments are time independent and $\limsup_{t\to\infty}d_t<\infty$, we can easily verify condition \eqref{boundPhi}. So, Lemma \ref{justifyB}\ref{holding-condition1} applies and condition (B) holds.\\
%
%
{\bf Situation 2}. Since $u,v$ are continuous and $\frac{u^{\prime}(\underline{e})}{ v^{\prime}(e^o)}<R_m\equiv \min\big\{\limsup_{t\to\infty}D_t^{1/t},  G_n\big\}$,  there exist $R$, $\bar{\epsilon}_1$ and $\bar{\epsilon}_2$ such that 
 \begin{align}\frac{u^{\prime}(\underline{e}(1-\epsilon_1)}{ v^{\prime}(e^o+\bar{e}\epsilon_2)}<R<R_m \text{ } \forall \epsilon_1\in (0,\bar{\epsilon}_1), \forall \epsilon_2\in (0,\bar{\epsilon}_2).
\end{align}

By the assumption $e^y_t\in [\underline{e},\bar{e}]$, we have 
\begin{align}\frac{V^t_1\left(1-\epsilon_1,g_{e,t+1}+\epsilon_2\right)}{V^t_2\left(1-\epsilon_1,g_{e,t+1}+\epsilon_2\right)}&=\frac{u^{\prime}\big(e^y_t(1-\epsilon_1)\big)}{v^{\prime}\big(e^o+e^y_t\epsilon_2)}\leq \frac{u^{\prime}\big(\underline{e}(1-\epsilon_1)\big)}{v^{\prime}\big(e^o+\bar{e}\epsilon_2)}<R<R_m.
\end{align}So, condition \eqref{B-holding} holds. Condition \eqref{boundPhi} is satisfied thanks to the boundedness of endowments and dividends. So, Lemma \ref{justifyB}\ref{holding-condition1} applies and condition (B) holds.
\end{proof}

We now allow for growth in endowments.

\begin{corollary}\label{conditionB-example}
\begin{enumerate}
\item\label{conditionB-example1} Let Assumption \ref{assum0} be satisfied and the function $U^t$ be quasi-concave, continuously differentiable, and strictly increasing in each component.  Assume also that $\limsup\limits_{t\to\infty}D_t^{1/t}, \limsup_{t\to\infty} \frac{d_{t+1}+e^y_{t+1}}{e^y_t}<\infty$ and the function $\frac{V^t_1(x_1,x_2)}{V^t_2(x_1,x_2)}=f(x_1,x_2)$  $\forall t,x_1,x_2$. Then Condition (B) holds if one of the following conditions is satisfied.

\begin{enumerate}
\item \label{conditionB-example11} There exist $R>0$, $\bar{\epsilon}_1, \bar{\epsilon}_2>0$ and $T_0$ such that 
\begin{align}f(1-\epsilon_1,g_{e,t+1}+\epsilon_2)< R\leq \frac{G_ne^y_{t+1}}{e^y_t} \text{ and } \sum_{s\geq 1}\frac{D_s}{R^s}=\infty  
\end{align}
$\forall t\geq T_0,$ $\forall \epsilon_1\in (0,\bar{\epsilon}_1)$, $\forall  \epsilon_2\in (0,\bar{\epsilon}_2)$.

\item \label{conditionB-example12}  Assumption \ref{derivative-ij} holds and 
\begin{align}\label{conditionB-example-condition1}
f\big(1,\limsup_{t\to\infty}\frac{e^o_{t+1}}{e^y_t}\big)<R<\min\Big\{\limsup_{t\to\infty}D_t^{1/t},\liminf_{t\to\infty}\frac{G_ne^y_{t+1}}{e^y_t}\Big\}.
\end{align}
\end{enumerate}

\item \label{conditionB-example2}
Assume that $U^t(x_1,x_2)=\frac{x_1^{1-\sigma_1}}{1-\sigma_1}+\frac{x_2^{1-\sigma_2}}{1-\sigma_2}$ where $\sigma_1,\sigma_2>0$.
Condition (B) holds if  $\limsup_{t\to\infty}\frac{D_{t+1}}{D_t}<\infty$, $\limsup_{t\to\infty} \frac{d_{t+1}+e^y_{t+1}}{e^y_t}<\infty$, and there exist $\bar{\epsilon}_1, \bar{\epsilon}_2>0$ and $T$ such that
\begin{align}\label{B-holding-2}
(e^{y}_t)^{\sigma_2-\sigma_1}\frac{(\frac{e^o_{t+1}}{e^y_t}+\epsilon_2)^{\sigma_2}}{(1-\epsilon_1)^{\sigma_1}}\leq \min\Big\{\frac{D_{t+1}}{D_t},\frac{G_ne^y_{t+1}}{e^y_t}\Big\} \text{ }\forall t\geq T, \forall \epsilon_1\in (0,\bar{\epsilon}_1), \forall \epsilon_2\in (0,\bar{\epsilon}_2).
\end{align}
\end{enumerate}
\end{corollary}


Let us clarify some points. If the endowment sequences $(e^y_t),(e^o_t)$ are uniformly bounded away from zero and from above, condition (\ref{B-holding-2}) can be easily satisfied (if ${D_{t+1}}/{D_t}$ and $G_n$ are high).

 However, if $e^y_t\to\infty$ (for instance, $e^y_t=G^t$ with $G>1$), then the preferences' parameters $\sigma_1, \sigma_2$ play an important role.  Let us consider two cases. First, if $\sigma_2\leq \sigma_1$, condition (\ref{B-holding-2}) can hold in many cases. Second, if $\sigma_2>\sigma_1$, then  $(e^{y}_t)^{\sigma_2-\sigma_1}$ diverges to infinity. So, condition (\ref{B-holding-2}) is in general not satisfied.
}
Moreover, we observe that the statement \ref{onlybubblypart2} of Theorem \ref{onlybubbly} may not hold. Indeed, assume that $\limsup_{t\geq \infty}\frac{D_{t+1}}{D_t}<\infty$, and  $\frac{e^o_{t+1}}{e^y_t}=G_e>0$ $\forall t$. Then, the benchmark interest rate $R^*_{t+1}=(e^{y}_t)^{\sigma_2-\sigma_1}(\frac{e^o_{t+1}}{e^y_t})^{\sigma_2}=(e^{y}_t)^{\sigma_2-\sigma_1}G_e^{\sigma_2}$. This implies that $\frac{D_{t+1}}{D_tR^*_{t+1}}$ converges to zero. Therefore, $\sum_{t\to\infty}\frac{D_t}{R^*_1\cdots R^*_t}<\infty$. By Proposition \ref{necessitycondition}\ref{existence}, there exists a bubbleless equilibrium.

\begin{proof}[{\bf Proof of Corollary \ref{conditionB-example}}]
Condition $\limsup_{t\to\infty} {(d_{t+1}+e^y_{t+1})}/{e^y_t}<\infty$ ensures \eqref{boundPhi}.

The second statement of Corollary \ref{conditionB-example}  is a direct consequence of Lemma \ref{justifyB}\ref{holding-condition2}.  Let us prove the first one. 
Point \ref{conditionB-example11} of Corollary \ref{conditionB-example} is a direct consequence of Lemma \ref{justifyB}\ref{holding-condition1}. We prove here point \ref{conditionB-example12}.  Condition (\ref{conditionB-example-condition1}) implies that there exists $R>0$ and $t_0$ such that 
\begin{align}
f\big(1,\limsup_{s\to\infty}\frac{e^o_{s+1}}{e^y_s}\big)<R<G_n\frac{e^y_{t+1}}{e^y_t} \text{ } \forall t\geq t_0, \text{ and } R<\limsup_{t\to\infty}D_t^{1/t}
\end{align}
Denote $G_e\equiv \limsup_{t\to\infty}\frac{e^o_{t+1}}{e^y_t}$.  

Since $f$ is continuous, there exist $\bar{\epsilon}_1,\bar{\epsilon}_2>0$ such that $f(1-\epsilon_1,G_e+2\epsilon_2\big)<R$ $\forall \epsilon_1\in (0,\bar{\epsilon}_1), \forall \epsilon_2\in (0,\bar{\epsilon}_2)$.

Take any $\epsilon_1\in (0,\bar{\epsilon}_1), \epsilon_2\in (0,\bar{\epsilon}_2)$. 
By definition of $G_e$, we can choose $t_1$ such that $\frac{e^o_{t+1}}{e^y_t}<G_e+\epsilon_2$ $\forall t\geq t_1$. Observe that, by Assumption \ref{derivative-ij}, the function $f(x_1,x_2)$ is increasing in $x_2$, we have 
$f(1-\epsilon_1,\frac{e^o_{t+1}}{e^y_t}+\epsilon_2)\leq f(1-\epsilon_1,G_e+2\epsilon_2)<R$. To sum up, we have $R<\limsup_{t\to\infty}D_t^{1/t}$ and  
\begin{align}\frac{V^t_1\left(1-\epsilon_1,g_{e,t+1}+\epsilon_2\right)}{V^t_2\left(1-\epsilon_1,g_{e,t+1}+\epsilon_2\right)}=f(1-\epsilon_1,\frac{e^o_{t+1}}{e^y_t}+\epsilon_2)< R\leq \frac{G_n e^y_{t+1}}{e^y_t}
\end{align}
for all  $t\geq T\equiv \max\{t_0,t_1\}$. According to Lemma \ref{justifyB}\ref{holding-condition1}, Condition $B$ holds.
\end{proof}

\subsubsection*{An extension of Theorem \ref{onlybubbly}}

We provide a generalized version of Theorem \ref{onlybubbly}, where the saving rate is not necessarily bounded away from zero.

\begin{theorem}[a generalized version of Theorem \ref{onlybubbly}]
\label{onlybubbly-general}
Let Assumptions \ref{assum0}, \ref{assum1new} be satisfied.  Assume Condition (B'): there exist $\bar{\epsilon}\in (0,1)$, positive sequences $(\gamma_t)_t, (X_t)_t$, and  a date $T$  satisfying $\gamma_t\in (0,1]$ and the following conditions:
\begin{enumerate}
\item \label{onlybubbly2general}({\it Not-too-low dividend condition})   $\sum_{t=1}^{\infty}\dfrac{D_t}{X_1\cdots X_t}=\infty$.
\item \label{onlybubbly3general} {\it (Low interest rate conditions)}: 2(a): $X_{t+1}\leq G_n\dfrac{e^y_{t+1}\gamma_{t+1}}{e^y_t\gamma_t} $ $\forall t\geq T$. (2b):  For any $t\geq T$, if $\epsilon\in (0,\bar{\epsilon})$ and  $X\in \mathcal{R}_t(\epsilon)$, then $X\leq X_{t+1}$, where  $\mathcal{R}_t(\epsilon)$ is defined by 
\begin{align}\label{euler4}
\mathcal{R}_t(\epsilon)\equiv \Big\{X>0: X=\frac{V^t_1\left(1-\epsilon\gamma_t,g_{e,t+1}+X\epsilon\gamma_t\right)}{V^t_2\left(1-\epsilon\gamma_t,g_{e,t+1}+X\epsilon\gamma_t\right)},  0\leq e^y_tX\epsilon\gamma_t/G_n-d_{t+1}\leq e^y_{t+1}\Big\}.
\end{align}
\end{enumerate}
Then, for any equilibrium, we have $\liminf_{t\to\infty}\frac{a_t}{\gamma_te^y_t}>0$. If we assume, in addition, that $\sum_{t=1}^{\infty}\frac{D_t}{G_n^te^y_t\gamma_t}<\infty$, then every equilibrium is bubbly.  


\end{theorem}
{When $\gamma_t=1$ $\forall t$, Condition (B') in Theorem \ref{onlybubbly-general} becomes Condition (B) in  Theorem \ref{onlybubbly}.  To illustrate Theorem \ref{onlybubbly-general}, assume that $U^t(x_1,x_2)=(1-\delta_t)\ln(x_1)+\delta_t \ln(x_2)$, where $\delta_t\in (0,1)$, and $e^o_t=0$ $\forall t$. Then, there exists a unique equilibrium and it is determined by $a_t=\delta_te^y_t$.  This unique equilibrium is bubbly if and only if 
$
\sum_{t\geq 1}{d_t}/{(\delta_te^y_t)}<\infty.$ 
This condition can be fulfilled even if the saving rate $\delta_t$ converges to zero. 


\begin{proof}[{\bf Proof of Theorem \ref{onlybubbly-general}}]

{\bf Part 1}. We need to prove that $\liminf_{t\to\infty}\frac{a_t}{\gamma_te^y_t}>0$ for any equilibrium. 

Let $\bar{\epsilon}\in (0,1)$, positive sequences $(\gamma_t), (X_t)$, and  a date $T$ be in Condition (B').

Take an equilibrium. Denote $\epsilon_t\equiv \frac{a_t}{\gamma_te^y_t}.$  Suppose that $\liminf_{t\to\infty}\frac{a_t}{\gamma_te^y_t}=0$. Then there exists $t_0\geq T$ such that $ \frac{a_{t_0}}{\gamma_te^y_{t_0}}<\bar{\epsilon}<1$. Since $0<\bar{\epsilon}<1,0<\gamma_t\leq 1$, we have $0<a_{t_0}<e^y_{t_0}$. By consequence, we have the Euler condition
\begin{align}R_{t_0+1}
&=\frac{V^{t_0}_1\left(1-\frac{a_{t_0}}{e^y_{t_0}},g_{e,t_0+1}+R_{t_0+1}\frac{a_{t_0}}{e^y_{t_0}}\right)}{V^{t_0}_2\left(1-\frac{a_{t_0}}{e^y_{t_0}},g_{e,t_0+1}+R_{t_0+1}\frac{a_{t_0}}{e^y_{t_0}}\right)}=\frac{V^{t_0}_1\left(1-\epsilon_{t_0}\gamma_{t_0},g_{e,t_0+1}+R_{t_0+1}\epsilon_{t_0}\gamma_{t_0}\right)}{V^{t_0}_2\left(1-\epsilon_{t_0}\gamma_{t_0},g_{e,t_0+1}+R_{t_0+1}\epsilon_{t_0}\gamma_{t_0}\right)}.
\end{align}
By the asset pricing equation, we have $a_{t_0+1}=R_{t_0+1}e^y_{t_0}\gamma_{t_0}\epsilon_{t_0}/G_n-d_{t_0+1}$. Since $a_{t_0+1}\in [0,e^y_{t_0+1}]$, we have $0\leq e^y_{t_0} R_{t_0+1}\gamma_{t_0}\epsilon_{t_0}/G_n-d_{t_0+1}\leq e^y_{t_0+1}$. 

By our condition (B'), we have $R_{t_0+1}\leq X_{t_0+1}$. Since $X_{t_0+1}\leq  G_n\frac{e^y_{t_0+1}\gamma_{t_0+1}}{e^y_{t_0}\gamma_{t_0}}$, we have $R_{t_0+1}\leq   G_n\frac{e^y_{t_0+1}\gamma_{t_0+1}}{e^y_{t_0}\gamma_{t_0}}$. Combining with the non-arbitrage condition $a_{t_0+1}+d_{t_0+1} =a_{t_0}\frac{R_{t_0+1}}{G_n}$, we get that
\begin{align}
a_{t_0+1}\leq a_{t_0}\frac{R_{t_0+1}}{G_n}&\leq a_{t_0}\frac{e^y_{t_0+1}\gamma_{t_0+1}}{e^y_{t_0}\gamma_{t_0}} 
\Rightarrow \epsilon_{t_0+1}= \frac{a_{t_0+1}}{\gamma_{t_0+1}e^y_{t_0+1}}&\leq  \frac{a_{t_0}}{\gamma_{t_0}e^y_{t_0}}= \epsilon_{t_0}<\bar{\epsilon}.
\end{align}
Therefore, by induction, we have
$\epsilon_{t+1}\leq   \epsilon_{t} <\bar{\epsilon},  R_{t+1}\leq X_{t+1} $ $\forall t\geq t_0$. Consequently, $R_{t_0+1}\cdots R_t\leq X_{t_0+1}\cdots X_t$ $\forall t>t_0$. 
We now look at the fundamental value
\begin{align}
F_0&=\sum_{t\geq 1}\frac{D_t}{R_1\cdots R_t}=\sum_{t=1}^{t_0}\frac{D_t}{R_1\cdots R_t}+\sum_{t=t_0+1}^{\infty}\frac{D_{t}}{R_1\cdots R_{t_0}R_{t_0+1}\cdots R_{t}}.
\end{align}
Consider the second term $A_0\equiv \sum_{t=t_0+1}^{\infty}\frac{D_{t}}{R_1\cdots R_{t_0}R_{t_0+1}\cdots R_{t}}$.
We have 
\begin{align*}
A_0&= \frac{1}{R_1\cdots R_{t_0}}\sum_{t=t_0+1}^{\infty}\frac{D_{t}}{R_{t_0+1}\cdots R_{t}}\geq \frac{1}{R_1\cdots R_{t_0}}\sum_{t=t_0+1}^{\infty}\frac{D_{t}}{X_{t_0+1}\cdots X_t}\\
&=\frac{X_{1}\cdots X_{t_0}}{R_1\cdots R_{t_0}}\sum_{t=t_0+1}^{\infty}\frac{D_{t}}{X_{1}\cdots X_t}=\infty
\end{align*}
because of our assumption (B'\ref{onlybubbly2general}), i.e., $\sum_{t=1}^{\infty}\frac{D_t}{X_{1}\cdots X_t}=\infty$. This implies that $F_0=\infty$. Since $q_0\geq F_0$, we have  $q_0=\infty$, a contradiction. So, we have $\liminf_{t\to\infty}\frac{a_t}{\gamma_te^y_t}>0$. \\
{\bf Part 2}. 
 Let the condition in the first statement be satisfied. According to  part \ref{onlybubbly2general} of  Theorem \ref{onlybubbly-general}, any equilibrium satisfies $\liminf_{t\to\infty}\frac{a_t}{\gamma_t e^y_t}>0$.  Combining this with our assumption  $\sum_{t=1}^{\infty}\frac{D_t}{\gamma_tG_n^te^y_t}<\infty$  and  Lemma \ref{prop1}\ref{prop1-5}, we conclude that this equilibrium is bubbly.



\end{proof}

\subsection*{\bf  Online Appendix 3: Detailed proofs for intermediate results used in the proof of Theorem \protect\ref{allsets}}

\begin{proof}[{Proof of Lemma \ref{tl2g2}}]

Since $R_{t}\leq G_n$, we have  $
a_{t}=\frac{R_{t}}{G_n}a_{t-1}-d_{t}<a_{t-1}.$ This implies that $R_{t+1}=g(a_{t})<g(a_{t-1})=R_{t}\leq G_n.$ Hence, $R_{t+1}<R_{t}$. It means that we have $R_{t+1}<R_{t}$ and $R_{t+1}<G_n$. By induction, we get  $
G_n\geq R_{t}>R_{t+1}>\cdots$, 
which implies that $\left( R_{t},a_{t}\right) $ converges to $\left(
R,a\right) $ and $a(R-G_n)=0$. We have ${R}<G_n$ because $\lim_{s\rightarrow \infty }R_{s}\leq
R_{t+1}<R_{t}\leq G_n$. This implies that $a=0$. So, $R_{t+1}=g(a_{t})$
converges to $R^{\ast }$.
 \end{proof}
\begin{proof}[{Proof of Lemma \ref{tl5g2}}]Consider $a_{0}^{\prime }<a_{0}$ two bubbly equilibria. Suppose that  for both values,
the interest rates $R_{t}, R_t^{\prime }$ converge to $G_n$. 
  By Lemma \ref{tl3g2}, we have  $\lim_{t\rightarrow \infty }a_{t}=\lim_{t\rightarrow \infty }a_{t}^{\prime }=\hat{a}$. Moreover, $\hat{a}>0$ because $R^*<G_n$. 
 Since $a_{0}^{\prime }<a_{0}$, we have $R_{t}^{\prime }<R_{t},a_{t}^{\prime
}<a_{t} \text{  } \forall  t\geq 1$. Therefore, we have  
\begin{equation*}
\frac{a_{t}^{\prime }}{a_{t}}=\frac{\frac{R_{t}^{\prime }}{G_n}a_{t-1}^{\prime
}-d_{t}}{\frac{R_{t}}{G_n}a_{t-1}-d_{t}}<\frac{R_{t}^{\prime }a_{t-1}^{\prime }%
}{R_{t}a_{t-1}}<\frac{a_{t-1}^{\prime }}{a_{t-1}}<\cdots <\frac{%
a_{0}^{\prime }}{a_{0}}<1.
\end{equation*}%
So, ${a_{t}^{\prime }}/{a_{t}}$ does not converge to $1$. \end{proof}

\begin{proof}[{Proof of Lemma \ref{tl7g}}] Let $(a_{t}^{b})$ be an equilibrium
satisfying $a_{t}^{b}\rightarrow a^{b}\in \lbrack 0,e^{y})$ with $%
R^{b}=\lim_{t\rightarrow \infty }g(a_{t}^{b})<G_n$. We have $u'(e^y-a^b_t)=  R^b_{t+1}v'(e^o+R^b_{t+1}a^b_t)$ and  $u'(e^y-a^b)=  R^b v'(e^o+R^ba^b)$.
By Lemma \ref{5}, we have $a^bu'(e^y-a^b)<   \lim_{c\to \infty}cv'(c)$. Indeed, this is trivial if $a^b=0$. If $a^b>0$, we apply Lemma  \ref{5}.

Therefore, there exist $T,x,x_a$ such that $x\in (0,G_n)$,  $x_{a}\in (0,e^{y})$, $g(a_{t}^{b})<x<G_n$, $%
a_{t}^{b}<x_{a}$ $\forall t\geq T$, and $x_au'(e^y-x_a)<  \lim_{c\to \infty}cv'(c)$.

We can choose $b_0>0$ small enough and define the sequence $(a_{t},R_{t})_{t=0}^T$ by 
\begin{align}
a_0&=a^b_0+b_0, \quad
R_{t+1}=g(a_t), \quad
a_{t+1}=\frac{R_{t+1}}{G_n}a_t- d_{t+1}
\end{align}
such that $ a_T<x_a, g(a_T)<x$, $a_tu'(e^y-a_t)<  \lim_{c\to \infty}cv'(c)$ $\forall t\in \{0,1,\ldots, T\}$.

Then, it is easy to see that $a_t>a^b_t,R_t>R^b_t$ $\forall t\in \{0,1,\ldots, T\}$.


We now define $R_{t+1},a_{T+1}$. Since $a_Tu'(e^y-a_T)<  \lim_{c\to \infty}cv'(c)$, Lemma \ref{5} allows us to define $R_{T+1}=g(a_T)$ and  then $a_{T+1}=\frac{R_{T+1}}{G_n}a_T- d_{T+1}$. We have 
$$
R_{T+1}=g(a_T)<x<G_n, \quad 
a_{T+1}=\frac{R_{T+1}}{G_n}a_T- d_{T+1}\leq \frac{x}{G_n}a_T<a_T<x_a.$$
Since $a_{T+1}<x_a$ and $x_au'(e^y-x_a)<  \lim_{c\to \infty}cv'(c)$, we have $a_{T+1}u'(e^y-a_{T+1})<  \lim_{c\to \infty}cv'(c)$. Then, by induction, we can construct $(a_t,R_t)_{t\geq 0}$ such that $R_{T+s}<x<G_n, a_{T+s}<x_a \text{  } \forall  s\geq 1$. This implies that $$a_{T+s+1}=\frac{R_{T+s+1}}{G_n}a_{T+s}-d_{T+s+1}< \frac{G_n}{G_n}a_{T+s}=a_{T+s}.$$ Hence, $R_{T+s}<R_{T+s-1}$ $\forall s\geq 1$, which implies that  $\lim_{t\rightarrow \infty }R_{t}<R_{T+1}<G_n$. 

Summing up, the sequence $(a_{t})$ is an equilibrium because $0<a_{t}^{b}<a_{t}<x_{a}<e^{y}$. According to Lemma \ref{interval1}'s  point 3, this is bubbly because $a_{0}>a_{0}^{b}$.\end{proof}

\subsection*{\bf  Online Appendix 4: An explicit model with asset bubbles}
\label{explicitmodel}
 We now provide a model, based on Example 3 in \cite{blp21_appendix}, where we can explicitly compute the equilibrium prices with bubbles and it completely fulfills Theorem \ref{allsets}.  According to  (\ref{system1}), the Euler condition becomes  $
u'(e_{t}^{y}-a_t)=  G_n \frac{a_{t+1}+d_{t+1}}{a_t} v'\big(e_{t+1}^{o}+G_n (a_{t+1}+d_{t+1})\big).$  

Let us consider a special setup where $u(c)=\ln(c), v(c)=\beta \ln(c)$, where $\beta>0$, and assume that $e^o_{t}>0$ $\forall t$. We have the following non-autonomous system 
\begin{align*}
a_{t+1}+d_{t+1}=\frac{a_t}{\frac{G_n\beta e^y_t}{e^o_{t+1}}-\frac{G_n(1+\beta)}{e^o_{t+1}}a_t}, \text{ or, equivalently, }
\frac{1}{a_{t+1}+d_{t+1}}=\frac{G_n\beta e^y_t}{e^o_{t+1}}\frac{1}{a_t}-\frac{G_n(1+\beta)}{e^o_{t+1}}.
\end{align*}

Assume a stationary endowment: $e^y_t=e^y>0,e^o_t=e^o>0$ $\forall t$. Note that the interest rate $R^*=\frac{e^o}{\beta e^y}$.  

Let the interest rate be lower than the population growth rate: $R^*<G_n$. 

Let $x>0$ be such that 
$\frac{x+1}{x}\frac{R^*}{G_n}>1$, or, equivalently, $1-x(\frac{G_n}{R^*}-1)>0$.

Denote $h\equiv \frac{G_n(1+\beta)}{e^o}$. Define the dividend sequence $(d_t)$ by
\begin{align}
\frac{1}{d_t}-\frac{hx(1+x)}{1-x(\frac{G_n}{R^*}-1)}&=\Big(\frac{x+1}{x}\frac{R^*}{G_n}\Big)^t\Big(\frac{1}{d_0}-\frac{hx(1+x)}{1-x(\frac{G_n}{R^*}-1)}\Big)\\
0<d_0&<\frac{1-x(\frac{G_n}{R^*}-1)}{hx(1+x)}.
\end{align}
We can check that $\frac{1}{d_{t+1}}=\frac{x+1}{x}\frac{R^*}{G_n}\frac{1}{d_t}-\frac{(x+1)hR^*}{G_n}$. Moreover, $\lim_{t\to \infty}d_t^{1/t}=\frac{x G_n}{(x+1)R^*}$ which is, by our assumption, lower than $1$.


In the economy with above specifications, we can check that the following sequence is an equilibrium 
\begin{align}\label{example}
a_t=\big(\frac{G_n}{R^*}-1\big)\frac{1}{h}+xd_t \text{ for any  } t\geq 0. 
\end{align}
Since $\frac{x+1}{x}\frac{R^*}{G_n}>1$, we have $\sum_{t\geq 1}d_t<\infty$ and hence $\sum_{t\geq 1}\frac{d_t}{a_t}<\infty$. Therefore, this equilibrium price is bubbly. Moreover, we have $\lim_{t\to \infty}a_t=\big(\frac{G_n}{R^*}-1\big)\frac{1}{h}.$ According to Theorem \ref{allsets}\ref{part2}, this is the unique equilibrium satisfying $\lim_{t\to\infty}a_t>0$. By applying Claims 1 and 2 in Theorem \ref{allsets}, we see that:
\begin{itemize}
\item  If $R^*>\lim_{t\to \infty}D_t^{1/t}= G_n\lim_{t\to \infty}d_t^{1/t}=\frac{xG_n^2}{(x+1)R^*}$ (i.e., $\big(\frac{e^o}{\beta e^y}\big)^2\frac{x+1}{x}> G_n^2$), then Claim 1 in Theorem \ref{allsets} holds. We have a continuum of equilibria and the maximal equilibrium is $(a_t)$ defined by (\ref{example}).

\item  If $R^*<\lim_{t\to \infty}D_t^{1/t}=G_n\lim_{t\to \infty}d_t^{1/t}=\frac{xG_n^2}{(x+1)R^*}$ (i.e., $\big(\frac{e^o}{\beta e^y}\big)^2\frac{x+1}{x}<G_n^2$), then Claim 2 in Theorem \ref{allsets} holds. There exists a unique equilibrium and the equilibrium asset value $(a_t)$ is defined by (\ref{example}).
\end{itemize}

\subsection*{\bf Online Appendix 5: Additional results and Proofs for Sections \ref{section-pareto} and \ref{section-bubble-pareto}}
\label{section-pareto-proof}

\subsubsection*{Justifying the uniform strictness and smoothness conditions}
The following results justify  the uniform strictness and smoothness conditions by proving that they can be satisfied in a large class of models. Observe that (\ref{strict-uniform}) is equivalent to
$$
U^t_2(c^{o'}_{t+1}-c^{o}_{t+1})+U^t_1(c^{y'}_t-c^{y}_t)\geq\frac{\mu}{U^t_1c^y_t}\big(U^t_1(c^{y'}_t-c^{y}_t)\big)^2 \text{ }\forall (c^{y'}_t,c^{o'}_{t+1})\in B_t(c), \text{  } c^{y'}_t<c^{y}_t$$
while the second inequality in (\ref{smooth-uniform-1}) becomes 
$$
v'(c^{o}_{t+1})(c^{o'}_{t+1}-c^{o}_{t+1})+u'(c^y_t)(c^{y'}_t-c^{y}_t)\geq\frac{{\theta}_2(x)}{v'(c^{o}_{t+1})c^o_{t+1}}\big(v'(c^{o}_{t+1})(c^{o'}_{t+1}-c^{o}_{t+1})\big)^2 +\frac{{\theta}_1(x)}{u'(c^y_t)c^y_t}\big(u'(c^y_t)(c^{y'}_t-c^{y}_t)\big)^2.
$$

\begin{lemma}[checking the uniform strictness condition]
\label{check-strict}Assume that $U^t(x_1,x_2)= u_t(x_1)+v_t(x_2)$ where the two functions $u_t,v_t:\rr_+\to \rr$ are in $C^2$, strictly concave, strictly increasing. 
\begin{enumerate}
\item \label{check-strict-part1} Any allocation  $(c^y_t,c^o_t)_t$ with $c^y_t>0,c^o_t>0$ $\forall t$, satisfies  the uniform strictness condition if there exists $h\in (0,1]$ such that $$\inf_{t\geq 0}\Big\{\frac{c^y_t}{u_t^{\prime}(c^y_t)}\inf_{x\in [(1-h)c^y_t,c^y_t]} \big(-\frac{1}{2}u_t^{\prime\prime}(x)\big)\Big\}>0$$
\item \label{check-strict-part2}If $u_t'(c)=c^{-\sigma}$ with $\sigma>0$, then any allocation $(c^y_t,c^o_t)_t$ with $c^y_t>0,c^o_t>0$ $\forall t$, satisfies  the uniform strictness condition.

\end{enumerate}
\end{lemma}

\begin{lemma}[checking the uniform smoothness condition]\label{check-smooth}
Assume that $U^t(x_1,x_2)=u_t(x_1)+v_t(x_2)$ where $u_t,v_t:\rr_+\to \rr$ are in $C^2$, strictly concave, strictly increasing, and $u_t^{\prime\prime}<0$. 

\begin{enumerate}
\item Any allocation $(c^y_t,c^o_t)_t$ with $c^y_t>0,c^o_t>0$  $\forall t$, satisfies  the uniform smoothness condition if for each $x\in (0,1)$, we have
$$\bar{M}_1\equiv \sup_{t\geq 0}\Big\{\frac{c^y_t}{u_t^{\prime}(c^y_t)}\sup_{c\in [xc^y_t,c^y_t]} \big(-\frac{1}{2}u_t^{\prime\prime}(c)\big)\Big\}, \bar{M}_2\equiv \sup_{t\geq 0}\Big\{\frac{c^o_{t+1}}{v_t^{\prime}(c^o_{t+1})}\sup_{c\in [c^o_{t+1},G_ne_{t+1}]} \big(-\frac{1}{2}v_t^{\prime\prime}(c)\big)\Big\}<\infty.$$
\item Assume that $u_t(c)=\frac{c^{1-\sigma}}{1-\sigma}$ and $v_t'(c)=\gamma_tc^{-\sigma}$ with $\sigma>0,\gamma_t>0$ $\forall t$.
Then any allocation $(c^y_t,c^o_t)_t$ with $c^y_t>0,c^o_t>0$ $\forall t$, satisfies  the uniform smoothness condition.

\end{enumerate}
\end{lemma}
The proofs of Lemmas \ref{check-strict} and \ref{check-smooth} are basically based on the Taylor's expansion and the concavity of the functions $u_t,v_t$. 

\begin{proof}[{\bf Proof of Lemma \ref{check-strict}}]
The allocation $(c^y_t,c^o_t)_t$ satisfies the uniform strictness condition if there exists $\mu>0$ such that 
\begin{align}
\frac{U^t_2(c^{o'}_{t+1}-c^{o}_{t+1})+U^t_1(c^{y'}_t-c^{y}_t)}{\big(U^t_1(c^{y'}_t-c^{y}_t)\big)^2}U^t_1c^y_t\geq \mu
\end{align}
for any $t$ and for any couple $(c^{y'}_t,c^{o'}_{t+1})$ satisfying
\begin{align}
U^t(c^{y'}_t,c^{o'}_{t+1})\geq  U^t(c^{y}_t,c^{o}_{t+1}), \quad \
0<(1-h)c^y_t<c^{y'}_t<c^{y}_t,\quad c^{o'}_{t+1}>c^{o}_t
\end{align}
By the Taylor's expansion, there exists $s_{yt},s_{ot}\in (0,1)$ such that 
\begin{align*}
&U^t(c^{y'}_t,c^{o'}_{t+1})-  U^t(c^{y}_t,c^{o}_{t+1})=u_t(c^{y'}_t)-u_t(c^{y}_t)+v_t(c^{o'}_{t+1})-v_t(c^{o}_{t+1})\\
=&u^{\prime}_t(c^{y}_t)(c^{y'}_t-c^{y}_t)+\frac{1}{2}u^{\prime\prime}_t\big(c^{y}_t+s_{yt}(c^{y'}_t-c^{y}_t)\big)(c^{y'}_t-c^{y}_t)^2\\
&+v_t^{\prime}(c^{o}_{t+1})(c^{o'}_{t+1}-c^{o}_{t+1})+\frac{1}{2}v_t^{\prime\prime}\big(c^{o}_{t+1}+s_{ot}(c^{o'}_{t+1}-c^{o}_{t+1})\big)(c^{o'}_{t+1}-c^{o}_{t+1})^2.
\end{align*}
Since $U^t(c^{y'}_t,c^{o'}_{t+1})\geq  U^t(c^{y}_t,c^{o}_{t+1})$, we have 
\begin{align*}
&u_t^{\prime}(c^{y}_t)(c^{y'}_t-c^{y}_t)+\frac{1}{2}u_t^{\prime\prime}\big(c^{y}_t+s_{yt}(c^{y'}_t-c^{y}_t)\big)(c^{y'}_t-c^{y}_t)^2\\
&+v_t^{\prime}(c^{o}_{t+1})(c^{o'}_{t+1}-c^{o}_{t+1})+\frac{1}{2}v_t^{\prime\prime}\big(c^{o}_{t+1}+s_{ot}(c^{o'}_{t+1}-c^{o}_{t+1})\big)(c^{o'}_{t+1}-c^{o}_{t+1})^2\geq 0.
\end{align*}
Since $v_t^{\prime\prime}\leq 0$, we have 
\begin{align*}
u_t^{\prime}(c^{y}_t)(c^{y'}_t-c^{y}_t)+v_t^{\prime}(c^{o}_{t+1})(c^{o'}_{t+1}-c^{o}_{t+1})\geq -\frac{1}{2}u_t^{\prime\prime}\big(c^{y}_t+s_{yt}(c^{y'}_t-c^{y}_t)\big)(c^{y'}_t-c^{y}_t)^2.
\end{align*}
By consequence, 
\begin{align*}
\frac{U^t_2(c^{o'}_{t+1}-c^{o}_{t+1})+U^t_1(c^{y'}_t-c^{y}_t)}{\big(U^t_1(c^{y'}_t-c^{y}_t)\big)^2}U^t_1c^y_t&\geq \frac{-1}{2}u_t^{\prime\prime}\big(c^{y}_t+s_{yt}(c^{y'}_t-c^{y}_t)\big)\frac{c^y_t}{u_t^{\prime}(c^y_t)}\\
&\geq\frac{c^y_t}{u_t^{\prime}(c^y_t)}\inf_{x\in [(1-h)c^y_t,c^y_t]} \frac{-1}{2}u_t^{\prime\prime}(x).
\end{align*}
\begin{enumerate}
\item  If $\bar{x}\equiv \inf_{t\geq 0}\Big\{\frac{c^y_t}{u_t^{\prime}(c^y_t)}\inf_{x\in [(1-h)c^y_t,c^y_t]} \frac{-1}{2}u_t^{\prime\prime}(x)\Big\}>0$, we define $\mu \equiv \bar{x}.$ We have proved Lemma \ref{check-strict}\ref{check-strict-part1}.

\item  Consider the case where $u_t^{\prime}(c)=c^{-\sigma}$. Then,  $u_t^{\prime\prime}(c)=-\sigma c^{-(\sigma+1)}$.
Therefore, 
$$\frac{-1}{2}u_t^{\prime\prime}\big(c^{y}_t+s_{yt}(c^{y'}_t-c^{y}_t)\big)\frac{c^y_t}{u_t^{\prime}(c^y_t)}=\frac{\sigma}{2}\Big(\frac{c^y_t}{c^{y}_t+s_{yt}(c^{y'}_t-c^{y}_t)}\Big)^{1+\sigma}>\frac{\sigma}{2}$$
because $c^{y'}_t-c^{y}_t<0$. Therefore, we get
\begin{align}
\frac{U^t_2(c^{o'}_{t+1}-c^{o}_{t+1})+U^t_1(c^{y'}_t-c^{y}_t)}{\big(U^t_1(c^{y'}_t-c^{y}_t)\big)^2}U^t_1c^y_t>\frac{\sigma}{2}.
\end{align}
So, the uniform strictness condition is satisfied.
\end{enumerate}
\end{proof}

\begin{proof}[{\bf Proof of Lemma \ref{check-smooth}}]

We will prove that, for each $x>0$, there exists ${\theta}_1(x),{\theta}_2(x)>0$ such that, for any $t$, if the couple
$(c^{y'}_t,c^{o'}_{t+1})$ satisfies 
\begin{align}
&\begin{cases}
 xc^{y}_t<c^{y'}_t<c^{y}_t, c^{o}_{t+1}<c^{o'}_{t+1}<G_ne_{t+1}\\
P_{t+1}(c^{o'}_{t+1}-c^{o}_{t+1})+G_n P_t(c^{y'}_t-c^{y}_t)\geq\frac{{\theta}_2(x)}{P_{t+1}c^o_{t+1}}\big(P_{t+1}(c^{o'}_{t+1}-c^{o}_{t+1})\big)^2\\
\quad \quad +\frac{{\theta}_1(x)}{G_n P_tc^y_t}\big(G_n P_t(c^{y'}_t-c^{y}_t)\big)^2
 \end{cases}
\end{align}
then $(c^{y'}_t,c^{o'}_{t+1})\in B_t(c)$. i.e., $U^t(c^{y'}_t,c^{o'}_{t+1})\geq  U^t(c^{y}_t,c^{o}_{t+1})$.

By the Taylor's expansion, there exists $s_{yt},s_{ot}\in (0,1)$ such that 
\begin{align*}
U^{t'}-U^t\equiv &U^t(c^{y'}_t,c^{o'}_{t+1})-  U^t(c^{y}_t,c^{o}_{t+1})=u_t(c^{y'}_t)-u_t(c^{y}_t)+v_t(c^{o'}_{t+1})-v_t(c^{o}_{t+1})\\
=&u_t^{\prime}(c^{y}_t)(c^{y'}_t-c^{y}_t)+\frac{1}{2}u_t^{\prime\prime}(\tilde{c}^y_t)(c^{y'}_t-c^{y}_t)^2+v_t^{\prime}(c^{o}_{t+1})(c^{o'}_{t+1}-c^{o}_{t+1})+\frac{1}{2}v_t^{\prime\prime}(\tilde{c}^o_{t+1})(c^{o'}_{t+1}-c^{o}_{t+1})^2,
\end{align*}
where $\tilde{c}^y_t\equiv c^{y}_t+s_{yt}(c^{y'}_t-c^{y}_t)\in (c^{y'}_t,c^{y}_t)$ and $\tilde{c}^o_{t+1}\equiv c^{o}_{t+1}+s_{ot}(c^{o'}_{t+1}-c^{o}_{t+1})\in (c^{o}_{t+1},c^{o'}_{t+1})$.

From (\ref{smooth-uniform-1}) and $\frac{P_{t+1}}{P_t}=\frac{G_n}{R_{t+1}}=G_n\frac{U^t_2}{U^t_1}=G_n\frac{v_t'(c^{o}_{t+1})}{u_t'(c^y_t)}$, we have 
$$
v_t'(c^{o}_{t+1})(c^{o'}_{t+1}-c^{o}_{t+1})+u_t'(c^y_t)(c^{y'}_t-c^{y}_t)\geq\frac{{\theta}_2(x)}{v_t'(c^{o}_{t+1})c^o_{t+1}}\big(v_t'(c^{o}_{t+1})(c^{o'}_{t+1}-c^{o}_{t+1})\big)^2 +\frac{{\theta}_1(x)}{u_t'(c^y_t)c^y_t}\big(u'(c^y_t)(c^{y'}_t-c^{y}_t)\big)^2.
$$
Therefore
\begin{align*}
U^{t'}-U^t\geq &\Big(\theta_2(x)\frac{v_t'(c^{o}_{t+1})}{c^{o}_{t+1}}+\frac{1}{2}v_t^{\prime\prime}(\tilde{c}^o_{t+1})\Big)(\epsilon^o_{t+1})^2+ \Big(\theta_1(x)\frac{u_t'(c^{y}_{t})}{c^{y}_{t}}+\frac{1}{2}u_t^{\prime\prime}(\tilde{c}^y_{t})\Big)(\epsilon^y_t)^2.
\end{align*}
Point 1 of Lemma \ref{check-smooth} is clear since $xc^{y}_t<c^{y'}_t<c^{y}_t, c^{o}_{t+1}<c^{o'}_{t+1}<G_n e_{t+1}$. Indeed, if we choose  $\theta_1(x)>\bar{M}_1$ and $\theta_2(x)>\bar{M}_2$, where $\bar{M}_1, \bar{M}_2$ are defined in  Lemma \ref{check-smooth}, we have $U^{t'}-U^t>0$.

Let us check point 2. In this case, we have 
\begin{align*}
U^{t'}-U^t\geq &\Big(\theta_2(x)\frac{v_t'(c^{o}_{t+1})}{c^{o}_{t+1}}+\frac{1}{2}v_t^{\prime\prime}(\tilde{c}^o_{t+1})\Big)(\epsilon^o_{t+1})^2+ \Big(\theta_1(x)\frac{u_t'(c^{y}_{t})}{c^{y}_{t}}+\frac{1}{2}u_t^{\prime\prime}(\tilde{c}^y_{t})\Big)(\epsilon^y_t)^2\\
&=\gamma_t\big(\frac{\theta_2(x)}{(c^{o}_{t+1})^{1+\sigma}}-\frac{\sigma}{2(\tilde{c}^{o}_{t+1})^{1+\sigma}}\big) (\epsilon^o_{t+1})^2+ \big(\frac{\theta_1(x)}{(c^{y}_{t})^{1+\sigma}}-\frac{\sigma}{2(\tilde{c}^{y}_{t})^{1+\sigma}}\big)(\epsilon^y_t)^2.
\end{align*}
Choose $\theta_2(x)>\frac{\sigma}{2}$, we have $\frac{\theta_2(x)}{(c^{o}_{t+1})^{1+\sigma}}-\frac{\sigma}{2(\tilde{c}^{o}_{t+1})^{1+\sigma}}>0$ because $\tilde{c}^{o}_{t+1}>{c}^{o}_{t+1}$.

Choose $\theta_1(x)$ such that $\theta_1(x)x^{1+\sigma}-\frac{\sigma}{2}>0$, we have
\begin{align*}
\frac{\theta_1(x)(\tilde{c}^{y}_{t})^{1+\sigma}}{(c^{y}_{t})^{1+\sigma}}-\frac{\sigma}{2}>\theta_1(x)x^{1+\sigma}-\frac{\sigma}{2}>0.
\end{align*}
Therefore, we have $U^{t'}-U^t>0$. 

\end{proof}

\label{checking-proofs}
\begin{proof}[{\bf Proof of Lemma \ref{lemmaPareto}}]
We follow the classical idea of support prices \citep{malinvaud53_appendix, cass72_appendix}. 
Let us consider a feasible allocation path $(c_{t}^{y\prime },c_{t}^{o\prime
})_{t}$. We have $c_{t}^{y\prime }+\frac{c_{t}^{o\prime }}{G_n}=e_{t}^{y}+\frac{e_{t}^{o}}{G_n}%
+d_{t} \text{  } \forall t.$ 

Denote $U_{t}\equiv U^t\left( c_{t}^{y},c_{t+1}^{o}\right) $ and $
U_{t}^{\prime }\equiv U^t\left( c_{t}^{y\prime },c_{t+1}^{o\prime }\right) $.
Since the function $U^t$ is concave, we have that 
\begin{align}
\label{difu}
U_{t}-U_{t}^{\prime }\geq U^t_1(c_{t}^{y},c_{t+1}^{o})\left( c_{t}^{y}-c_{t}^{y\prime }\right) +U^t_2\left( c_{t}^{y},c_{t+1}^{o}\right)\left(
c_{t+1}^{0}-c_{t+1}^{o\prime }\right) \text{  } \forall  t\geq -1.
\end{align}
By the feasibility of allocations, we have $
c_{t}^{o}-c_{t}^{o\prime }=-G_n(c_{t}^{y}-c_{t}^{y\prime }) \text{  } \forall  t\geq 0.$ Combining with (\ref{difu}) and the no-arbitrage condition (\ref{00131}), we get 
\begin{align*}
\left( U_{t}-U_{t}^{\prime }\right) \frac{1}{R_{t+1}U^t_2(c_{t}^{y},c_{t+1}^{o})}& \geq c_{t}^{y}-c_{t}^{y\prime }+\frac{1}{R_{t+1}}\left(
c_{t+1}^{0}-c_{t+1}^{o\prime }\right)\\
&= 
c_{t}^{y}-c_{t}^{y\prime }-
\frac{G_n}{R_{t+1}}\left( c_{t+1}^{y}-c_{t+1}^{y\prime }\right) \text{  } \forall  t\geq 0.
\end{align*}
For the households born at date $-1$, we have
\begin{align*}
U_{-1}-U_{-1}^{\prime }&\geq U^{-1}_1
(c_{-1}^{y},c_{0}^{o})\left( c_{-1}^{y}-c_{-1}^{y\prime }\right) +U^{-1}_2(c_{-1}^{y},c_{0}^{o})\left(
c_{0}^{o}-c_{0}^{o\prime }\right)\\
&=-G_n U^{-1}_2(c_{-1}^{y},c_{0}^{o})\left(
c_{0}^{y}-c_{0}^{y\prime }\right) \text{ because $c_{-1}^{y}=c_{-1}^{y\prime }$,}\\
\Rightarrow \frac{U_{-1}-U_{-1}^{\prime }}{G_n U^{-1}_2(c_{-1}^{y},c_{0}^{o})}&\geq -\left(
c_{0}^{y}-c_{0}^{y\prime }\right)
\end{align*}

Denote $X_{t}\equiv \frac{G_n^{t}}{R_{1}\cdots R_{t+1}U^t_2(c_{t}^{y},c_{t+1}^{o})} $ $\forall t\geq 0$,  $X_{-1}\equiv \frac{1}{G_n U^{-1}_2(c_{-1}^{y},c_{0}^{o})}$ and $P_{t}\equiv 
\frac{G_n^{t}}{R_{1}\cdots R_{t}} \text{  } \forall  t\geq 1$, $P_0=1$. We have 
\begin{align}
X_t\left( U_{t}-U_{t}^{\prime }\right) & \geq P_{t}\left( c_{t}^{y}-c_{t}^{y\prime }\right)
-P_{t+1}(c_{t+1}^{y}-c_{t+1}^{y\prime }) \text{  } \forall  t\geq 0.
\end{align}%
Therefore, we get $\sum_{t=-1}^{T}X_t(U_{t}-U_{t}^{\prime })\geq
-P_{T+1}(c_{T+1}^{y}-c_{T+1}^{y^{\prime }})\geq -P_{T+1}c_{T+1}^{y}.$

Combining with our assumption  $\liminf\limits_{t\rightarrow \infty }\frac{%
G_n^{t}}{R_{1}\cdots R_{t}}c_{t}^{y}=0$, we get  
\begin{equation*}
\limsup_{T\rightarrow \infty }\sum\limits_{t=-1}^{T}X_t
\left( U_{t}-U_{t}^{\prime }\right)\geq \limsup_{T\rightarrow \infty }\big(-P_{T+1}c_{T+1}^{y}\big)=-\liminf_{T\rightarrow \infty }P_{T+1}c_{T+1}^{y}=0.
\end{equation*}

So, $(c_{t}^{y},c_{t}^{o})_{t}$ is Pareto optimal. Indeed, take another
feasible allocation $(c_{t}^{y^{\prime }},c_{t}^{o^{\prime }})_{t}$. Suppose
that $U_{t}^{\prime }\geq U_{t} \text{  } \forall  t$ and there exists $t_{0}$ such
that $U_{t_{0}}^{\prime }>U_{t_{0}}$. Then, $\sum\limits_{t=-1}^{T}X_t(U_{t}-U_{t}^{\prime })\leq X_{t_{0}}(U_{t_{0}}-U_{t_{0}}^{\prime
})<0 \text{  } \forall  t\geq t_{0}$. By consequence, $\limsup_{t\rightarrow \infty
}\sum\limits_{t=-1}^{T}X_t(U_{t}-U_{t}^{\prime })\leq
X_{t_{0}}(U_{t_{0}}-U_{t_{0}}^{\prime })<0$, a contradiction. Therefore, $%
(c_{t}^{y},c_{t}^{o})_{t}$ is Pareto optimal. \end{proof}

\subsubsection*{\bf Proof of Theorem \ref{pareto-theorem} and other results} 
To prove Theorem \ref{pareto-theorem}, we follow the strategy of the proofs of Theorem 3A  in \cite{OkunoZilcha1980_appendix} and Proposition 5.6 in \cite{bs80_appendix}. We need an intermediate step.

\begin{lemma}\label{pareto2}
Let Assumptions \ref{assum0}, \ref{assumption-pareto} be satisfied. 

Consider an equilibrium. Denote, for each $t\geq 1$, $
Q_t\equiv \frac{1}{R_1\cdots R_t},  P_t\equiv \frac{ G_n^t}{R_1\cdots R_t}.$ 
This equilibrium is not Pareto optimal if and only if there exist a feasible allocation $(c^{y'}_t,c^{o'}_t)_t$ which Pareto dominates the allocation  $(c^y_t,c^o_t)_t$ and a date $t_0$ such that
\begin{subequations}\label{condition-epsilon}
\begin{align}
&\epsilon^y_t\equiv c^{y'}_t-c^{y}_t<0,  \epsilon^o_t\equiv c^{o'}_t-c^{o}_t=-G_n\epsilon^y_t>0   \text{ } \forall t\geq t_0 \text{ and }  \epsilon^y_t=\epsilon^o_t=0 \text{ }\forall t< t_0,\\
&Q_{t+1}\epsilon^o_{t+1}>\frac{1}{G_n}Q_{t}\epsilon^o_t  \quad (i.e., P_{t+1}\epsilon^o_{t+1}>P_{t}\epsilon^o_t) \text{ }\forall t\geq t_0-1.
\end{align}
\end{subequations}
\end{lemma}
\begin{proof}
The sufficient condition $(\Leftarrow)$ is obvious.  We prove the necessary condition: $(\Rightarrow)$.  

Suppose that $(c^y_t,c^o_t)_t$ is not Pareto optimal. Then, there exists a feasible allocation $(c^{y'}_t,c^{o'}_t)_t$ which Pareto dominates the allocation  $(c^y_t,c^o_t)_t$. By consequence, there exists a date $s$ such that $U^s(c^{y'}_s,c^{o'}_{s+1})>U^s(c^{y}_s,c^{o}_{s+1})$. This allows us to define 
\begin{align}
t_0\equiv \min\{s: c^{o'}_s\not=c^{o}_s\}.
\end{align}
By the definition of $t_0$, we have $c^{y'}_t=c^{y}_t, c^{o'}_t=c^{o}_t$ $\forall t\leq t_0-1$. 

Since $(c^{y'}_t,c^{o'}_t)_t$ Pareto dominates the allocation  $(c^y_t,c^o_t)_t$, we have 
\begin{align}
U^{t_0-1}(c^{y'}_{t_0-1},c^{o'}_{t_0})\geq U^{t_0-1}(c^{y}_{t_0-1},c^{o}_{t_0}).
\end{align}
Recall that $c^{y'}_{t_0-1}=c^{y}_{t_0-1}$, $c^{o'}_{t_0}\not=c^{o}_{t_0}$, and the function $U^{t_0-1}$ is strictly increasing in each component,  we have $c^{o'}_{t_0}>c^{o}_{t_0}$ and, hence, $c^{y'}_{t_0}<c^{y}_{t_0}$ because $c_{t}^{y}+\frac{c_{t}^{o}}{G_n}=c_{t}^{y'}+\frac{c_{t}^{o'}}{G_n}=e_{t}^{y}+\frac{e_{t}^{o}}{G_n}+d_{t}$.

Now, since $c^{y'}_{t_0}<c^{y}_{t_0}$ and $U^{t_0}(c^{y'}_{t_0},c^{o'}_{t_0+1})\geq U^{t_0}(c^{y}_{t_0},c^{o}_{t_0+1})$, we get that $c^{o'}_{t_0+1}>c^{o}_{t_0+1}$. By induction, we obtain conditions (\ref{condition-epsilon}).

It remains to prove that $Q_{t+1}\epsilon^o_{t+1}>\frac{1}{G_n}Q_{t}\epsilon^o_t  \text{ }\forall t\geq t_0-1.$

Observe that, by the non-arbitrage condition $R_{t+1}=\frac{q_{t+1}+D_{t+1}}{q_t}$, the budget constraint of households $t$ can be rewritten as
$Q_tc^y_t+Q_{t+1}c^o_{t+1}=Q_te^y_t+Q_{t+1}e^o_{t+1}$. Therefore, we have $
Q_tc^{y'}_t+Q_{t+1}c^{o'}_{t+1}=
Q_te^y_t+Q_{t+1}e^o_{t+1}+ (Q_t\epsilon^y_t+Q_{t+1}\epsilon^o_{t+1}).$

Consider $t\geq t_0-1$. We have $U^t(c^{y'}_t,c^{o'}_{t+1})\geq U^t(c^{y}_t,c^{o}_{t+1})$.  Suppose that $Q_t\epsilon^y_t+Q_{t+1}\epsilon^o_{t+1}\leq 0$, we have $Q_tc^{y'}_t+Q_{t+1}c^{o'}_{t+1}\leq 
Q_te^y_t+Q_{t+1}e^o_{t+1}$. 
Recall that $U^t(c^{y}_t,c^{o}_{t+1})$ is the maximum value of the maximization problem of household $t$. By consequence, $U^t(c^{y'}_t,c^{o'}_{t+1})$ is also the maximum value. This implies that $(c^{y'}_t,c^{o'}_{t+1})$ is a solution to the maximization problem of agent $t$. However, since the function $U^t$ is strictly quasi-concave, the solution is unique. Therefore, we have $(c^{y'}_t,c^{o'}_{t+1})=(c^{y}_t,c^{o}_{t+1})$, a contradiction. To sum up, we get that $Q_t\epsilon^y_t+Q_{t+1}\epsilon^o_{t+1}>0$. 
\end{proof}

\begin{proof}[{\bf  Proof of Theorem \ref{pareto-theorem}}]
{\bf Part 1 ("if" part)}. Suppose that  $(c^y_t,c^o_t)_t$ is not Pareto optimal. Applying Lemma \ref{pareto2},  there exist a feasible allocation $(c^{y'}_t,c^{o'}_t)_t$ which Pareto dominates the allocation  $(c^y_t,c^o_t)_t$ and a date $t_0$ such that
\begin{subequations}
\begin{align*}
\epsilon^y_t&\equiv c^{y'}_t-c^{y}_t<0,\quad  \epsilon^o_t\equiv c^{o'}_t-c^{o}_t=-G_n\epsilon^y_t>0  \text{ } \forall t\geq t_0 \text{ and } \epsilon^y_t=\epsilon^o_t=0 \text{ }\forall t< t_0,\\
Q_{t+1}\epsilon^o_{t+1}&>\frac{1}{G_n}Q_{t}\epsilon^o_t  \quad (i.e., P_{t+1}\epsilon^o_{t+1}>P_{t}\epsilon^o_t) \text{ }\forall t\geq t_0-1.
\end{align*}
\end{subequations}
So, $(c^{y'}_t,c^{o'}_{t+1})\in B_t(c)$ and $c^{y'}_t<c^{y}_t.$ 

Let $h$ as in Definition \ref{boz-condition}(i). Without loss of generality, we can assume that $h\in (0,1)$. We define the sequence $(x^y_t,x^o_t)_t$ by $x^y_t\equiv c^y_t+h\epsilon^y_t, x^o_{t}\equiv c^o_t+h\epsilon^o_t.$ Then, we have $x^y_t=(1-h)c^y_t+hc^{y'}_t>(1-h)c^y_t$, and $x^o_{t}=(1-h)c^o_t+hc^{o'}_t.$ Since the function $U^t$ is strictly concave, we have 
$$
U^t(x^y_t,x^o_{t+1})>(1-h)U^t(c^y_t,c^o_{t+1})+hU^t(c^{y'}_t,c^{o'}_{t+1})\geq U^t(c^y_t,c^o_{t+1}).
$$
By the uniform strictness condition in Theorem \ref{pareto-theorem}, there  exists $\mu$ such that 
\begin{align*}
P_{t+1}(x^{o}_{t+1}-c^{o}_{t+1})+G_nP_t(x^{y}_t-c^{y}_t)&\geq\frac{\mu}{P_tc^y_t}\big(G_n P_t(x^{y}_t-c^{y}_t)\big)^2\\
\Leftrightarrow P_{t+1}\epsilon^{o}_{t+1}&\geq P_{t}\epsilon^{o}_{t}+\frac{h\mu}{P_tc^y_t}\big(P_t\epsilon^{o}_{t}\big)^2 \Rightarrow P_{t+1}\epsilon^{o}_{t+1}\geq P_{t}\epsilon^{o}_{t}\big(1+h\mu\frac{P_t\epsilon^{o}_{t}}{P_te_t}\big),
\end{align*}
where recall that $e_t\equiv e^y_t+\frac{e^o_t}{G_n}+d_t>c^y_t$. By consequence, we have 
\begin{align*}
\frac{1}{P_{t+1}\epsilon^{o}_{t+1}}\leq \frac{1}{P_t\epsilon^{o}_{t}\big(1+\frac{h\mu P_t\epsilon^{o}_{t}}{P_te_t}\big)}=\frac{1}{P_t\epsilon^{o}_{t}}\Big(1-\frac{\frac{h\mu P_t\epsilon^{o}_{t}}{P_te_t}}{1+\frac{h\mu P_t\epsilon^{o}_{t}}{P_te_t}}\Big)\\
\Rightarrow \frac{1}{P_t\epsilon^{o}_{t}}-\frac{1}{P_{t+1}\epsilon^{o}_{t+1}}\geq  \frac{1}{P_t\epsilon^{o}_{t}}\frac{\frac{h\mu P_t\epsilon^{o}_{t}}{P_te_t}}{1+\frac{h\mu P_t\epsilon^{o}_{t}}{P_te_t}}=\frac{h\mu}{P_te_t\big(1+h\mu\frac{P_t\epsilon^{o}_{t}}{P_te_t}\big)}.
\end{align*}
Since $\epsilon^o_t\leq G_n e_t$ $\forall t$, we get that $\frac{P_t\epsilon^{o}_{t}}{P_te_t}\leq G_n$ and hence
\begin{align*}
\frac{1}{P_t\epsilon^{o}_{t}}-\frac{1}{P_{t+1}\epsilon^{o}_{t+1}}\geq \frac{1}{P_te_t}\frac{h\mu}{1+h\mu G_n}.
\end{align*}
Taking the sum over $t$, we have $\sum_{t\geq 1}\frac{1}{P_te_t}<\infty$, a contradiction. Therefore, the equilibrium allocation is Pareto optimal.\\
{\bf Part 2 ("only if" part)}.   Let conditions in part 2 be satisfied. Then there exist $\underline{x}>0, \bar{x}\in (0,1),y>0 $   such that 
$c^y_t>\underline{x}e_t,  c^o_t<\bar{x}G_n e_t, P_{t+1}c^o_{t+1}>yP_{t}e_{t}.$

Suppose that $\sum_{t\geq 1}\frac{1}{P_te_t}<\infty$. Then, there exists $M$ such that $\sum_{t=1}^T\frac{1}{P_te_t}<M$ $\forall T$.

For $h>0$, define the sequence $(\epsilon_t)_t$ by 
\begin{align}\label{epsilont_definition}
\epsilon_t\equiv \frac{P_1e_1\epsilon_1}{P_te_t}+\frac{h}{P_te_t}\Big(\frac{1}{P_1e_1}+\cdots +\frac{1}{P_{t-1}e_{t-1}}\Big) \text{ }\forall t\geq 2,
\end{align}
and $\epsilon_1>0$. Since  $\sum_{t\geq 1}\frac{1}{P_te_t}<\infty$, we have $\lim_{t\to\infty}\frac{1}{P_te_t}=0$ and $\lim_{t\to\infty}\epsilon_t=0$. So, we can take $\epsilon_1>0$ and $h>0$ small enough and $x\in (0,\bar{x})$ so that
\begin{align}
c^y_t-\frac{1}{G_n}\epsilon_te_t>xe_t, \quad c^o_t+\epsilon_te_t<G_n e_t \text{ } \forall t.
\end{align}
Let $\lambda\in (0,1)$. Define 
\begin{align}
\epsilon^o_t&\equiv \lambda \epsilon_te_t, \quad &c^{o'}_t&\equiv c^o_t+\epsilon^o_t, 
&\epsilon^y_t&\equiv -\frac{1}{G_n}\epsilon^o_t, \quad &c^{y'}_t&\equiv c^y_t+\epsilon^y_t.
\end{align}
It is clear that the allocation $(c^{y'}_t,c^{o'}_{t})_t$ is feasible. We have 
$$
c^{y'}_t=c^y_t+\epsilon^y_t=c^y_t-\frac{1}{G_n}\lambda\epsilon_te_t>xe_t >xc^y_t\text{ and } 
c^{o'}_t\equiv c^o_t+\lambda\epsilon_te_t<c^o_t+\epsilon_te_t<G_n e_t \text{ } \forall t.
$$
 
By Definition (\ref{epsilont_definition}) of $\epsilon_t$, we have $P_t\epsilon^o_t-P_1\epsilon^o_1=\lambda h\Big(\frac{1}{P_1e_1}+\cdots +\frac{1}{P_{t-1}e_{t-1}}\Big)$ $\forall t\geq 2$. This implies that 
$$
P_{t+1}\epsilon^o_{t+1}-P_t\epsilon^o_t=\frac{\lambda h}{P_te_t} \text{ and }\frac{P_{t+1}\epsilon^o_{t+1}-P_t\epsilon^o_t}{\frac{(P_t\epsilon^o_t)^2}{P_tc^y_t}}=\lambda h\frac{P_tc^y_t}{P_te_t}\frac{1}{(P_t\epsilon^o_t)^2}.
$$
We have $\frac{P_tc^y_t}{P_te_t}=\frac{c^y_t}{e_t}\geq \underline{x}$ and 
\begin{align*}
P_t\epsilon^o_t=\lambda P_1e_1\epsilon_1+\lambda h\Big(\frac{1}{P_1e_1}+\cdots +\frac{1}{P_{t-1}e_{t-1}}\Big)<\lambda(P_1e_1\epsilon_1+hM).
\end{align*}
Therefore, we get that
\begin{align}
\notag \frac{1}{2}\frac{P_{t+1}\epsilon^o_{t+1}-P_t\epsilon^o_t}{\frac{(P_t\epsilon^o_t)^2}{P_tc^y_t}}&\geq \frac{1}{2}\lambda h\underline{x}\frac{1}{(\lambda(P_1e_1\epsilon_1+hM))^2}= \frac{1}{2\lambda}\frac{h\underline{x}}{(P_1e_1\epsilon_1+hM)^2}\\
\label{1/21}\Rightarrow \frac{1}{2}\big(P_{t+1}\epsilon^o_{t+1}-P_t\epsilon^o_t\big)&\geq \frac{1}{2\lambda}\frac{h\underline{x}}{(P_1e_1\epsilon_1+hM)^2} \frac{(P_t\epsilon^o_t)^2}{P_tc^y_t} \text{ } \forall t.
\end{align}
We also have
\begin{align}\notag
\frac{1}{2}\frac{P_{t+1}\epsilon^o_{t+1}-P_t\epsilon^o_t}{\frac{(P_{t+1}\epsilon^o_{t+1})^2}{P_{t+1}c^o_{t+1}}}&=\frac{1}{2}\lambda h\frac{P_{t+1}c^o_{t+1}}{P_te_t}\frac{1}{(P_{t+1}\epsilon^o_{t+1})^2}\geq \frac{1}{2}\lambda h y\frac{1}{\lambda^2(P_1e_1\epsilon_1+hM)^2}\\
\label{1/22}\Rightarrow \frac{1}{2}\big(P_{t+1}\epsilon^o_{t+1}-P_t\epsilon^o_t\big)&\geq  \frac{1}{2\lambda} \frac{h y}{(P_1e_1\epsilon_1+hM)^2} \frac{(P_{t+1}\epsilon^o_{t+1})^2}{P_{t+1}c^o_{t+1}}.
\end{align}
Since we  can choose $\lambda>0$ arbitrarily small, we choose $\lambda$ so that 
\begin{align}
\frac{1}{2\lambda} \frac{h y}{(P_1e_1\epsilon_1+hM)^2}>\theta_2(x), \quad \frac{1}{2\lambda}\frac{h\underline{x}}{(P_1e_1\epsilon_1+hM)^2}>\frac{\theta_1(x)}{G_n}.
\end{align}
where $\theta_1(x), \theta_2(x)$ are defined in Definition \ref{boz-condition}(ii).

From (\ref{1/21}), (\ref{1/22}), by noting that $c^{y'}_t-c^y_t=-\epsilon^o_t/G_n$, we have \begin{align*}
  &P_{t+1}(c^{o'}_{t+1}-c^{o}_{t+1})+G_nP_t(c^{y'}_t-c^{y}_t)\\
  =&P_{t+1}\epsilon^o_{t+1}-P_t\epsilon^o_t\geq\frac{{\theta}_2(x)}{P_{t+1}c^o_{t+1}}\big(P_{t+1}(c^{o'}_{t+1}-c^{o}_{t+1})\big)^2+\frac{{\theta}_1(x)}{G_nP_tc^y_t}\big(G_nP_t(c^{y'}_t-c^{y}_t)\big)^2.  
\end{align*}

Recall that $0<xc^y_t<c^{y'}_t<c^{y}_t$ and $c^{o}_{t+1}<c^{o'}_{t+1}<G_ne_{t+1}$. By the uniform smoothness condition, we have $(c^{y'}_t,c^{o'}_{t+1})\in B_t(c)$, i.e., $U^t(c^{y'}_t,c^{o'}_{t+1})\geq U^t(c^{y}_t,c^o_{t+1})$ $\forall t$.

For $\lambda\in (0,1)$, we define the allocation $(x^y_t,x^o_t)_t$ by $x^y_t\equiv \lambda c^{y'}_t+(1-\lambda)c^{y}_t>0 $ and $x_{t+1}^o\equiv \lambda c^{o'}_{t+1}+(1-\lambda)c^o_{t+1}>0$. Of course, $(x^y_t,x^o_t)$ is feasible. Since the function $U^t$ is strictly concave, we have $
U^t\big(\lambda c^{y'}_t+(1-\lambda)c^{y}_t,\lambda c^{o'}_{t+1}+(1-\lambda)c^o_{t+1}\big)>\lambda U^t(c^{y'}_t,c^{o'}_{t+1})+(1-\lambda)U^t(c^{y}_t,c^o_{t+1})\geq U^t(c^{y}_t,c^o_{t+1}).$
Therefore, the allocation $(x^{y}_t,x^{o}_{t})_t$ dominates $(c^{y}_t,c^{o}_{t})_t$ in the sense of Pareto, a contradiction. So, we have $\sum_{t\geq 1}\frac{1}{P_te_t}=\infty$.

\end{proof}

\begin{remark}\label{optimal-propertyc'} Part \ref{pareto-theorem-part1} of Theorem \ref{pareto-theorem} remains valid when the uniform strictness condition is replaced by Property (C'), using the terminology of  \citet[Proposition 5.6]{bs80_appendix}.
\\
{\bf Property (C')}. We say that the allocation $(c^{y}_t,c^{o}_{t+1})$ satisfies the property (C') if there exists $\alpha>0$ such that, for any $t$, if the couple $(c^{y'}_t,c^{o'}_{t+1})\in \rr_{++}^2$ satisfies 
\begin{subequations}
\begin{align}&U^t(c^{y'}_t,c^{o'}_{t+1})\geq U^t(c^{y}_t,c^{o}_{t+1}), \quad 
\epsilon^y_t\equiv c^{y'}_t-c^{y}_t<0,  \epsilon^o_{t+1}\equiv c^{o'}_{t+1}-c^{o}_{t+1}>0,\\
&\frac{c^{o'}_{t+1}}{G_n}<e_{t+1}^{y}+\frac{e_{t+1}^{o}}{G_n}+d_{t+1}, \quad P_{t+1}\epsilon^o_{t+1}-P_{t}\epsilon^o_{t}>0, \text{ where  } \epsilon^o_{t}\equiv -G_n\epsilon^y_t,   
\end{align}
\end{subequations}
then $(P_{t+1}\epsilon^o_{t+1})^2\leq \alpha P_{t+1}c^{o}_{t+1} \big(P_{t+1}\epsilon^o_{t+1}-P_{t}\epsilon^o_{t}\big).$
\end{remark}

\begin{proof}
Suppose that  $(c^y_t,c^o_t)_t$ is not Pareto optimal. We can take the allocation $(c^{y'}_t,c^{o'}_t)_t$ as in the proof of Theorem \ref{pareto-theorem}. So, the couple $(c^{y'}_t,c^{o'}_t) \in \rr_{++}^2$ satisfies Property (C'). Therefore, we have
 \begin{align}(P_{t+1}\epsilon^o_{t+1})^2\leq \alpha P_{t+1}c^{o}_{t+1} \big(P_{t+1}\epsilon^o_{t+1}-P_{t}\epsilon^o_{t}\big).
\end{align}
Denote  $\mu_t\equiv P_{t+1}\epsilon^o_{t+1}-P_{t}\epsilon^o_{t}>0$. Since $\mu_t=0$ $\forall t<  t_0-1$, and $\mu_t>0$ $\forall t\geq t_0-1$, we have $\mu_{t_0-1}+\cdots +\mu_{t}=P_{t+1}\epsilon^o_{t+1}$. This implies that
\begin{align}
\epsilon^o_{t+1}=\frac{\mu_{t_0-1}+\cdots +\mu_{t}}{P_{t+1}} \text{ }\forall t\geq t_0-1.
\end{align}

We have
\begin{align*}\frac{(P_{t+1}\epsilon^o_{t+1})^2}{P_{t+1}\epsilon^o_{t+1}-P_{t}\epsilon^o_{t}}=\frac{(P_{t+1}\epsilon^o_{t+1})^2}{\mu_t}=\frac{(\mu_{t_0-1}+\cdots +\mu_{t}) ^2}{\mu_t}.
\end{align*}
This implies that, for any $t\geq t_0,$
\begin{align*}
\frac{1}{P_{t+1}c^o_{t+1}}&\leq \alpha \frac{P_{t+1}\epsilon^o_{t+1}-P_{t}\epsilon^o_{t}}{(P_{t+1}\epsilon^o_{t+1})^2}=\alpha\frac{\mu_t}{(\mu_{t_0-1}+\cdots +\mu_{t}) ^2}\\
&<\alpha \frac{\mu_t}{(\mu_{t_0-1}+\cdots +\mu_{t})(\mu_{t_0-1}+\cdots +\mu_{t-1})}\\
&=\alpha\Big(\frac{1}{\mu_{t_0-1}+\cdots +\mu_{t-1}}-\frac{1}{\mu_{t_0-1}+\cdots +\mu_{t}}\Big).
\end{align*}
By taking the sum over $t$ from $t_0-1$ until $T-1$ of this inequality, we have 
\begin{align}
\frac{1}{P_{t_0}c^o_{t_0}}+\cdots + \frac{1}{P_{T}c^o_{T}}\leq \alpha \Big(\frac{1}{\mu_{t_0-1}}-\frac{1}{\mu_{t_0-1}+\cdots + \mu_{T-1}}\Big)< \frac{\alpha}{\mu_{t_0-1}}.
\end{align}Therefore, we have $\sum_{t\geq 1}\frac{1}{P_{t}c^o_t}<\infty$. Combining with $c^o_t\leq G_ne_t$ $\forall t$,  we get that $\sum_{t\geq 1}\frac{1}{P_{t}e_t}<\infty$, a contradiction. As a result, the equilibrium allocation  $(c^y_t,c^o_t)_t$ is Pareto optimal.

\end{proof}



\begin{proof}[{\bf Proof of Proposition \ref{explicit-1}}]
First, since $e^o_t=0$ $\forall t$, the benchmark interest rate equals zero, i.e., $R^*_t=0 \text{  } \forall  t.$  It is easy to see that there exists a unique equilibrium determined by $a_t=\frac{\beta}{1+\beta} e^y_t$. The interest rate sequence $(R_t)$  is determined by
\begin{align}
R_{t+1}&=G_n\frac{a_{t+1}+d_{t+1}}{a_t}=G_n\frac{\frac{\beta}{1+\beta} e^y_{t+1}+d_{t+1}}{\frac{\beta}{1+\beta} e^y_t}.
\end{align}
According to Definition \ref{definition-interest}, $R_{t+1}\equiv \frac{1}{\beta}\frac{c_{t+1}^{o}}{c_{t}^{y}}$. Then, we can find 
\begin{align}\label{allocation-gamma-t}
c^y_{t}=\frac{1}{1+\beta} e^{y}_t, \quad c^{o}_{t+1}= G_n(\frac{\beta}{1+\beta}  e^y_{t+1}+d_{t+1}).
\end{align}
By Lemma \ref{prop1}\ref{prop1-5}, the equilibrium is bubbly if and only if $\sum_{t\geq 1}\frac{d_t}{a_t}<\infty$, which is equivalent to $\sum_{t\geq 1}\frac{d_t}{e^y_t}<\infty$. We consider two cases.
\begin{enumerate}

\item If $\sum_{t\geq 1}\frac{d_t}{e^y_t}<\infty$ (i.e., $\sum_{t=1}^{\infty}\frac{D_t}{G_n^te^y_t}<\infty$),
this equilibrium is bubbly.  Then, by using Lemma \ref{prop1}\ref{prop1-4}, we have $\lim_{t\to\infty}\frac{b_t}{e^y_t}=\lim_{t\to\infty}\frac{b_t}{a_t}\frac{a_t}{e^y_t}=\frac{\beta}{1+\beta}$. It means that the equilibrium is asymptotically bubbly.

By Lemma \ref{check-strict}\ref{check-strict-part2}, the equilibrium allocation given by (\ref{allocation-gamma-t}) satisfies the uniform strictness condition.  Moreover, 
since $d_t/e^y_t\to 0$, we have $a_t/e_t\to \beta/(1+\beta)>0$. Thus, by applying Proposition \ref{efficient-bubbleless-saving}\ref{efficient-bubbleless-saving2}, this equilibrium is Pareto optimal.

\item If $\sum_{t\geq 1}\frac{d_t}{e^y_t}=\infty$ (i.e., $\sum_{t=1}^{\infty}\frac{D_t}{G_n^te^y_t}=\infty$), this equilibrium is bubbleless. By Proposition \ref{efficient-bubbleless-saving}\ref{efficient-bubbleless-saving1}, this equilibrium is Pareto optimal.
\end{enumerate}
\end{proof}

\end{document}